\documentclass[10pt,journal,twoside]{IEEEtran}

\usepackage{multicol}
\usepackage{multirow, array}
\usepackage{amsthm,amsmath,amssymb}
\usepackage{mathrsfs}
\usepackage{cases}

\usepackage{amssymb}
\usepackage[dvips]{graphicx}
\usepackage{amsmath}
\usepackage{amsthm}
\usepackage{latexsym,bm}
\usepackage{color}
\usepackage{subfigure}
\usepackage{longtable}
\usepackage{accents}
\usepackage{cite}
\usepackage{enumerate}
\usepackage{arydshln}
\usepackage[ruled]{algorithm}
\usepackage{algorithmic}
\usepackage{diagbox}

\def\bc{{\mathbf{c}}}   
\def\bd{{\mathbf{d}}}   \def\bx{{\mathbf{x}}}
  \def\bs{{\mathbf{s}}} \def\by{{\mathbf{y}}}
   
\def\bg{{\mathbf{g}}} 

 \def\bH{{\mathbf{H}}}  
\def\bB{{\mathbf{B}}}   
   \def\bW{{\mathbf{W}}}
   \def\bX{{\mathbf{X}}}
 \def\bL{{\mathbf{L}}}

\makeatletter
\def\widebar{\accentset{{\cc@style\underline{\mskip10mu}}}}
\def\Widebar{\accentset{{\cc@style\underline{\mskip8mu}}}}
\makeatother

\allowdisplaybreaks

\newcommand{\tabincell}[2]{\begin{tabular}{@{}#1@{}}#2\end{tabular}}
\theoremstyle{plain}

\theoremstyle{definition}
\theoremstyle{definition}

\begin{document}

\title{{\Huge Protograph Bit-Interleaved Coded Modulation: A Bandwidth-Efficient Design Paradigm for 6G Wireless Communications\vspace{-0mm}}
\thanks{Y.~Fang is with the School of Information Engineering, Guangdong University
of Technology, Guangzhou 510006, China, and also with the State Key Laboratory of Integrated Services Networks, Xidian University, Xi'an 710126, China~(email: fangyi@gdut.edu.cn).}
\thanks{P.~Chen is with the Department of Electronic Information, Fuzhou
University, Fuzhou 350116, China (e-mail: ppchen.xm@gmail.com). {\em Corresponding author: Pingping Chen.}}
\thanks{Y.~L.~Guan is with the School of Electrical and Electronic
Engineering, Nanyang Technological University, Singapore 639798~(e-mail:
eylguan@ntu.edu.sg).}
\thanks{F.~C.~M.~Lau is with the Department of Electronic and Information
Engineering, Hong Kong Polytechnic University, Hong Kong~(e-mail:
encmlau@polyu.edu.hk).}
\thanks{Y.~Li Li is with the School of Electrical and Information Engineering,
The University of Sydney, Sydney, NSW 2006, Australia~(e-mail:
yonghui.li@sydney.edu.au).}
\thanks{G.~Chen is with the Department of Electronic Engineering, City University
of Hong Kong, Hong Kong~(e-mail: eegchen@cityu.edu.hk).}}
\author{Yi Fang, {\em Senior Member, IEEE}, Pingping Chen, {\em Member, IEEE}, Yong Liang Guan, {\em Senior Member, IEEE}, \\Francis C. M. Lau, {\em Fellow, IEEE}, Yonghui Li, {\em Fellow, IEEE}, and Guanrong Chen, {\em Life Fellow, IEEE}\vspace{-9mm}}

%\markboth{Proceedings of the IEEE, Vol.~XX, No.~XX, Month Year}%
%{Fang \MakeLowercase{\textit{et al.}}: PB-LDPC-Coded BICM and Its Spatially-Coupled Counterpart: State-of-the-Art and Future Challenges}

\maketitle

\begin{abstract}

Bit-interleaved coded modulation (BICM) has attracted considerable attention from the research community in the past three decades, because it can achieve desirable error performance with relatively low implementation complexity for a large number of communication and storage systems.
By exploiting the iterative demapping and decoding (ID), the BICM is able to approach capacity limits of coded modulation over various channels.
In recent years, protograph low-density parity-check (PLDPC) codes and their spatially-coupled (SC) variants have emerged to be a pragmatic forward-error-correction (FEC) solution for BICM systems due to their tremendous error-correction capability and simple structures, and found widespread applications such as deep-space communication, satellite communication, wireless communication, optical communication, and data storage.
This article offers a comprehensive survey on the state-of-the-art development of PLDPC-BICM and its innovative SC variants over a variety of channel models, e.g., additive white Gaussian noise (AWGN) channels, fading channels, Poisson pulse position modulation (PPM) channels, and flash-memory channels. Of particular interest is code construction, constellation shaping, as well as bit-mapper design, where the receiver is formulated as a serially-concatenated decoding framework consisting of a soft-decision demapper and a belief-propagation decoder. Finally, several promising research directions are discussed, which have not been adequately addressed in the current literature.
\end{abstract}

\providecommand{\keywords}[1]{\textbf{\textit{Index terms---}} #1}

\vspace{-1mm}

%\newpage

\section{Introduction}\label{sect:I}

\subsection{Overview of PLDPC-BICM}\label{sect:I-A}\vspace{-0.1mm}

To provide high-reliability and high-throughput transmission and storage is an extremely challenging task in modern bandwidth-limited communication and storage systems \cite{8403963,7403840}. In $1948$, Shannon published a piece of milestone work, which revealed and proved that information (data) can be reliably transmitted over a noisy channel by using forward-error-correction (FEC) codes with an arbitrarily high rate below the {\em channel capacity} (also called the {\it maximum achievable rate}) \cite{6773024}. %\footnote{In this article, both ``channel capacity" and ``maximum achievable rate" are indiscriminately used to represent the fundamental upper-limit on the spectral efficiency of an arbitrary coded-modulation scheme over a noisy channel.}
Since then, the research field of channel coding has been created, and a great deal of effort has been devoted to finding coding mechanisms that can strongly resist channel impairments and closely approach the channel capacity \cite{6772729,Gallager63}. In fact, FEC codes have stood out as a more promising choice to implement reliable communication systems compared with other error-correction techniques (e.g., data retransmission techniques), especially for low-latency and low-power applications \cite{8673568,4282117,4390033}.

Aiming to improve the spectral efficiency, there has been increasing demand for incorporating high-order modulation into FEC code design because it can transform several binary bits to a non-binary symbol through a high-order constellation \cite{1053909}. As a pioneering work on the coded modulation (CM), the channel code and high-order modulation were properly combined into a single entity to improve the spectral efficiency of communication systems \cite{massey1974coding}. In this direction, great research efforts have been made to promote the development of the joint coding-and-modulation design. Specifically, trellis-coded modulation (TCM), which combines the trellis codes with phase-shift-keying/quadratic-amplitude modulations (PSK/QAM), was conceived in \cite{1056454} to improve the transmission efficiency over additive white Gaussian noise (AWGN) channels. Later on, a symbol-level interleaver was added between the encoder and the modulator to improve the performance of TCM over fading channels \cite{7512}. As an alternative to TCM, multi-level CM (MLCM) was developed, which exploits different convolutional/turbo codes to protect a modulated symbol \cite{1055718,771140,8616866}. In particular, an individual binary code in the MLCM scheme was utilized to enhance the transmission reliability of each labeling bit within a modulated symbol. However, both TCM and MLCM schemes cannot maximize the minimum Hamming distance of an FEC code, which results in a non-trivial gap to the channel capacity in fading scenarios.

To overcome the aforementioned weakness, a great deal of research effort has been devoted to developing more robust combinational solutions that can adequately exploit the potential advantages of coding and modulation. In $1993$, a breakthrough in coding theory was marked by the inception of turbo codes \cite{397441}. The turbo code, which concatenates two individual convolutional codes through a bit-level interleaver in parallel, is capable of accomplishing performance very close to the channel capacity with iterative decoding \cite{johnson2010iterative}. Inspired by the fundamental work in \cite{397441}, it was re-discovered \cite{748992} that the low-density parity-check (LDPC) codes can also approach the channel capacity under iterative decoding. More importantly, the notable success of ``{\em turbo principle}" opened up an important direction to design near-capacity bandwidth-efficient CM systems. In $1992$, a novel CM scheme was proposed in \cite{141453}, referred to as {\em bit-interleaved coded modulation (BICM)}, which is a serial concatenation of a channel code, a bit-level interleaver, and a high-order modulation. It can achieve higher diversity order and better performance than the conventional TCM over fading channels. Then, a thorough information-theoretic study on BICM was carried out in \cite{669123} in terms of channel-capacity and asymptotic-performance analysis. It was also demonstrated in \cite{669123} that using soft-decision demodulation (i.e., demapping) could help reasonably compensate the performance loss incurred by separating coding and modulation in BICM. In fact, the intrinsic serial concatenated structure of BICM resembles a turbo code, so using iterative demapping and decoding (ID) may yield an additional performance gain. Motivated by this feature, effective ID algorithms with hard-decision and soft-decision feedbacks for convolutional-based BICM, were developed in \cite{649929} and \cite{706217} respectively, leading to BICM-ID frameworks, which reduce the performance loss caused by independent demapping. In general, the soft-decision-aided BICM-ID performs better than the hard-decision-aided BICM-ID because soft information carries more reliable message of the transmitted symbols compared with hard information. More precisely, it is possible to achieve capacity-approaching performance for BICM-ID with the aid of a powerful channel code, an optimized bit interleaver (i.e., bit mapper), and a carefully chosen constellation \cite{schreckenbach2007iterative,i2008bit,franceschini2009ldpc}.

However, the convolutional-based BICM is unable to have good near-capacity performance even with a sophisticated design \cite{8053803}. As a remedy, more powerful FEC codes (e.g., turbo codes and LDPC codes \cite{richardson2008modern,957394,8616900}) were developed for BICM systems, showing that they can operate very close to the channel capacity under some specific constellations \cite{6522460,boutros2003turbo,haykin2003turbo,975756,hou2002design,1226601,1194444,1291808}.\footnote{``BICM" will be used to collectively represent both ``BICM-ID" and ``BICM with non-iterative decoding (BICM-NI)" in this article, although it was initially utilized to represent ``BICM-NI" only.} In fact, LDPC codes possess better error performance and lower decoding complexity than turbo codes, and thus were recognized as one of the most powerful FEC codes that enable error-free transmissions with rates up to the channel capacity for communication and storage systems since the late $1990$s \cite{910577,1267048,liva2006design,6517051,4383367,7112076, 8269289}. Due to its flexible design and excellent performance, BICM has been further deployed in a wide range of practical applications and industry standards, e.g., wireless local area networks, digital video broadcasting-satellite-second generation (DVB-S$2$), and the fourth/fifth-generation (i.e., $4$G/$5$G) mobile networks, thereby marking a paradigmatic shift in the CM field \cite{7383286,924878,4787602,7534851}.

With the continuous development of analysis tools and construction methodologies \cite{1347354}, LDPC codes have become a prospective type of FEC codes to formulate pragmatic BICM systems, which not only can achieve high throughput but also enable satisfactory performance over different types of channels, including AWGN channels, fading channels, Poisson pulse position modulation (PPM) channels, multi-level-cell (MLC) flash memory channels, and underwater acoustic channels \cite{7047772,7801862,7105928,4558590,6663748,8873472,8013174,8314735,8052124,5073425,5887355,9056068}. According to the salient feature of BICM, a large number of theoretic analysis methodologies were developed, such as density evolution (DE) \cite{1226601,910577,1532194}, extrinsic information transfer (EXIT) function \cite{6359874,8338131}, asymptotic weight enumerator (AWE) \cite{schreckenbach2007iterative,5174517,6785991,7152893}, and harmonic mean of minimum squared Euclidean distance (HMMSED) \cite{669123,924878,6064854,6515491}, to characterize the asymptotic performance and to facilitate optimization of the system. Based on these theoretical advancements, great endeavor has been dedicated to improving the performance and reducing the complexity of the LDPC-BICM from the perspectives of code construction \cite{6663748,8873472,8024788,8740906,6205592,8708250,8620336,8906157}, constellation shaping \cite{6359874,5598322,8878166,1413232,8186234,6241383}, bit-mapper optimization \cite{5191353,8170954,6133952,7354659,4525723,8999517,5199551,7801862,8611290,6952165,5960810}, as well as receiver design \cite{i2008bit,5887355,7095513,7956256,7589683,6942236,6204017,tan2016bicm,7887724}. Although several capacity-approaching LDPC-BICM systems have been devised in recent years, the irregular structures of the designed codes suffer from difficult implementation, which becomes a major obstacle in practical applications. %Besides, the traditional LDPC codes impose a performance trade-off between the low and high signal-to-noise ratio (SNR) regions. For example,
Besides, the capacity-approaching LDPC codes and its BICM relatives always have undesirable high-SNR performance, making them unfavorable for the applications with extremely low bit-error-rate (BER) requirement, such as data storage and optical communication.

To circumvent the above-mentioned limitations, structured LDPC codes, such as protograph LDPC (PLDPC) codes and quasi-cyclic (QC) LDPC codes, have emerged as a competitive component in BICM systems, which can achieve outstanding performance with simple design and easy hardware implementation \cite{6497022,7080854,7157697,8291725,8798970,6552958,7398027,8651338,8750852}. Particularly, these two types of LDPC codes possess linear encoding and decoding complexity, and hence attracted increasing interest from both academic and industrial communities \cite{8314100,8594605,8477009,8316763,secretariat2017ccsds,access2008multiplexing,TcomRCA}. In fact, the QC-LDPC codes can be considered as a special version of the PLDPC codes, for which the macroscopic structure can be described by a small-size graph, called {\em protograph} \cite{7308964,6145509}. An advantage of the PLDPC codes is that they are able to achieve desirable performance in both low- and high-SNR regions after proper designs \cite{7112076}. Thereby, two rate-compatible PLDPC codes has been already selected as the channel coding schemes for 5G new-radio standard \cite{8316763,access2008multiplexing}. Nowadays,
the research work on PLDPC-BICM and its variants has grown rapidly to improve the performance of modern communication and storage systems \cite{8314735,8740906,8708250}. For example, PLDPC codes and bit mappers have been considerably advanced for $M$PSK/$M$QAM-aided BICM systems over different types of channels \cite{1605713,van2012design,5613828,6777400,6133952,8740906,6777401} during the past fifteen years.

To extend BICM for other emerging scenarios, PLDPC codes have been combined with other types of modulations, such as code-phase-shift keying (CPSK) modulation \cite{esplugabinary}, on-off-keying (OOK) modulation \cite{7320948} and PPM \cite{6663748,6955112,7990043,zhaojietvt}, to support more diverse transmission and storage services. Moreover, exploiting the differential chaos-shift keying (DCSK) modulation \cite{7442517,cai2020design}, several novel PLDPC-BICM schemes were devised for block-fading channels, which are of particular interest for low-power and low-complexity short-range wireless communication applications \cite{7093149,7935433,7968491,7956256,8338131}.
In addition, in the non-orthogonal multiple access (NOMA) scenarios \cite{DCai1,DCai2,DCai3}, the optimization of PLDPC-BICM schemes has been investigated to achieve desirable performance in 5G systems \cite{newref1}.
Inspired by the appealing advantages of PLDPC-BICM, research efforts have been devoted to the joint design of this technique and physical-layer network coding (PNC) in order to realize higher-throughput wireless communications \cite{7723910,8544026}.

More recently, spatially-coupled (SC) PLDPC codes, which are generated by serially coupling a sequence of PLDPC codes into a single coupled chain, have emerged as a powerful FEC scheme \cite{6852099}. As a type of convolutional-like codes, the SC-PLDPC codes not only can obtain additional ``{\em convolutional gains}" over the conventional PLDPC codes, but also can keep a better balance between the decoding threshold and minimum free or Hamming distance \cite{7152893,5695130,6374679}. Accordingly, research interest turned to optimization and analysis of different types of SC-PLDPC codes, such as terminated (TE) and tail-biting (TB) SC-PLDPC codes under different transmission environments \cite{9056068,8398231,7353121,7593071,7086074,8281448,7339427,8449217,8272426}. Based on carefully designed SC-PLDPC codes, several spectral-efficiency BICM systems were proposed for satellite broadcasting communication, wireless communication, and optical communication \cite{8798970,6883627,hager2014improving,7005396,hager2016analysis,7593089,7460483,8883091,9057491,8012533}. Moreover, SC-PLDPC codes have been employed to improve the performance of modern dense flash-memory devices, including MLC, triple-level-cell (TLC), and quadruple-level-cell (QLC) flash memory, which can be regarded as high-order pulse-amplitude modulation (PAM)-aided discrete memoryless channels \cite{8013174,8007178,8278036,6804933,8277940,9032092}. In parallel with the transmitter design progress, the low-complexity message passing decoding algorithms were designed for the SC-PLDPC-BICM systems, so as to ease the hardware implementation \cite{8847350,TbroadcastSC-LDPC}. Today, research on PLDPC-BICM and its SC variant has opened up a promising direction in the CM field, which has gained tremendous attention from industry.%\footnote{Throughout this article, ``protograph-based LDPC (PB-LDPC) codes" will be used to collectively represent both ``PLDPC codes" and ``SC-PLDPC codes".}

During the first decade of this century, there have been several tutorial and survey-type of articles regarding the BICM \cite{schreckenbach2007iterative,i2008bit,franceschini2009ldpc}. On the one hand, comprehensive overviews of the convolutional- and turbo-BICM-ID
were presented in \cite{schreckenbach2007iterative} and \cite{i2008bit}, which also analyzed the feasibility of LDPC-BICM-ID. On the other hand, a general review on the attainable LDPC coded modulations, including both non-binary CM and binary BICM techniques, was presented in \cite{franceschini2009ldpc}. These articles summarized the research progress before $2009$, more than a decade ago. Moreover, they did not discuss the research progress regarding PLDPC-BICM. Since $2010$, the study of BICM has moved into a new stage and experienced a rapid development, in which the PLDPC-BICM plays an important role as an alternative in addressing new challenges from modern communication and storage applications. In fact, a systematic design guideline for the PLDPC-BICM is still lacking in the literature today. Therefore, it is timely needed to put forward a thorough overview of the up-to-date research advancements in PLDPC-BICM systems to meet the stringent reliability-and-throughput demand of modern communication and storage equipment \cite{8705373,9026873,6875180,7249024,6920539,8290977,7383250,7339431,7282623,6334510}.
For ease of understanding this article, Table~\ref{tab:II} provides a brief glimpse at the
%During the past decade, there was a growing number of studies on the PB-LDPC-BICM systems, aiming to refine the design and analysis, improve the transmission performance, and ease the system implementation . In this pursuit, PB-LDPC-BICM has invoked a new research direction in the CM field and offered a promising opportunity for future communication and storage applications. T
the major contributions in the historical evolution of PLDPC-BICM paradigm during the past two decades, with a special highlight on the encoder, constellation and interleaver design. %\Cite{8740906,6205592,8708250,8620336,8906157,6133952,8999517,7308964,1605713,Van2012Design,
%5613828,6777400,6777401,Esplugabinary,7320948,6955112,7990043,7442517,Cai2020Design,7093149,
%7935433,7968491,7723910,8544026,7956256,8398231,6883627,Hager2014Improving,7005396,Hager2016Analysis,7593089,7460483,8883091,9057491, 8012533,6875180,7249024,6920539,8290977,7383250,7339431,7282623,6334510}.

%Table 2
\begin{table*}[!t]\scriptsize
\caption{Major contributions in PLDPC-BICM design}\vspace{-1.5mm}
\centering
\begin{tabular}{|l|l|l|}
\hline
{\bf Year} & $\qquad~~~$ {\bf Author(s)} & $\qquad\qquad\qquad\qquad\qquad\qquad\qquad\qquad\qquad\qquad${\bf Contribution} \\ \hline
$1992$ & Zehavi \cite{141453} & \tabincell{l}{Conceived the concept of BICM to achieve bandwidth-efficient transmissions.} \\ \hline
$1997$ & Li {\it et al.} \cite{649929} & Introduced a hard-decision ID framework for BICM to establish a BICM-ID system. \\ \hline
$1998$ & {Brink \cite{706217}} & \tabincell{l}{Designed a soft-decision BICM-ID framework and illustrated that the anti-Gray
labeling outperforms the Gray labeling in such a scenario.} \\ \hline
$1998$ & Caire {\it et al.} \cite{669123} & Provided a systematic information-theoretic study on BICM in terms of channel capacity and HMMSED. \\ \hline
$2000$ & Knopp {\it et al.} \cite{2000On} & Formulated a performance-analysis methodology for BICM-NI over non-ergodic block-fading channels. \\ \hline
$2003$ & Thorpe \cite{2003IPNPR.154C...1T} & Invented the PLDPC codes and introduced their representation methods. \\ \hline
$2003$ & Brink {\it et al.} \cite{2003Design} & \tabincell{l}{Constructed capacity-approaching repeat-accumulate (RA) codes, i.e., a special type of PLDPC codes, for QPSK-aided BICM-ID systems.} \\ \hline
%$2004$ & Brink {\it et al.} \cite{1291808} & Developed the EXIT chart for the design and analysis of LDPC codes in MIMO BICM-ID systems. \\ \hline
$2005$ & Divsalar {\it et al.} \cite{1605713} & \tabincell{l}{Explored the applicability of PLDPC codes in BICM-NI systems and presented a novel bit-mapping scheme under the $16$QAM.} \\ \hline
$2006$ & Liva {\it et al.} \cite{liva2006design} & Discussed the spectral efficiency of RA-coded BICM systems with $M$PSK/$M$APSK. \\ \hline
%$2007$ & Schreckenbach \cite{schreckenbach2007iterative} & \tabincell{l}{Provided a comprehensive tutorial on the design of classic convolutional- and turbo-coded BICM-ID systems.} \\ \hline
$2009$ & Xie {\it et al.} \cite{5073425} & \tabincell{l}{Investigated the channel capacity and BER performance of the LDPC-BICM-ID systems with Gray-labeled and pseudo-Gray-labeled $M$APSK.} \\ \hline
$2010$ & Jin {\it et al.} \cite{5613828} & \tabincell{l}{Proposed an optimal VDMM scheme, which can obtain desirable threshold and WER enhancement over the water-filling VDMM scheme \\in PLDPC-BICM-NI systems with $16$QAM modulation.} \\ \hline
$2010$ & Barsoum {\it et al.} \cite{5629496} & \tabincell{l}{Formulated a protograph EXIT (PEXIT)-chart-aided PLDPC-code design method for BICM-ID systems over Poisson PPM channels.} \\ \hline
$2011$ & Nguyen {\it et al.} \cite{6133952} & \tabincell{l}{Developed a type of {\em enhanced AR4JA (EAR4JA) codes} and a two-stage-lifting-aided mapping (TSLM)
to constitute a novel PLDPC-BICM \\system, which is suitable for a wide range of code rates and modulation orders.} \\ \hline
%$2011$ & Liu {\it et al.} \cite{5960810} & \tabincell{l}{Introduced a novel irregular mapping (IM) technique based on the modified adaptive binary-switch  algorithm \\(ABSA) to offer near-capacity performance for LDPC-BICM-ID systems over AWGN channels.} \\ \hline
$2011$ & Kudekar {\it et al.} \cite{5695130} & \tabincell{l}{Conceived the concept of SC-PLDPC codes and revealed their inherent ``threshold saturation" feature.} \\ \hline
%$2012$ & Duyck {\it et al.} \cite{6205592} & \tabincell{l}{Developed an EXIT-aided optimization method for LDPC codes in BICM-ID systems over block-fading channels.} \\ \hline
$2013$ & Schmalen {\it et al.} \cite{6469365} & \tabincell{l}{Proposed a simple coding scheme that is universally suited for TE-SC-PLDPC-BICM systems with different constellation labelings.}  \\ \hline
$2013$ & Zhou {\it et al.} \cite{6663748} & \tabincell{l}{Optimized PLDPC codes for BICM-ID systems over Poisson PPM channels by using a modified EXIT algorithm.} \\ \hline
$2014$ & Tang {\it et al.} \cite{5613828} & \tabincell{l}{Generalized the water-filling VDMM scheme to combine any protograph with any modulation in BICM systems.} \\ \hline
$2014$ & H{\"a}ger {\it et al.} \cite{hager2014improving} & \tabincell{l}{Proposed an EXIT-aided bit-mapping optimization scheme for PLDPC-BICM systems with polarization-multiplexed (PM) QAM.} \\ \hline
$2014$ & Benaddi {\it et al.} \cite{6875180} & Introduced a general design framework for PLDPC-BICM with continuous phase modulation (CPM). \\ \hline
%$2015$ & Mitchell {\it et al.} \cite{7152893} & Investigated the performance of SC-PLDPC codes in a comprehensive and thorough manner. \\ \hline
$2015$ & Tang {\it et al.} \cite{7320948} & \tabincell{l}{Optimized the PLDPC-BICM with OOK modulation in light-emitting diode (LED)-based visible light communication (VLC) systems.} \\ \hline
$2015$ & Lyu {\it et al.} \cite{7093149} & Proposed a PLDPC-BICM-ID system with Walsh-code-based $M$DCSK modulation. \\ \hline
$2015$ & H{\"a}ger {\it et al.} \cite{7005396} & \tabincell{l}{Optimized the bit-mapping schemes for SC-PLDPC-BICM-NI to achieve high spectral-efficiency
fiber-optical communications.} \\ \hline
%$2015$ & Benaddi {\it et al.} \cite{7249024} & \tabincell{l}{Constructed a type of {\em direct truncation (DT) SC-PLDPC codes} for CPM-aided BICM-ID systems.}  \\ \hline
$2016$ & Cammerer{\it~et\,al.} \cite{7593089} & \tabincell{l}{Proposed a hybrid mapping scheme for TB-SC-PLDPC-BICM-ID systems with $16$QAM modulation, which simultaneously adopts Gray- \\and SP-labeled constellations for a codeword to trigger the wave-like decoding.} \\ \hline
$2016$ & Steiner {\it et al.} \cite{7339431} & \tabincell{l}{Presented a PLDPC-code design method for ASK-aided BICM systems under shaped bit-metric decoding.} \\ \hline
$2017$ & Zhan {\it et al.} \cite{7956256} & \tabincell{l}{Proposed an ARJA BICM system with constellation-based $M$DCSK modulation, which benefits from higher
spectral efficiency compared \\with the Walsh-code-based counterpart.} \\ \hline
$2017$ & Chen {\it et al.} \cite{7723910} & \tabincell{l}{Estimated the performance of AR$3$A-coded PNC BICM-ID systems over block-fading channels.} \\ \hline
$2017$ & Fang {\it et al.} \cite{8019817}& \tabincell{l}{Designed a family of multi-layer root-PLDPC codes for multi-relay coded-cooperative systems and extended
their application to the QPSK-\\aided BICM scenario over block-fading channels.} \\ \hline
%$2018$ & Chen {\it et al.} \cite{8338131} & \tabincell{l}{Conceived a new PLDPC-CM system with Walsh-code-based $M$DCSK modulation, which outperforms the \\conventional PLDPC-BICM-ID relative.} \\ \hline
$2018$ & Chen {\it et al.} \cite{8314735} & \tabincell{l}{Constructed a family of RC-PLDPC codes and a novel bit-mapping scheme for MLC BICM-NI flash-memory systems.} \\ \hline
$2019$ & Fang {\it et al.} \cite{8740906} & Presented an outage-limit-approaching PLDPC-BICM system over block-fading channels. \\ \hline
$2019$ & Bu {\it et al.} \cite{8708250} & \tabincell{l}{Designed a family of high-rate PLDPC codes, called {\em optimized ARA (OARA)} {\em codes}, for BICM-ID flash-memory systems.} \\ \hline
%$2019$ & Espluga {\it et al.} \cite{esplugabinary} & Optimized the family of root-PLDPC codes for the CSK-aided BICM systems over AWGN channels. \\ \hline
%$2020$ & Zhao {\it et al.} \cite{8999517} & \tabincell{l}{Proposed an MI-aided bit-mapping scheme for TE-SC-PLDPC-BICM systems over fast-fading channels.} \\ \hline
$2020$ & Zhang {\it et al.} \cite{8798970}& \tabincell{l}{Developed a QC-SC-PLDPC-BICM scheme, which is able to achieve near-capacity performance with low implementation complexity.} \\ \hline
$2020$ & Yang {\it et al.} \cite{8883091} & Presented novel constellation and bit-mapping schemes for TB-SC-PLDPC-BICM-ID systems. \\ \hline
%$2020$ & Yang {\it et al.} \cite{9057491} & \tabincell{l}{Constructed a type of {\em improved TB-SC-PLDPC (I-TB-SC-PLDPC) codes} for $16$QAM-aided BICM-NI systems \\under shuffled BP decoding.} \\ \hline
$2021$ & Yang {\it et al.} \cite{9210097} & \tabincell{l}{Constructed a type of structural quadrant (SQ) constellations for TB-SC-PLDPC-hierarchical modulated (HM)-BICM-ID systems.} \\ \hline
$2021$ & Dai {\it et al.} \cite{9519519} & Designed a novel irregular-mapped (IM) PLDPC-BICM framework for VLC systems. \\ \hline
\end{tabular}
\label{tab:II}\vspace{-2.5mm}
\end{table*}%

%Table 1
\begin{table*}[t]\scriptsize
\caption{List of acronyms used in this article.}\vspace{-0.15cm}
\centering
\begin{tabular}{|l|l||l|l|}
\hline
{\bf Acronym} & {\bf Full Name} &  {\bf Acronym}  & {\bf Full Name} \\ \hline
$3$D & three-dimensional &  MSB/LSB & most/least significant bit \\ \hline
$4$G/$5$G & fourth/fifth-generation & MSEW  &  maximum squared Euclidean weight \\ \hline
ABSA & adaptive binary-switch algorithm & OOK &	on-off-keying \\ \hline
% & OREC & octal rectangular \\ \hline
AR$3$A/AR$4$A & accumulate-repeat-$3/4$-accumulate & OREC & octal rectangular \\ \hline
AR$4$JA/EAR$4$JA & \tabincell{l}{accumulate-repeat-by-$4$-jagged-accumulate/enhanced AR4JA} &  PAM & pulse-amplitude modulation  \\ \hline
ARA/IARA & accumulate-repeat-accumulate/improved ARA & PAS & probabilistic amplitude shaping \\ \hline
%{\color{red}ASTC$3.0$} & advanced television systems committee $3.0$ & PE/PEG & program-and-erase/progressive-edge-growth \\ \hline
AWE & asymptotic weight enumerator & PDF & probability density function \\ \hline
AWGN & additive white Gaussian noise  & PE/PEG & program-and-erase/progressive-edge-growth \\ \hline
BER/WER & bit/word error rate & PLDPC/DPLDPC & protograph LDPC/double-PLDPC \\ \hline
BICM & bit-interleaved coded modulation & PN/PM & pseudo-noise/polarization-multiplexed  \\ \hline
BP/S-BP & belief propagation/shuffled BP & PNC& physical-layer network coding \\ \hline
HDD/SSD & hard disk drive/solid state drive  & PPM & pulse position modulation \\ \hline
CM/MLCM & coded-modulation/multi-level CM  & PSK/QAM & \tabincell{l}{phase-shift-keying/quadratic-amplitude modulation} \\ \hline
%CD/HDD/SSD & compact disc/hard disk drive/solid state drive & {\color{red}QMP} & quaternary message passing \\ \hline
CPM & continuous phase modulation   & QC & quasi-cyclic \\ \hline
CPSK & code-phase-shift keying  & QRC & quaternary raised-cosine \\ \hline
CSI & channel state information   & RA/RJA & repeat-accumulate/repeat-jagged-accumulate \\ \hline
DCSK & differential chaos-shift keying  & RC & rate-compatible \\ \hline
%{\color{red}DC} & differentially coherent & {\color{red}RF} & radio-frequency \\ \hline
DE & density evolution & RGB & red-green-blue \\ \hline
DT & direct-truncated  & RP & root-protograph \\ \hline
DSSS & direct-sequence spread-spectrum  & SC & spatially-coupled \\ \hline
DVB & digital video broadcasting  & SISO & soft-input soft-output \\ \hline
EXIT/PEXIT & extrinsic information transfer/protograph EXIT  & SLC/MLC & single/multi-level cell \\ \hline
FEC & forward error correction & SNR & signal-to-noise-ratio \\ \hline
GML & generalized-maximum-likelihood  & SP & set-partitioning \\ \hline
GMSK & Gaussian minimum shift keying  & SPMM & spatial-position matched mapping \\ \hline
HM & hierarchical modulated  & SQ & structural quadrant \\ \hline
HMMSED & harmonic mean of the minimum squared Euclidean distance  & TB/TE & tail-biting/terminated \\ \hline
ID/NI & iterative demapping and decoding/non-ID  & TCM & trellis-coded modulation \\ \hline
i.i.d. & independent and identically distributed  & TLC/QLC & triple/quadruple-level cell \\ \hline
IM & irregular mapping  & TSLM & two-stage-lifting-aided mapping \\ \hline
LED & light-emitting diodes  & TWR & two-way relay  \\ \hline
%ID/NI & iterative demapping and decoding/non-ID & {\color{red}TTD} & time to data \\ \hline
LDPC/PLDPC & low-density parity-check/protograph-based LDPC  & UEP & unequal-error-protection  \\ \hline
LLR & log-likelihood ratio & VDMM & variable degree matched mapping \\ \hline
MAP/MDS & maximum a-posteriori/maximum-distance separable &  VLC/VS & visible light communication/voltage-sensing \\ \hline
MED & minimum Euclidean distance  & VNDM & variable-node-degree-based mapping \\ \hline
MHDGR/MFDGR & minimum Hamming/free-distance growth rate  & VNFAM & variable-node fractional-allocation mapping \\ \hline
MI/MMI & mutual information/maximum MI  & VNMM & variable node matched mapping \\ \hline
%MSB/LSB & most/least significant bit & VNMM & variable node matched mapping \\ \hline
%  & {\color{red}WLAN} & wireless local area network \\ \hline
%MSB/LSB & most/least significant bit   & WiMax & worldwide interoperability for microwave access \\ \hline
% & {\color{red}WPAN/WSN} & wireless personal area/sensor network \\ \hline
\end{tabular}\vspace{-3mm}
\label{tab:I}
\end{table*}
\vspace{-0.3mm}
\subsection{Organization and Structure of This Survey}\label{sect:I-B}\vspace{-0.3mm}

This article provides a survey of PLDPC-BICM on the most recent works investigating joint design of binary PLDPC codes and high-order modulations to support high-reliability and high-rate digital communications and data storage. Specifically, it will start by introducing the fundamental configuration, transmission mechanism, and the information-theoretic limit of the PLDPC-BICM. Then, some preliminary basis of PLDPC codes will be discussed, which can be seamlessly combined with high-order modulations to construct bandwidth-efficient BICM systems. Further, some typical theoretical-analysis tools will be reviewed, which have been extensively adopted for designing PLDPC-BICM. On the basis of the above preliminaries, the enabling design methodologies will be classified for PLDPC-BICM systems over different transmission channels, such as AWGN channels, fading channels, Poisson PPM channels, and NAND flash-memory channels. In particular, state-of-the-art technologies will be described, including code construction, constellation shaping and bit-mapper design, which are useful for enabling the relevant systems with capacity-approaching performance. Additionally, several promising research directions will be introduced, which may further enhance the robustness of the PLDPC-BICM and expand its application domain. Since the aim is to present a focal review on PLDPC-BICM techniques rather than presenting an extensive coverage of all BICM techniques, some irrelevant topics are excluded. This is the first attempt to systematically provide design guidelines for a concatenated framework comprising PLDPC codes, bit mappers and high-order modulations. Although the references are not exhaustive, this survey could serve as a starting point for further studies on the topics of interest.

The structure of this article is organized as follows. Section~\ref{sect:II} describes the system model and transmission mechanism of the PLDPC-BICM with emphasis on its transceiver architecture, channel model and channel capacity. %This section also summarizes the major advancement in the design of such a technique.
Section~\ref{sect:III} presents basic concepts of PLDPC codes and reviews some theoretical-analysis tools, which are always exploited to assess and optimize the performance of PLDPC-BICM. Sections~\ref{sect:V}$\sim$\ref{sect:VIII} details the latest design paradigms of PLDPC-BICM over AWGN, fading, Poisson PPM and NAND flash-memory channels, respectively, which are used to characterize the statistic features of some classic transmission and storage environments. Sections~\ref{sect:IX} and \ref{sect:IXI} recommend some valuable future research directions and draw some conclusions, respectively. To facilitate reading, the acronyms used in the paper are listed in Table~\ref{tab:I}.\vspace{-0.3mm}
%Fig.2
\begin{figure}[t]
\center\vspace{-0.1cm}
\includegraphics[width=3.5in,height=1.43in]{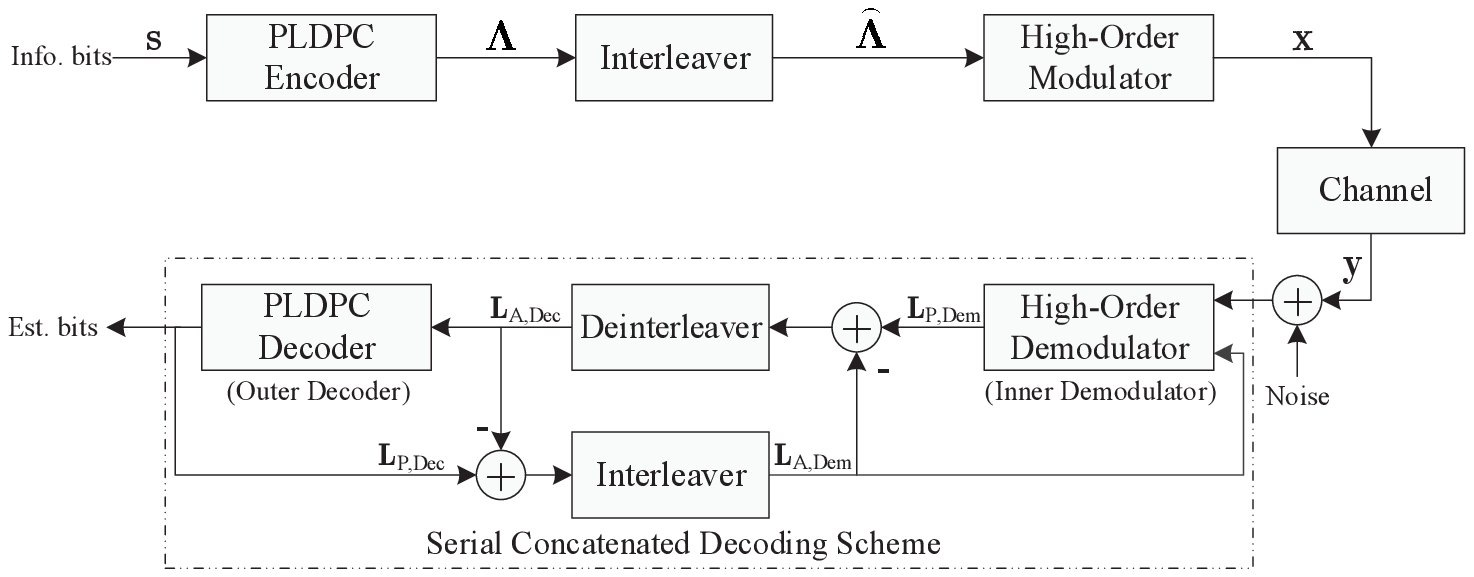}
\vspace{-0.3cm}
\caption{Block diagram of a general PLDPC-BICM-ID system.}\vspace{-0.4cm}
\label{fig:Fig.2}\vspace{-0.1cm}
\end{figure}

\section{System Model and Transmission Mechanism of PLDPC-BICM}\label{sect:II}\vspace{-0.3mm}

We first describe the general PLDPC-BICM system architecture, and then present some preliminaries that will be useful throughout the article.
%Moreover, major progresses in the historical evolution of PLDPC-BICM design are summarized.

\subsection{System Architecture }\label{sect:II-A}

Fig.~\ref{fig:Fig.2} depicts a general PLDPC-BICM system model. In this system, the source information bits ${\bs}=(s_1, s_2, \ldots, s_K)$ are first encoded into a rate-$r$ binary PLDPC code ${\bm \Lambda}=(v_1, v_2, \ldots, v_N)$, where $K$ is the information length, $N$ is the codeword length, $r=\frac{K}{N-N_{\rm E}} = \frac{K}{N_{\rm T}}$ is the code rate, and $N_{\rm E}$ and $N_{\rm T}$ are the punctured and transmitted codeword lengths, respectively.
%\footnote{Unless otherwise stated, only binary PLDPC codes are considered because only BICM systems are considered in this paper.}
The coded sequence is permuted by a bit-level interleaver to yield an interleaved bit sequence $\widehat{\bm \Lambda}=(\hat{v}_1, \hat{v}_2, \ldots, \hat{v}_N)$, which is subsequently modulated to an $M$-ary symbol sequence $\bx=(x_1, x_2, \ldots, x_{N'})$, where $M > 2$ is the modulation order, $w=\log_2 M$ is the number of bits carried by a modulated symbol, and $N'= \frac{N}{w}$ is the length of symbol sequence. It is usually assumed that $N$ is a multiple of $w$. In particular, every $w$ interleaved coded bits are mapped to an $M$-ary symbol through a specific signal constellation. The modulated symbol sequence is then transmitted through a noisy channel.

At the receiver terminal, the ``impaired" signal output from the channel is decoded by a serially concatenated decoding scheme, which contains a soft-input soft-output (SISO) inner demodulator (or {\em demapper}) and an SISO outer decoder \cite{schreckenbach2007iterative}. To compensate the performance loss arising from separated demapping and decoding, the soft information, such as probabilistic and log-likelihood ratio (LLR) information, of the demodulator and decoder in the BICM system are updated in an iterative fashion, which is similar to the turbo decoding \cite{i2008bit,franceschini2009ldpc}.
Specifically, given the received signal $\by$ and {\it a priori} LLR $\bL_{\rm A,Dem}$, the demodulator can compute its extrinsic LLR (i.e., $\bL_{\rm P,Dem} - \bL_{\rm A,Dem}$), which will be used as the {\it a priori} LLR $\bL_{\rm A,Dec}$ of the decoder. Subsequently, the decoder can calculate its extrinsic LLR (i.e., $\bL_{\rm P,Dec} - \bL_{\rm A,Dec}$), which will be fed back to the demodulator as the updated {\it a priori} LLR for the next iteration. Some key components of this system are further elaborated as follows.

\subsubsection{PLDPC Encoder}\label{sect:II-A-1}

In the BICM-ID system, the PLDPC codes, are used as the channel coding scheme to enhance the reliability of the source information prior to the transmission over noisy channels. It has been demonstrated that the source information can be well protected when transmitting over a noisy and interference channel if the FEC codes are carefully designed \cite{8474959,7307154}. However, similarly to conventional FEC codes, the PLDPC code is constructed by appending some redundancy (i.e., parity-check bits) to the information bits, which leads to low throughput \cite{9042862}. To partially overcome this weakness, puncturing techniques were developed to increase the code rate of such codes \cite{7152893,8708250}. In contrast to conventional LDPC codes, a well-designed puncturing scheme not only increases the transmission rate but also improves the error performance of PLDPC codes \cite{7112076,8398231}. For simplicity, the puncturing module is omitted in Fig.~\ref{fig:Fig.2} as it can be embedded into the encoder.

\subsubsection{Interleaver~(Bit Mapper)}\label{sect:II-A-2}

As PLDPC codes possess an intrinsic interleaving feature, there is no need to add an interleaver into a PLDPC-coded system with BPSK modulation. However, an individual bit interleaver (or {\em bit mapper}) is of great importance to guarantee the performance of the PLDPC-BICM system, especially over a fading channel, because it can optimize the mapping rule between the coded bits in a codeword and labeling bits within a modulated symbol. A bit-level interleaver not only enables the PLDPC-BICM to achieve a larger diversity order over a fading channel, but also can better protect the high-priority coded bits \cite{i2008bit}. To realize the unequal-error-protection (UEP) property and to acquire a desirable interleaving gain for a particular channel model, several bit-mapping schemes have been designed, such as water-filling-like mapping scheme \cite{1605713}, variable-degree matched mapping (VDMM) scheme \cite{6133952,5613828,6777400,7320948}, variable-node matched mapping (VNMM) scheme \cite{8883091}, and UEP-based bit-mapping scheme \cite{8740906}, to optimize the interface between PLDPC-coded bits and modulated symbols.

\subsubsection{High-order Modulator}\label{sect:II-A-3}

To achieve high spectral efficiency, high-order modulations (i.e., with $M>2$) instead of binary modulations are utilized in the PLDPC-BICM systems. In particular, the interleaved bit sequence is first uniformly divided into $N'$ sub-sequences of length $w$, i.e.,\vspace{-2mm}
%$\widehat{\bm \Lambda}=(\hat{v}_1, \hat{v}_2, \ldots, \hat{v}_N)=(\hat{v}_{1,1}, \hat{v}_{1,2}, \ldots, \hat{v}_{1,w}; \hat{v}_{2,1}, \hat{v}_{2,2}, \ldots, \hat{v}_{2,w}; \ldots; \hat{v}_{N',1}, \hat{v}_{N',2}, \ldots, \\\hat{v}_{N',w})$.
\begin{equation}\vspace{-2mm}
\begin{split}
\widehat{\bm \Lambda}=(\hat{v}_1, \hat{v}_2, \ldots, \hat{v}_N)=&(\hat{v}_{1,1}, \hat{v}_{1,2}, \ldots, \hat{v}_{1,w}; \hat{v}_{2,1}, \hat{v}_{2,2}, \ldots, \hat{v}_{2,w}; \nonumber\\
&\ldots; \hat{v}_{N',1}, \hat{v}_{N',2}, \ldots, \hat{v}_{N',w}).
\end{split}\label{eq:new-eq-1}
\end{equation}Then, the $N'$ sub-sequences are mapped to $N'$ transmitted symbols $\bx=(x_1, x_2, \ldots, x_{N'})$, where $x_k \triangleq (\hat{v}_{k,1},\hat{v}_{k,2},\ldots, \hat{v}_{k,w})$ is chosen from an $M$-ary modulated signal constellation (i.e., signal set) ${\cal X}$ with a labeling rule $\phi: \{ 0, 1\}^w \to {\cal X}$, $w=\log_2 M$, $|{\cal X}|=M$, and $k=1,2,\ldots,N'$. Therefore, the transmitted symbol output from an $M$-ary modulator is associated with $w$ binary coded bits (also called {\em labeling bits}). More precisely, the performance of a PLDPC-BICM is dependent not only on the modulation scheme but also on the labeling scheme. In this article, the system design approaches are outlined for various modulation schemes tailored to different communication applications.%\cite{haykin1988digital,8765384}. %with emphasis on the classic $M$-ary PSK and QAM modulations.

%either binary or $q$-ary manner, where $q=2^\mu,~{\rm and}~ \mu=2,3,\ldots$ Generally, high-order modulations can achieve higher transmission throughput and be seamlessly combined with non-binary ECCs with respect to the binary ones, but suffer from relatively worse error performance \cite{7296682,6199941}. In particular, we restrict ourselves to phase-shift-keying (PSK) modulations in this paper. %In particular, the BPSK modulation is considered when illustrating the principles of LDPC-coded MR systems.

\subsubsection{Channel Models}\label{sect:II-A-4}

In the past two decades, BICM has been applied to various communication and storage systems, such as deep-space communication \cite{1605713,van2012design,5613828,6777400}, wireless communication \cite{7093149,7935433,7968491,8883091}, optical communication \cite{9519519,7320948,6955112,7990043,7005396}, and NAND-flash-based data storage \cite{8708250,7887724,6777401}. To facilitate applications in such transmissions, AWGN channel, fast-fading channel, block-fading channel, Poisson PPM channel, and NAND flash-memory channel have been considered in developing design and analysis methodologies for the PLDPC-BICM systems. Among all the above-mentioned channels, AWGN channel is one of the simplest memoryless noisy channel models that can reasonably describe the fundamental characteristics of many realistic communication systems, including deep-space communication systems, visible light communication (VLC) systems, fiber-optical communication systems, as well as wireless communication systems. Despite its extreme simplicity, AWGN channel serves as an important model for designing capacity-approaching BICM schemes. As illustrated in \cite{7112076,4026721}, the LDPC codes optimized for an AWGN channel can also exhibit excellent performance over other memoryless channel models, e.g., binary erasure channel and ergodic fast-fading channel. In this sense, the code-design criteria developed for PLDPC-BICM systems over AWGN channels are very promising for many practical applications and, therefore, have been intensively studied in recent years \cite{5073425,6515491,5598322,8878166,1413232}.

In addition to the AWGN, channel fading is another key factor that degrades the transmission reliability of wireless communications. At the early stage of BICM studies, a transmitted signal was always assumed to suffer from Rayleigh fading over wireless channels. However, it has been demonstrated that Rayleigh distribution cannot accurately capture the underlying physical properties of several wireless communication systems, such as complex indoor environments, land mobile systems and ionospheric radio links, and thus is not a good model for such transmission \cite{7112076}. As an improved and more general model, the Nakagami-$m$ distribution was introduced to match the fading statistical behavior of many wireless communication channels. Specifically, the Nakagami-$m$ distribution, which covers Rayleigh fading channel, Rician fading channel and AWGN channel as special cases, is able to offer the widest range of fading among all the existing distributions \cite{8674555,7954666,7296682}. Due to the above-mentioned advantage, a great deal of research effort has been devoted to studying the PLDPC-code construction and their respective BICM-system design over Nakagami-$m$ fading channels \cite{7442517,4573263,7112587,8019817,7080854}.

Besides, Poisson PPM channel \cite{6955112,7990043,5629496} and flash-memory channel \cite{8007178,8278036,6804933,8277940,8715674,8528505,9000906}, which are respectively utilized to characterize the free-space optical communication and non-volatile storage systems, may have quite different forms of transmitted signals (i.e., pulse position and read voltage), channel noises and gains. They will be further discussed in Sections~\ref{sect:VII} and \ref{sect:VIII} before getting into technical details.

As in the majority of existing works investigating the channel-coded BICM, the symbol-to-symbol interference is not considered here in the design and analysis for simplicity. Moreover, without loss of generality, for all the examples considered below, the scalar fading gains in a wireless fading channel are assumed to be subject to a Nakagami-$m$ distribution, denoted by $|\alpha| \sim {\cal NAK}(m)$, where $m \ge \frac{1}{2}$ is the fading depth \cite{8674555}.
%whose probability density function (PDF) is given by
%\begin{equation}
%f_{|\alpha|}(\tau)
%= \frac{2}{\Gamma(m)}\left( \frac{m}{\Omega}\right)^m
%\tau^{2m-1} \exp\left( - \frac{m}{\Omega} \tau^{2} \right),
%\label{eq:Nakagami-pdf}
%\end{equation}
%where $m \ge 1/2$ is the fading depth, $\Omega = \mathbb{E} [\tau^2] =1$ is the power of fading gain, $\mathbb{E}[\cdot]$ is the expectation operation, and $\Gamma(\cdot)$ is the Gamma function.

In general, the input-output relationship of an $M$-ary modulated PLDPC-BICM system over an AWGN/fading channel can be written as
\begin{equation}
y_{k} = \alpha_l x_{k} + n_{k},
\label{eq:y-BICM-AWGN-fading}
\end{equation}
where $x_k \in {\cal X}$ is the $M$-ary complex-valued transmitted symbol, ${\cal X }$ is the constellation set, $y_k$ is the received signal, $n_k$ is the complex Gaussian noise with zero mean and variance $\sigma_{n}^{2} = \frac{N_0}{2}$ per dimension, $N_0$ is the noise power spectral density, $k=1,2,\ldots,N'$; $\alpha_l = |\alpha_l| \exp(j \varphi_l)$ is the channel gain with $|\alpha_l|$ and $\varphi_l$ being the scalar fading gain and phase shift, respectively, which satisfy the following properties: (i) $|\alpha_l|= \alpha$ ($\alpha$ controls the average power per symbol), $ \varphi =0 $, and $l=k$ for an AWGN channel; (ii) $|\alpha_l| \sim {\cal NAK}(m)$, $ \varphi \sim {\cal U} [-\pi,\pi]$~(i.e., a uniform distribution), and $l=k$ for a fast-fading channel, (iii) $|\alpha_l| \sim {\cal NAK}(m)$, $ \varphi \sim {\cal U} [-\pi,\pi]$, and $l = 1+ \lfloor (k-1)/B' \rfloor$ ($B'$ is the symbol block length and $\lfloor \cdot \rfloor$ is the floor operation) for a block-fading channel \cite{8740906,1564429}. Moreover, the spectral efficiency of a PLDPC-BICM system is defined as $R_{\rm SE} = r \log_2 M = r w$. The average SNR per information bit is defined as $E_{b}/N_0=(E_{\rm s}/N_0)/R_{\rm SE}=\gamma_{\rm s}/R_{\rm SE}$,
where $\gamma_{\rm s}$ is the average symbol SNR and $E_{\rm s}$ is the average energy per transmitted symbol.

\subsubsection{Detector/Decoder}\label{sect:II-A-5}

At the receiver of the PLDPC-BICM-ID system, a turbo-like iterative decoding framework, which includes an inner demapper and an outer decoder, is employed to decode the ``impaired" signal. In this iterative decoding framework, the extrinsic information output from the decoder is fed back to the demodulator as its {\it a priori} information. Then, the {\it a posteriori} information of each labeling bit within an $M$-ary symbol output from the demodulator can be estimated by exploiting the {\it a priori} information of the remaining $w-1$ labeling bits, which will further help accelerate the convergence of the decoder. As such, the turbo-like decoding framework allows the BICM-ID system to extract more reliable soft information and to achieve better performance as compared with the BICM-NI system \cite{8740906}. There are a variety of soft-information-oriented demodulation algorithms designed for BICM systems in the literature, such as maximum {\it a posteriori} (MAP) algorithm \cite{7095513} and the max-sum approximation of its log-domain counterpart (i.e., Max-Log-MAP) \cite{6942236}. Here, the MAP algorithm and log-domain belief-propagation (log-BP) algorithm \cite{8269289,ryan2009channel,7254117} are adopted to implement the demodulator and decoder, respectively. The iteration between the inner demodulator and outer decoder is referred to as {\em global iteration}, while the iteration between the variable-node decoder and the check-node decoder in the PLDPC decoder is referred to as {\em local iteration} (or {\em BP iteration}). Especially, the BICM-ID is simplified to a BICM-NI if the number of global iterations is set to be one. The following discussions will focus on the transmitter design of the PLDPC-BICM, while treating the receiver architecture as a turbo-like decoder. It is assumed that the channel state information (CSI) can be perfectly captured by the receiver with the help of a well-designed channel estimator \cite{8019817,6094140}, as assumed in most studies \cite{1226601,7383286,5073425,1532194,6359874,8338131,8740906}.\vspace{-1mm}

\subsection{Channel Capacity}\label{sect:II-B}

For an ergodic memoryless channel, the channel capacity $C$ is defined as the maximum mutual information (MI) between the channel input $X$ and output $Y$ over all the channel input probability distributions $\{\Pr(x)\}$ \cite{tm1over}. As one of the most significant performance metrics for channel coding and BICM schemes over a given channel, the capacity $C$ is used to quantify the information rate that the channel can reliably convey with the use of an FEC code. %In implementation, channel capacity can be represented in two different units, i.e., information bits per channel symbol and information bits per channel bit. Here, the former is adopted unless otherwise indicated.
Nonetheless, the BICM system cannot continue to provide reliable transmission over the channel with any FEC code once the spectral efficiency exceeds the channel capacity (i.e., $R_{\rm SE}>C$). The capacity of PLDPC-BICM over an AWGN channel is further analyzed below.

\subsubsection{Constellation-Unconstrained Capacity}\label{sect:II-B-1}

%Consider an AWGN channel with the output signal shown in \eqref{eq:y-BICM-AWGN-fading}.
Suppose that the transmitted symbol $x_k$ is randomly chosen from an infinite-size complex-Gaussian constellation ${\cal X}$. The capacity for an AWGN channel is given by \cite{tm1over}\vspace{-0.1cm}
%$C_{\rm Shannon} = \log_2 (1+E_{\rm s}/N_0) = \log_ 2(1 + \gamma_{\rm s})$,
\begin{equation}\vspace{-0.1cm}
C_{\rm Shannon} = \log_2 (1+E_{\rm s}/N_0) = \log_ 2(1 + \gamma_{\rm s}),
\label{eq:new-eq-2}
\end{equation}where the power of transmitted symbol satisfies $\mathbb{E} [ |x_k|^2] = E_{\rm s}$ and the cardinality of the constellation satisfies $|{\cal X}| \to + \infty$. This equation is the well-known Shannon formula, which cannot be achieved by using a finite-size constellation.

\subsubsection{Constellation-Constrained Capacity}\label{sect:II-B-2}

From a practical point of view, it should be more realistic to map a PLDPC codeword to a symbol sequence chosen from a finite-size signal constellation $\chi$, such as $M$PSK and $M$QAM, which has a uniform distribution for all the $M$ component points $x_k \in \chi$. In this scenario, assuming that the $M$-ary transmitted symbol sequence $( x_1, x_2, \ldots, x_{N'} )$ is directly mapped from a length-$N'$ non-binary PLDPC codeword over $\mathbb{GF}(M)$, rather than from a length-$N$ binary PLDPC codeword. The channel capacity can be specified by the average input-output MI \cite{i2008bit}, i.e.,

\vspace{-3mm}
{\small\begin{equation}
C_{\rm CM} = I_{\rm CM} (X;Y \,|\, \gamma_{\rm s}) \hspace{-0.5mm}=\hspace{-0.5mm} w - \mathbb{E}_{x,y} \hspace{-1mm} \left[ \log_2 \frac{\sum_{x'\in \chi}
f(y | x')}{f(y | x)}\right],
\label{eq:capacity-CM}
\end{equation}}where $x \in \chi$, $|\chi|=M$, and $f(y|x')$ is the complex-Gaussian probability density function (PDF) \cite{8740906}. %is \vspace{-1mm}
%\begin{equation}
%f(y|x') = \frac{1} {2 \pi \sigma_{n}^2 }
%\exp \left(- \frac{| y - x' | ^2}{2 \sigma_{n}^2 } \right).\nonumber
%\label{eq:PDF-channel-output}
%\end{equation}
The capacity defined in \eqref{eq:capacity-CM} is referred to as the {\em CM capacity}. As shown in \cite{669123,i2008bit,biglieri2005coding}, CM is able to achieve the maximum achievable rate of an $M$-ary modulated channel, and thus is viewed as an optimal transmission scheme.

For a BICM-NI scheme, the $M$-ary modulated channel can be considered as $w$ independent parallel binary-input sub-channels. Thus, the {\em BICM capacity} is equal to the summation of the average MIs of the $w$ sub-channels \cite{669123}, i.e.,

\vspace{-3mm}
{\small\begin{align}
C_{\rm BICM-NI} &= \sum\nolimits_{\mu=1}^{w} I_{\rm BICM-NI}({\hat{v}}_\mu; Y \,|\, \gamma_{\rm s}) \nonumber\\
&= w - \sum\nolimits_{\mu=1}^{w} \mathbb{E}_{b ,y}
\left[ \log_2 \frac{\sum_{x' \in \chi} f(y | x')}
{\sum_{x' \in \chi_\mu^b } f(y | x')}\right],
\label{eq:capacity-BICM}
\end{align}}where $\hat{v}_{\mu}$ is the $\mu$-th~$ (\mu=1,2,\ldots, w)$ labeling bits within a transmitted symbol $x \in \chi$, and $\chi_\mu^b \subset \chi$ is the constellation subset of the signal points, in which the $\mu$-th labeling bit has value $ b \in \{0,1\}$. In a BICM-NI scheme, the $w$ labeling bits within the $M$-ary symbol are demapped independently, leading to some MI loss. As a result, BICM-NI is not able to maximize the achievable rate, and thus is a suboptimal transmission scheme. The suboptimality of the BICM-NI can be easily proved mathematically \cite{669123,i2008bit}, i.e., $ C_{\rm BICM-NI} \le C_{\rm CM}$.

To overcome the above drawback, in \cite{649929,706217}, a serially concatenated iterative decoding framework was proposed for the BICM-NI system, wherein the extrinsic information can be iteratively updated between the inner demodulator and outer decoder (see Fig.~\ref{fig:Fig.2}), which establishes a BICM-ID system. In the BICM-ID system, the demodulation of one labeling bit within an $M$-ary symbol is dependent on the remaining $w-1$ bits such that their soft information can be substantially exploited. The serially concatenated iterative decoding framework is beneficial for compensating the mutual-information loss of the traditional BICM-NI, allowing the BICM-ID system to maximize the achievable rate \cite{5073425,6064854,6241383,7095513,6469365,5669235}. Due to the above superiority, the BICM-ID system is able to achieve excellent performance as the CM system, and thus can be viewed as an optimal transmission scheme. Accordingly, the CM capacity can be directly treated as the fundamental capacity limit of BICM-ID, as highlighted in \cite{2018Bit,Xiaodong2002Bit,8740906}.\vspace{1mm}

\textit{Remark:} The CM capacity is only relevant to the signal constellation but irrelevant to the labeling scheme, while the BICM capacity is relevant to both the signal constellation and labeling scheme.
The CM and BICM capacities only indicate the fundamental performance limits of the BICM-ID and BICM-NI, respectively. In practice, the design of PLDPC codes, signal constellations, bit mappers and labeling schemes have significant influence on the system performance.
%\begin{itemize}
%\item
%The CM capacity is only relevant to the signal constellation but irrelevant to the labeling scheme, while the BICM capacity is relevant to both the signal constellation and labeling scheme.
%
%\item
%The CM and BICM capacities indicate the fundamental performance limits of the BICM-ID and BICM-NI, respectively. In practice, the design of PB-LDPC codes, signal constellations, bit mappers and labeling schemes have significant influence on the system performance.
% 	
%%\item
%%Gray labeling can maximize the BICM capacity over AWGN channels and, hence, is regarded as the optimal labeling scheme in these scenarios. However, it may not be the optimal choice in the BICM-ID and fading scenarios.
%\end{itemize}

There exist numerous labeling rules for high-order constellations, most of which are feasible for BICM systems. However, no labeling can achieve best performance over all scenarios \cite{1413232}. Among all the existing labelings, Gray labeling \cite{6942236}, anti-Gray labeling \cite{liew2015primer}, set-partitioning (SP) labeling \cite{7140767}, and maximum squared Euclidean weight (MSEW) labeling \cite{1413232} have attracted particular interest in the CM community.%, because they can show desirable performance in different transmission scenarios.

\begin{figure*}[!t]
\vspace{-0.3cm}
\center
\includegraphics[width=7.0in,height=9.3219in]{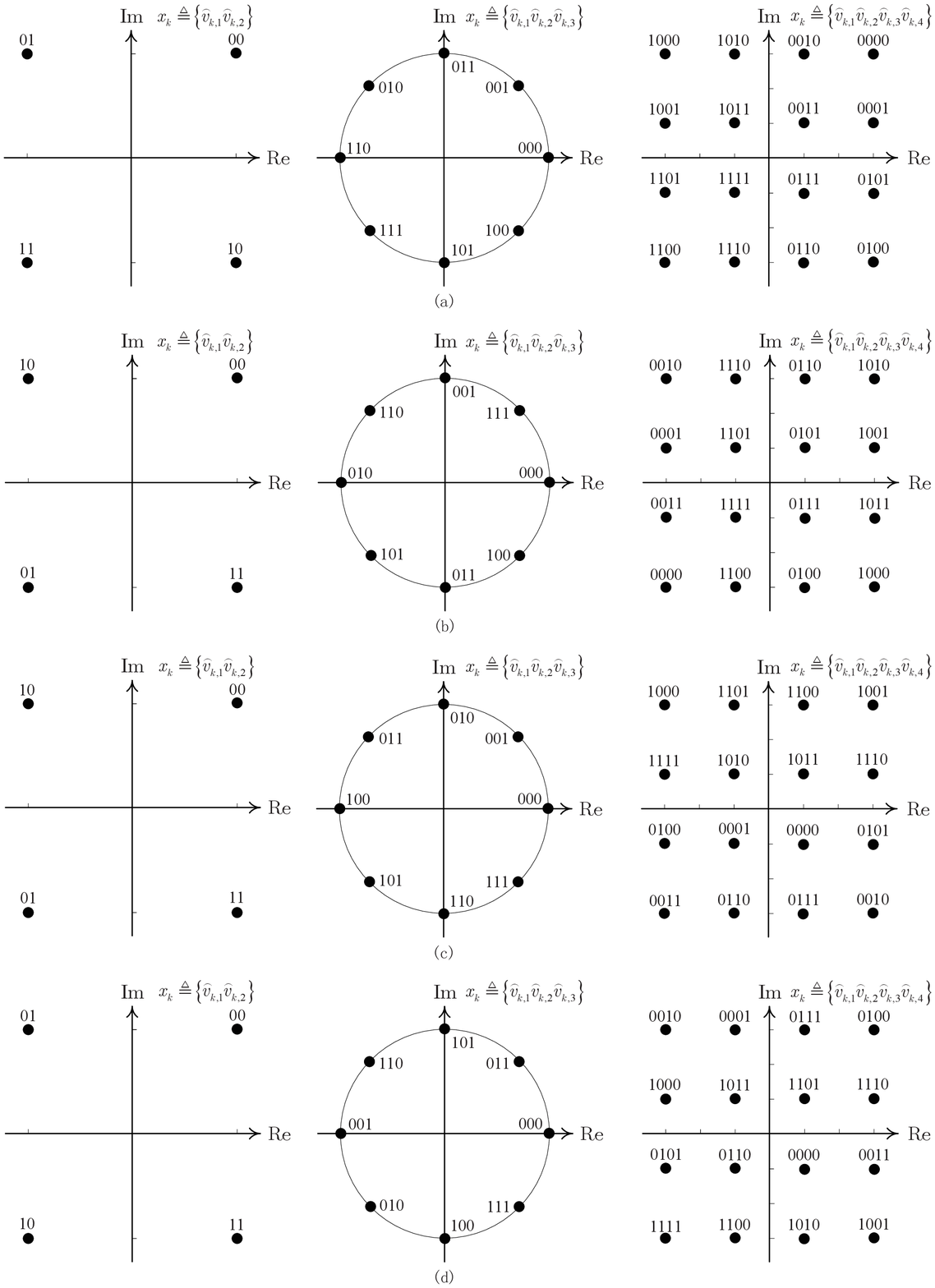}
\vspace{-0.2cm}
\caption{Constellations for the QPSK, $8$PSK, and $16$QAM with (a) Gray labeling, (b) anti-Gray labeling, (c) SP labeling, and (d) MSEW labeling.}
\label{fig:Fig.3V}%\vspace{-0.05cm}
\end{figure*}

%%Example-1
{\bf \textit{Example 1:}} Fig.~\ref{fig:Fig.3V} depicts the constellations of QPSK, $8$PSK, and $16$QAM, with Gray labeling, anti-Gray labeling, SP labeling, and MSEW labeling. Referring to this figure, the anti-Gray labeling, SP labeling, and MSEW labeling have identical constellation for QPSK modulation. %, which is also equivalent to that of MSEW-labeled QPSK.
Indeed, there are only two different possible labeling rules for QPSK constellation, since it consists of four different signal points. Thereby, anti-Gray labeling, SP labeling, and MSEW labeling naturally have the same performance (e.g., capacity) under QPSK modulation.

\begin{figure}[t]
\centering
\subfigure[\hspace{-0.5cm}]{ %% label for first subfigure
\includegraphics[width=3in,height=2.3in]{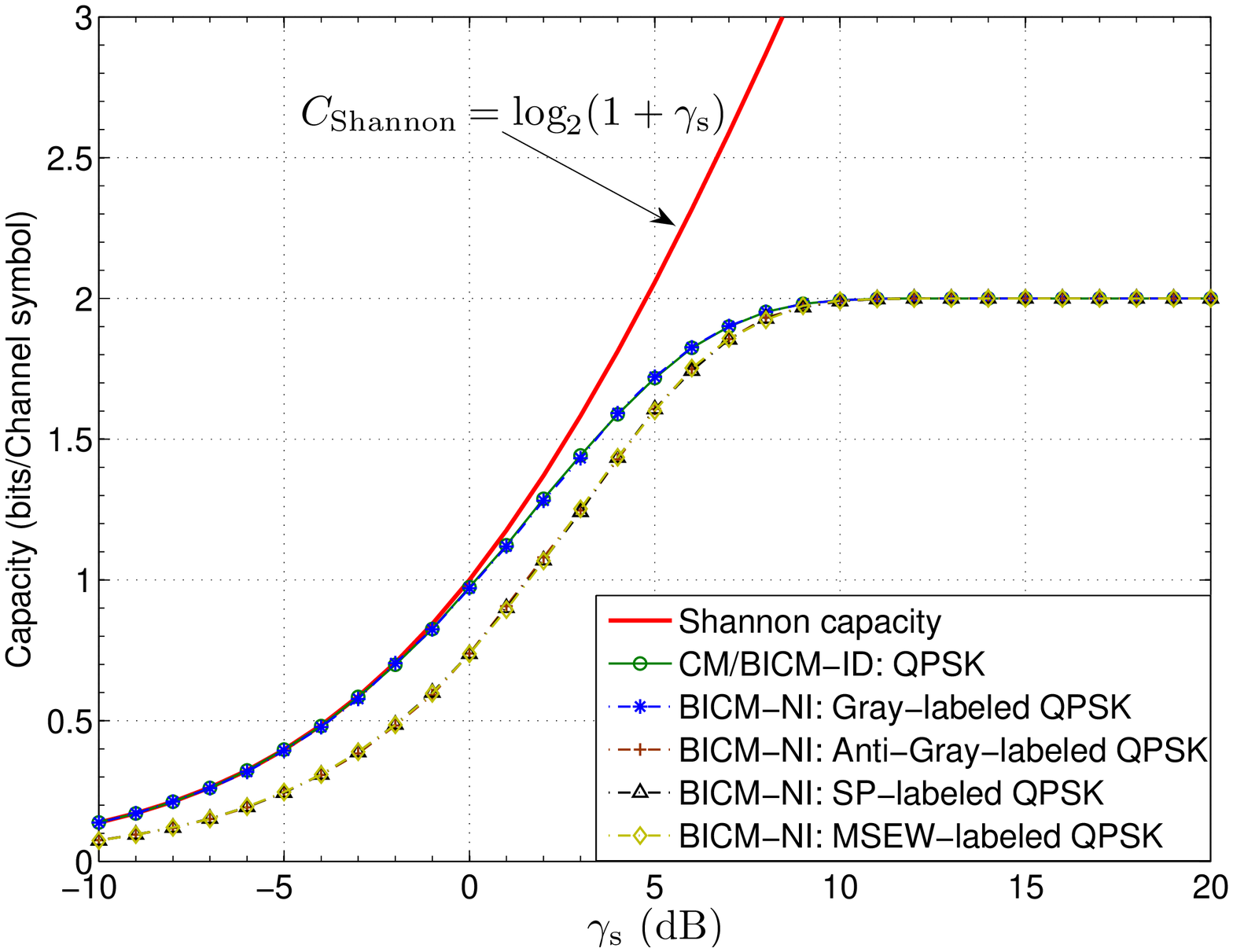}}\vspace{-2mm}
\subfigure[\hspace{-0.5cm}]{ %% label for first subfigure
\includegraphics[width=3.0in,height=2.3in]{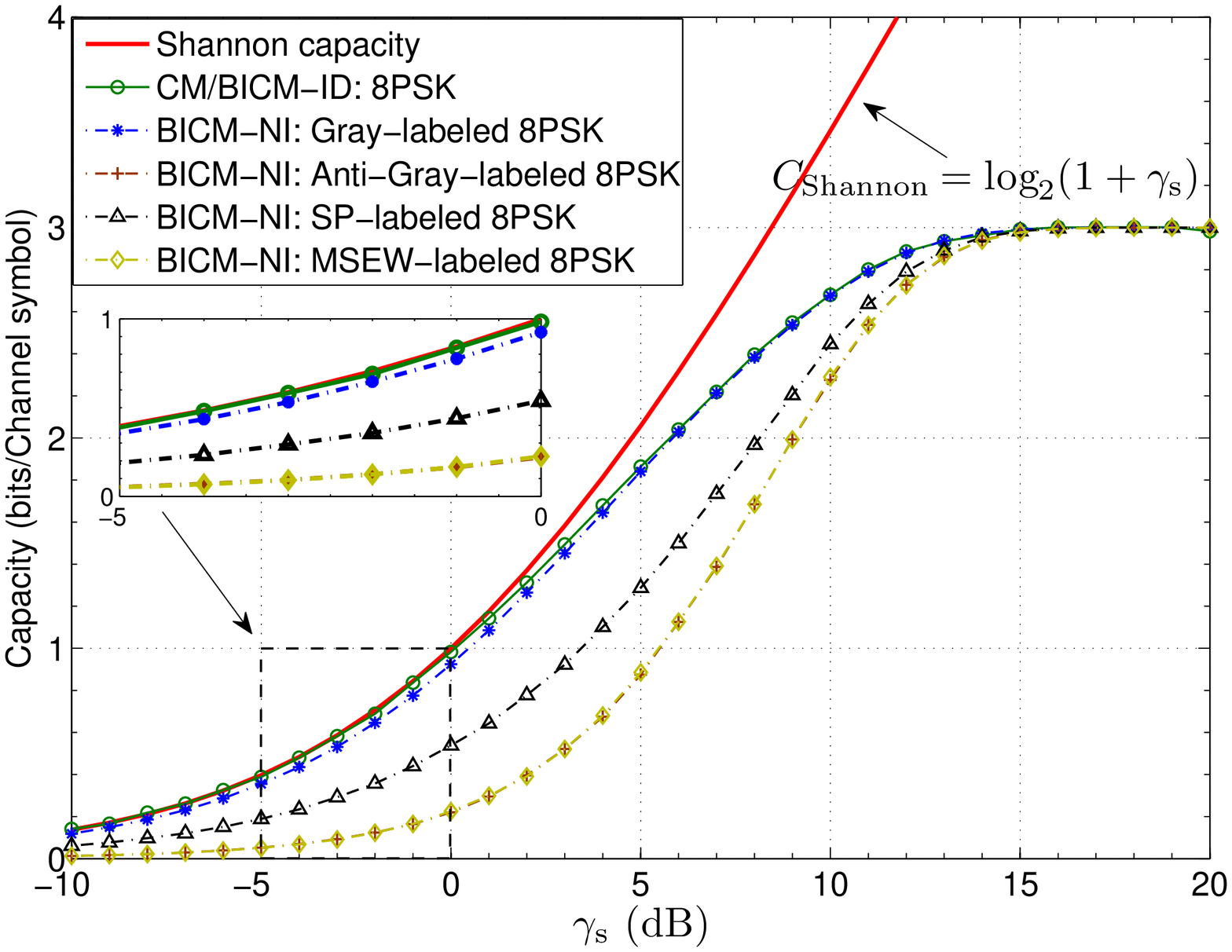}}\vspace{-2mm}
\subfigure[\hspace{-0.5cm}]{ %% label forsecond subfigure
\includegraphics[width=3.0in,height=2.3in]{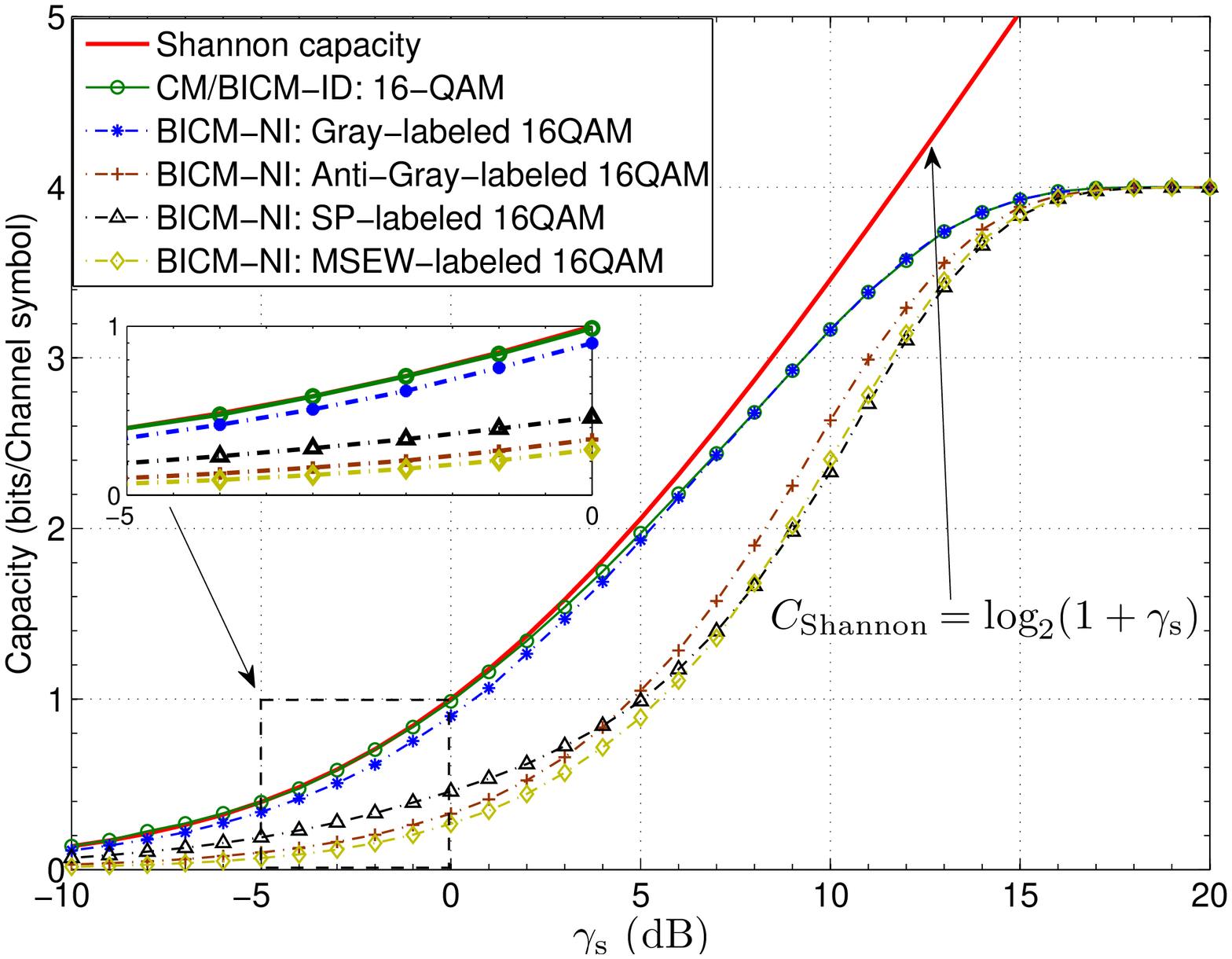}}\vspace{-2mm}
\caption{CM and BICM capacities of (a) QPSK, (b) 8PSK, and (c) 16QAM over an AWGN channel with Gray labeling, anti-Gray labeling, SP labeling, and MSEW labeling.}\vspace{-2mm}
\label{fig:Fig.4V}
\end{figure}

To further analyze the CM and BICM capacities of the QPSK, $8$PSK, and $16$QAM with the above four labelings, consider again Fig.~\ref{fig:Fig.4V} where the Shannon capacity is also included. It can be observed that the Gray labeling achieves the best capacity among all the four labelings in the BICM-NI system. Moreover, the Gray-labeled BICM-NI has capacities closely approaching its corresponding BICM-ID counterparts for QPSK, $8$PSK and $16$QAM modulations. Particularly, for the QPSK modulation, the Gray-labeled BICM-NI and BICM-ID have identical capacity, while the anti-Gray-labeled BICM-NI, SP-labeled BICM-NI, and MSEW-labeled BICM-NI have an obvious capacity gap from the BICM-ID. This indicates that Gray labeling may obtain trivial performance gain with the use of ID, while the other three types of labelings may obtain desirable performance gains in the ID scenario. Motivated by this conjecture, intensive research effort has been carried out to explore the applicability of the Gray labeling in BICM systems. Many evidences have been found to support the view that this labeling rule is only suitable for BICM-NI but is incapable of accomplishing satisfactory performance for BICM-ID \cite{706217,8052124,8740906,1413232,8883091,6809890}. %Apart from these labeling schemes, researchers have endeavored to develop other labeling designs tailored for BICM-ID systems in different transmission environments, which will be discussed later in this article.\vspace{-1mm}
%Fig.3

%%Fig.4
%\begin{figure*}[!t]
%%\vspace{-4mm}
%\centering
%\subfigure[\hspace{-0.6cm}]{ %% label for first subfigure
%\includegraphics[width=2.331in,height=1.92in]{Fig.4a.eps}}
%\subfigure[\hspace{-0.50cm}]{ %% label forsecond subfigure
%\includegraphics[width=2.33in,height=1.92in]{Fig.4b.eps}}
%\subfigure[\hspace{-0.50cm}]{ %% label forsecond subfigure
%\includegraphics[width=2.33in,height=1.92in]{Fig.4c.eps}}
%\vspace{-0.25cm}
%\caption{CM and BICM capacities of (a) QPSK, (b) 8PSK, and (c) 16QAM over an AWGN channel with Gray labeling, anti-Gray labeling, SP labeling, and MSEW labeling.}
%\label{fig:Fig.4V}  %label for entire figure
%\vspace{-2mm}
%\end{figure*}

\subsection{Summary}\label{sect:II-D}%\vspace{-0.2mm}

In this section, the system architecture and transmission mechanism of a general PLDPC-BICM is presented. Some relevant channel models employed in the study of PLDPC-BICM are reviewed and classified into four different categories: AWGN, Nakagami fading, Poisson PPM, and NAND flash-memory channels. Subsequently, the channel capacity for BICM is discussed, which is fundamental theoretical objective for systems design. Finally, major contributions in the development of PLDPC-BICM are summarized and commented.

%%Fig.7
%\begin{figure*}[!t]\vspace{1mm}
%\centering
%\subfigure[\hspace{-0.25cm}]{ %% label for first subfigure
%\includegraphics[width=1.3in,height=1.15in]{fig/Fig.7(a).eps}}\hspace{0.9cm}
%\subfigure[\hspace{-0.3cm}]{ %% label forsecond subfigure
%\includegraphics[width=1.3in,height=1.15in]{fig/Fig.7(b).eps}} \hspace{0.9cm}
%\subfigure[\hspace{-0.25cm}]{ %% label forsecond subfigure
%\includegraphics[width=1.3in,height=1.15in]{fig/Fig.7(c).eps}}
%\caption{The protographs of (a) a rate-$1/2~(3,6)$ regular protograph code, (b) a protograph code
%with a high-degree VN, and (c) an optimized protograph code with VN degrees
%at least $3$.}
%\label{fig:Fig.7}  %label for entire figure
%\end{figure*}

\section{Basic Principles of PLDPC Codes}\label{sect:III}

PLDPC codes represent a sophisticated class of LDPC codes, which have attracted considerable attention due to their excellent performance and simple structure \cite{7112076,5174517}. Motivated by these advantages, many meritorious variants of PLDPC codes have been formulated to satisfy the diverse requirements of modern communication and storage applications. Of most interest are the SC-PLDPC codes that inherit the advantages of both PLDPC codes and convolutional codes \cite{7152893,6852099,5695130}. During the past two decades, the PLDPC codes and their SC variants have been intensively investigated and become a promising practical alternative for spectral-efficiency BICM configurations \cite{cai2020design,7093149,7935433,7968491,7956256,8338131,7723910,8544026,
hager2014improving,7005396,hager2016analysis,7593089,7460483,8883091,9057491,8012533}. In the following, some basic preliminaries of PLDPC codes are presented.%, and then their corresponding theoretical-analysis methods are discussed. %As the attention is on BICM systems, only binary codes are considered.

\subsection{LDPC Codes}\label{sect:III-A}

%LDPC codes can provide capacity-approaching performance under iterative BP decoding in a large number of digital-communication and data-storage systems \cite{ryan2009channel}. %Owing to the appealing advantages, they have already been selected as the channel-coding solution for many communication and storage standards.
An LDPC code ${\bm \Lambda}$ can be represented by either a parity-check matrix or a Tanner graph \cite{ryan2009channel}. Specifically, an LDPC code ${\bm \Lambda}$ with $N$ variable nodes and $M$ check nodes can be represented by an $M \times N$ parity-check matrix $\bH = \{h_{i,j} \}$, where $h_{i,j} \in \{0, 1\}$ is the element in the $i$-th row and $j$-th column of $\bH$. Here, $K=N-M$ is the length of information bits, and $r= K/N$ is the code rate. In particular, each variable node corresponds to one column of $\bH$, while each check node corresponds to one row of $\bH$. If there is an edge between the $j$-th variable node $v_j$ and the $i$-th check node $c_i$, then $h_{i,j}=1$; otherwise, $h_{i,j}=0$. Furthermore, the degree $d_{v_j}$ (resp. $d_{c_i}$) of $v_j$ (resp. $c_i$) is defined as the number of its associated edges, which equals the weight of the $j$-th column (resp. $i$-th row) of $\bH$. According to \cite{richardson2008modern}, an LDPC code can be characterized by the degree-distribution pair, i.e., $(\{d_{v_j}\}, \{d_{c_i}\})$. If both the variable-node degree and the check-node degree are constants (i.e., $d_{v_1}=d_{v_2}=\ldots=d_{v_N}=d_v$ and $d_{c_1}=d_{c_2}=\ldots=d_{c_M}=d_c$), then the LDPC code is called {\em a regular LDPC code} with a rate of $r=1-d_v/d_c$; otherwise, the code is called {\em an irregular LDPC code}. A valid LDPC codeword must satisfy the condition of ${\bm \Lambda}_{\rm v} \bH ^{\rm T} = 0$, where the superscribe ``T" indicates the transpose operation.

There exist various realizations of the parity-check matrix $\bH$ of an LDPC code under a given parameter setting. Typically, one can exploit progressive-edge-growth (PEG) and approximate cycle extrinsic-message-degree algorithms, to produce a well-performing parity-check matrix $\bH$ based on a given degree-distribution pair and a given codeword length \cite{ryan2009channel}. On the basis of the parity-check matrix, a number of techniques have been proposed to construct finite-length LDPC codes for attaining capacity-approaching performance and low implementation complexity over the last half a century. Among all the existing techniques, the QC-LDPC-code-oriented construction algorithms, such as the circulant-based PEG algorithm \cite{7157697,van2012design} and the ``pre-lifted" algorithm \cite{6863699}, are of great interest due to the fact that they can perform using unstructured LDPC codes and possess linear-complexity encoding and higher-throughput decoding. Nonetheless, the detailed construction of QC-LDPC codes is beyond the scope of this article and hence is omitted.
\vspace{-0.5mm}

\subsection{Protograph LDPC Codes}\label{sect:III-B}
%Since the re-discovery of LDPC codes and the inception of their capacity-approaching instances, such codes have become a more mature class of FEC codes than convolutional and turbo codes. In past two decades, a great deal of research effort has been dedicated to improving the code structures towards easy encoding/decoding and simple design/analysis. With the above motivation

In $2003$, a novel type of structured LDPC codes, PLDPC codes, was introduced \cite{2003IPNPR.154C...1T}, which can be generated by expanding a sufficiently small graph/matrix (i.e., protograph/base matrix). %Based on proper expansion methods, the expanded graph (also called {\em derived graph}) corresponding to a PLDPC code can preserve all the features of the protograph.
Specifically, a protograph ${\cal G}_{\rm P} = {\cal (V, C, E)}$ is defined as a relatively small-size Tanner graph, which contains a set of $n_{\rm P}$ variable nodes ${\cal V}_{\rm P}=(v_1,v_2, \ldots, v_{n_{\rm P}})$, a set of $m_{\rm P}$ check nodes ${\cal C}_{\rm P}=(c_1,c_2, \ldots, c_{m_{\rm P}})$, and a set of $\varrho_{\rm P}$ edges ${\cal E}_{\rm P}=\{e_{i,j}\}$, where $e_{i,j}$ denotes the edge connecting the $j$-th variable node $v_j$ to the $i$-th check node $c_i$, $\varrho_{\rm P}=|{\cal E}_{\rm P}|$, and $r_{\rm P}=(n_{\rm P} - m_{\rm P})/n_{\rm P}$ denotes the code rate \cite{7112076,5174517}.\footnote{In this article, a protograph with $m_{\rm P}$ check nodes and $n_{\rm P}$ variable nodes is referred to as an $m_{\rm P} \times n_{\rm P}$ protograph, with size $m_{\rm P} \times n_{\rm P}$.} Moreover, the adjacency matrix of a protograph, whose columns and rows correspond to the $n_{\rm P}$ variable nodes and $m_{\rm P}$ check nodes, respectively, is defined by an $m_{\rm P} \times n_{\rm P}$ base matrix $\bB=\{b_{i,j} \}$. Also, the $(i,j)$-th element $b_{i,j}$ of $\bB$ denotes the number of edges connecting $v_j$ to $c_i$. To generate an $M \times N$ derived graph (i.e., Tanner graph) corresponding to an $M \times N$ parity-check matrix $\bH$, one should perform the ``copy-and-permute (also called {\em lifting})" operation on the protograph ${\cal G}_{\rm P}$, where $M=Z m_{\rm P}, N=Z n_{\rm P}$ and $Z$ is the lifting factor. In other words, the protograph is first copied for $Z$ times to obtain $Z$ individual protographs; the edges of the $Z$ individual replicas of ${\cal G}_{\rm P}$ are subsequently permuted within each type to produce a single derived graph under certain constraints. In particular, a type-$j$ variable node (i.e., a replica of $v_j$) can only be connected to a type-$i$ check node (i.e., a replica of $c_i$) via a type-$(i,j)$ edge (i.e., a replica of $e_{i,j}$) in the lifting procedure. The resultant derived graph or parity-check matrix gives rise to a length-$N$ PLDPC code. Unlike the conventional LDPC codes, parallel edges (i.e., multiple edges between a variable node and a check node) are permitted in a protograph, but they must be carefully permuted in the lifting process such that the derived graph only includes a single edge between any variable node and any check node \cite{2003IPNPR.154C...1T}. Besides, punctured variable nodes may exist in the protograph to either boost the code rate or improve the convergence performance. The code rate of a punctured PLDPC code becomes\vspace{-0.2cm}
%$r_{\rm P}=(n_{\rm P} - m_{\rm P})/(n_{\rm P} - n_{\rm E})=(n_{\rm P} - m_{\rm P})/n_{\rm T}$,
\begin{equation}\vspace{-0.2cm}
r_{\rm P}=\frac{n_{\rm P} - m_{\rm P}}{n_{\rm P} - n_{\rm E}}=\frac{n_{\rm P} - m_{\rm P}}{n_{\rm T}},
\label{eq:new-eq-3}
\end{equation}where $n_{\rm T}$ and $n_{\rm E}$ are the numbers of transmitted variable nodes and punctured variable nodes of the protograph, respectively.

To date, several methods have been proposed to expand a protograph to a derived graph, among which the modified PEG algorithm is the most popular one \cite{van2012design,2008Construction}. For a given lifting algorithm, the performance of a PLDPC code is determined only by the protograph, which tremendously reduces the degree of difficulty in code design and analysis. In accordance with the above reasons, the optimization of a PLDPC code is equivalent to the design of a capacity-approaching protograph associated with a small-size base matrix, which is more concise and reachable than the parity-check matrix of a conventional LDPC code. In this article, the modified PEG algorithm proposed in \cite[Appendix A]{van2012design} is adopted to implement the lifting procedure. %unless otherwise stated.

{\em Remark:} As illustrated in \cite{7308964,6145509}, the QC-LDPC codes can be treated as a special case of the PLDPC codes. A PLDPC code is exactly equivalent to a QC-LDPC code if a circulant-based PEG algorithm \cite{7157697,8291725,2008Construction} is used to perform the ``copy-and-permute" operation on the protograph.
%Fig.6
\begin{figure}[t]
\center
\includegraphics[width=3.61in,height=1.63in]{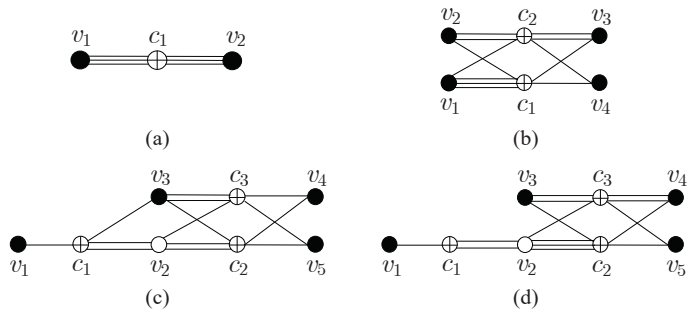}
\vspace{-0.3cm}
\caption{Protograph structures of the rate-$1/2$ (a) regular PLDPC code, (b) RJA PLDPC code, (c) AR$3$A PLDPC code, and (d) AR$4$JA PLDPC code.}
\label{fig:Fig.6V}\vspace{-0.2cm}
\end{figure}
%Fig.5
\begin{figure}[t]
\center
\includegraphics[width=3.25in,height=1.0in]{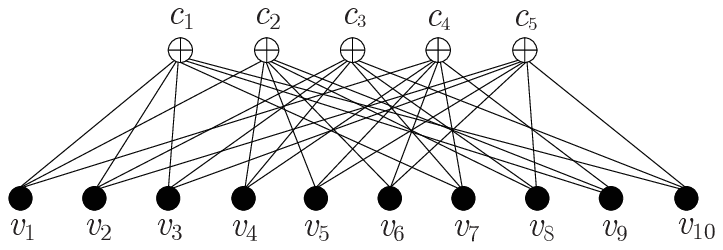}
\vspace{-0.25cm}
\caption{Derived graph of a rate-$1/2$ regular-$(3,6)$ PLDPC code with $m_{\rm P}=1, n_{\rm P}=2, Z=5, M=5,~{\rm and}~N=10$.}
\label{fig:Fig.5V}\vspace{-0.3cm}
\end{figure}

%%Example-2
{\bf \textit{Example 2:}}
The base matrices and protographs of four conventional rate-$1/2$ PLDPC codes, namely regular-$(3, 6)$ PLDPC code \cite{5174517}, repeat-jagged-accumulate (RJA) PLDPC code \cite{6262475}, accumulate-repeat-$3$-jagged-accumulate (AR$3$A) PLDPC code \cite{6253209}, and accumulate-repeat-by-$4$-jagged-accumulate (AR$4$JA) PLDPC code \cite{7152893}, are presented in \eqref{eq:B-Reg-RJA-AR3A-AR4JA} and Fig.~\ref{fig:Fig.6V}, respectively. In this figure, the blank circles represent the punctured variable nodes. Among the four codes, the AR$3$A and AR$4$JA codes are two prestigious types of PLDPC codes because they can respectively achieve excellent performance in the low- and high-SNR regions over AWGN channels. As shown in \cite{5174517,6253209}, introducing a small fraction of punctured variable nodes to the protographs can boost the transmission rates of PLDPC-systems and also improve their decoding thresholds in certain scenarios. The basic principle of puncturing for PLDPC codes is available in \cite{7112076,van2012design,7353121,6253209}.\vspace{-1mm}% and thus skipped here.
\begin{eqnarray}
&{\bB}_{\rm REG} \hspace{-0.5mm}=\hspace{-0.5mm}\left[\begin{array}{ll}
3 & 3 \cr
\end{array}\right]\hspace{-0.5mm},~
\bB_{\rm RJA} =
\left[\begin{array}{llll}
3 & 1 & 1 & 1  \cr
1 & 2 & 2 & 1  \cr
\end{array}\right]\hspace{-0.5mm},\nonumber\\\vspace{2.5mm}
&\hspace{-2mm}\bB_{\rm AR3A} \hspace{-0.6mm}=\hspace{-0.6mm} \left[\begin{array}{ccccc}
1 & 2 & 1 & 0 & 0 \cr
0 & 2 & 1 & 1 & 1 \cr
0 & 1 & 2 & 1 & 1 \cr
\end{array}\right]\hspace{-0.5mm},~
\bB_{\rm AR4JA} \hspace{-0.6mm}=\hspace{-0.6mm} \left[\begin{array}{ccccc}
1 & 2 & 0 & 0 & 0 \cr
0 & 3 & 1 & 1 & 1 \cr
0 & 1 & 2 & 2 & 1 \cr
\end{array}\right]\hspace{-0.5mm}.~~
\label{eq:B-Reg-RJA-AR3A-AR4JA}
\end{eqnarray}

As a further advance, with the lifting factor $Z=5$, one can promptly obtain the derived graph (i.e., Fig.~\ref{fig:Fig.5V}) corresponding to a length-$10$ regular-$(3, 6)$ PLDPC code by performing a randomly lifting operation \cite{5174517,8269289} on the protograph in Fig.~\ref{fig:Fig.6V}(a).

\subsection{Spatially-Coupled PLDPC Codes}\label{sect:III-C}

As a meritorious variant of PLDPC codes, SC-PLDPC codes are capable of achieving desirable convolutional gains without sacrificing the dominant feature of protographs \cite{7152893,6262475}. %Also, SC-PLDPC codes possess better minimum distance property compared with the PLDPC codes due to their naturally convolutional structure.
%%Indeed,
%In particular, SC-PLDPC codes seamlessly combine the best advantages of PLDPC codes and convolutional codes, e.g., near-capacity decoding thresholds, linear-minimum-distance-growth property, easy design and implementation \cite{6262475}.
%Thus, SC-PLDPC codes serve as a promising and practical coding candidate for spectral-efficiency communication systems.
In general, there are two major categories of SC-PLDPC codes, i.e., {\em TE-SC-PLDPC codes and TB-SC-PLDPC codes}.%, which are further discussed below.

\subsubsection{Terminated SC-PLDPC Codes}\label{sect:III-C-1}

A TE-SC-PLDPC code can be constructed by employing an edge-spreading rule to couple together $L_{\rm SC}$ disjoint replicas of a protograph into a single coupled chain, where $L_{\rm SC}$ is defined as the coupling length. One can proceed the construction by replicating a protograph $L_{\rm SC}$ times, where these individual protographs are indicated with a time index $t~(t=1,2,\ldots, L_{\rm SC})$. Suppose that the PLDPC code consists of $n_{\rm P}$ variable nodes, $m_{\rm P}$ check nodes, and $\varrho_{\rm P}$ edges. The $\varrho_{\rm P}$ edges emanating from the $n_{\rm P}$ variable nodes at time instant $t$ are spread over the $(\varsigma+1)m_{\rm P}$ check nodes at time instants $t, t+1, \ldots, t + \varsigma$, according to a predefined edge-spreading rule, where $0<\varsigma<L_{\rm SC}$ is the coupling width or edge-spreading factor. Owing to the termination effect, $\varsigma m_{\rm P}$ additional check nodes are inserted at time instants $ L_{\rm SC}+1, L_{\rm SC}+2, \ldots, L_{\rm SC}+\varsigma$ of the TE-SC protograph, which leads to a lower code rate with respect to the uncoupled PLDPC code. In other words, the code rate of a TE-SC-PLDPC code equals%\vspace{-0.2cm}
%$r_{\rm TE\text{-}SC} = 1 - ((L_{\rm SC}+ \varsigma) m_{\rm P})/(L_{\rm SC} n_{\rm P}) =  r_{\rm P} - (\varsigma m_{\rm P})/(L_{\rm SC} n_{\rm P})$,
\begin{equation}
r_{\rm TE\text{-}SC} = 1 - \frac{(L_{\rm SC}+ \varsigma) m_{\rm P}} {L_{\rm SC} n_{\rm P}} =  r_{\rm P} - \frac{\varsigma m_{\rm P}}{L_{\rm SC} n_{\rm P}},
\label{eq:new-eq-4}
\end{equation}where $r_{\rm P}$ is the code rate of the uncoupled PLDPC code. Moreover, $ \nu_{\rm SC}= (\varsigma+1)Zn_{\rm P} $ is defined as the constraint length of the TE-SC-PLDPC code, where $Z$ is the lifting factor. Alternatively, a TE-SC-PLDPC code can be obtained by directly terminating the coupled chain of a convolutional PLDPC code with a finite coupling length. Although TE-SC-PLDPC codes are able to asymptotically achieve the optimal MAP decoding thresholds of the underlying uncoupled PLDPC codes with BP decoding (called ``{\em threshold saturation}") due to the structured irregularity, they suffer from non-trivial code-rate loss with a finite coupling length \cite{7152893,6863699}. The rate loss of such codes can be an arbitrarily small value if the coupling length approaches infinity (i.e., $L_{\rm SC} \to +\infty $), but this leads to extremely long codeword lengths. The above drawback severely restricts the practical applications of TE-SC-PLDPC codes. %in practical communication systems.
%Fig.7
\begin{figure*}[t]\vspace{-0.2cm}
\center
\includegraphics[width=6.3in,height=2.15in]{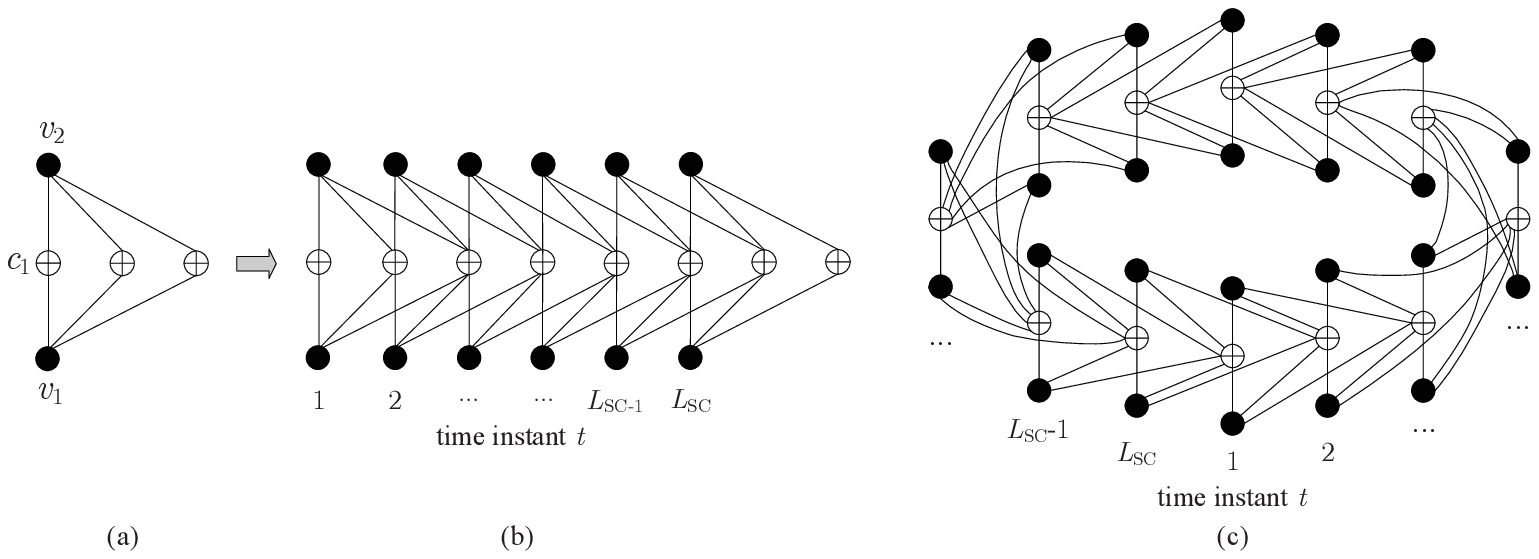}
\vspace{-0.2cm}
\caption{Protograph structures of the regular-$(3, 6)$ SC-PLDPC codes: (a) edge-spreading rule, (b) TE-SC-PLDPC code, and (c) TB-SC-PLDPC code, where the coupling width is $\varsigma =2 $ and coupling length is $L_{\rm SC}$.}
\label{fig:Fig.7V}\vspace{-0.25cm}
\end{figure*}

On the other hand, the edge spreading of a protograph is equivalent to the division of a base matrix. In particular, the base matrix $\bB$ of a protograph is split to $\varsigma+1$ sub-base matrices during the edge-spreading procedure, denoted by $\bB = \bB_{\rm S,1} +\bB_{\rm S,2}+\ldots+\bB_{\rm S, \varsigma+1}$, where $\bB_{\rm S,\mu}$ is the $\mu$-th~($\mu=1,2,\ldots, \varsigma+1$) sub-base matrix, and the sizes of all the sub-base matrices are identical to that of the original base matrix. Analogously to the protograph construction, combining $L_{\rm SC}$ groups of $\varsigma+1$ sub-base matrices results in an $m_{\rm P} (L_{\rm SC} + \varsigma) \times n_{\rm P} L_{\rm SC}$ base matrix of the TE-SC-PLDPC code, as shown in \cite[eq.\,(8)]{7152893}.
%\begin{equation}
%\bB_{\rm TE\text{-}SC}=
%  \begin{array}{c}
%                      L_{\rm SC}     \cr
%\begin{bmatrix}
% \overbrace{
%  \begin{array}{lllll}
%    \bB_{\rm S,1}        &                          &          &                           &                  \cr
%    \bB_{\rm S,2}        & \bB_{\rm S,1}            &          &                           &                   \cr
%    ~\vdots              & \bB_{\rm S,2}            &          &                           &                    \cr
%\bB_{\rm S, \varsigma+1} & ~\vdots                  &          &                           &                     \cr
%                         & \bB_{\rm S, \varsigma+1} & \ddots  &                           &                      \cr
%                         &                          & \ddots  &  \bB_{\rm S,1}            &                      \cr
%                         &                          & \ddots  &  \bB_{\rm S,2}            &  \bB_{\rm S,1}       \cr
%                         &                          &          & ~\vdots                   &  \bB_{\rm S,2}        \cr
%                         &                          &          &  \bB_{\rm S, \varsigma+1} & ~\vdots                \cr
%                         &                          &          &                           &  \bB_{\rm S, \varsigma+1} \cr
% \end{array}}
%\end{bmatrix}\label{eq:TE-base}
%  \end{array}.
%\end{equation}

\subsubsection{Tail-biting SC-PLDPC Codes}\label{sect:III-C-2}

To tackle the rate-loss weakness, a tail-biting method was introduced instead of the direct termination method to terminate the coupled chain \cite{7152893}, which yields a new type of SC-PLDPC codes, named {\em TB-SC-PLDPC codes}. Specifically, a TB-SC-PLDPC code can be constructed on the basis of a TE-SC-PLDPC code by combining the check nodes at time instants $t=L_{{\rm SC}}+1, L_{{\rm SC}}+2, \ldots, L_{\rm SC}+\varsigma$ with the check nodes at time instants $t=1, 2, \ldots, \varsigma$, into $\varsigma$ check nodes. As a result, the total number of check nodes decreases to $m_{\rm P} L_{\rm SC}$ and the code rate of the TB-SC-PLDPC code becomes $r_{\rm TB\text{-}SC} = 1 - m_{\rm P}/n_{\rm P} =  r_{\rm P}$. The TB-SC-PLDPC code shares an identical structure and degree-distribution pair as the uncoupled PLDPC code because of the tail-biting operation; but the former possesses a relatively better property of linear and minimum-distance growth, due to the spatially coupling operation. Yet, the ``threshold saturation" phenomenon is no longer retained for a TB-SC-PLDPC code as the irregularity feature vanishes \cite{7152893}.

Recently, some novel techniques have been proposed to enhance the convergence performance of TB-SC-PLDPC codes, making them a desirable choice for practical systems \cite{8281448}. %According to the above-described construction principle, on can easily get the $m_{\rm p} L_{\rm SC} \times n_{\rm p} L_{\rm SC}$ base matrix of the TB-SC-PLDPC code \cite[eq.\,(10)]{7152893}.
{\bf \textit{Example 3:}} Fig.~\ref{fig:Fig.7V}(a) shows a simple edge-spreading rule with $\varsigma=2$ for a rate-$1/2$ regular-$(3, 6)$ PLDPC code. As a further advance, the protograph structures of the regular TE-SC-PLDPC code and TB-SC-PLDPC code, constructed from the regular-$(3, 6)$ PLDPC code by using the above edge-spreading rule, are presented in Fig.~\ref{fig:Fig.7V}(b) and Fig.~\ref{fig:Fig.7V}(c), respectively. Supposing $L_{\rm SC}=5$, the regular TE-SC-PLDPC code and TB-SC-PLDPC code have rates $r_{\rm TE\text{-}SC}=3/10$ and $r_{\rm TB\text{-}SC}=1/2$, respectively. In comparison, TB-SC-PLDPC codes are more likely to realize high spectral-efficiency communication and storage than their TE relatives in finite-coupling-length scenarios.

\subsection{Theoretical-Analysis Tools}\label{sect:IV}

Designing and implementing capacity-approaching PLDPC-BICM is a major objective for communication and storage. To achieve this goal, it is important to develop efficient theoretical tools and performance metrics tailored for such serially concatenated schemes \cite{i2008bit}. Initially, some prestigious numerical analysis techniques, such as density evolution \cite{910577} and EXIT function \cite{1194444,1291808}, were deployed to characterize the convergence performance of all the components (e.g., encoder, constellation, interleaver, and decoder) of PLDPC-BICM systems in the low-SNR region. As a complementary analysis tool, AWE was designed to predict the convergence performance of PLDPC codes in BICM systems in the high-SNR region \cite{5174517,6262475}. Especially, the combination of the EXIT/DE and AWE acts as a complete analytical configuration for the PLDPC codes in BICM systems, which is able to accurately measure the asymptotic performance in the whole SNR region. Besides, harmonic mean of the minimum squared Euclidean distance (HMMSED) was introduced as a secondary analytical metric for the PLDPC-BICM systems, which is of great usefulness for estimating the attainable performance of constellations \cite{924878,8611290}. Although some other theoretical tools have been developed for BICM systems, the EXIT, AWE and HMMSED analyses have drawn perhaps the greatest attention from the field of PLDPC-BICM because they provide accurate performance estimates, and also facilitate the system design. %Here, we summarize the above three theoretical-analysis methodologies and their corresponding performance metrics.

\subsubsection{PEXIT Algorithm}\label{sect:IV-A}

In $2001$, the DE algorithm \cite{910577} was developed to analyze the convergence performance of LDPC codes in an iterative decoding processor. %The aim of DE algorithm is to determine the decoding thresholds of LDPC codes under iterative decoding by means of tracking the evolution of the PDF of their corresponding LLRs.\footnote{The decoding threshold is defined as the minimum SNR that guarantee a reliable transmission for an LDPC-coded system in an asymptotic fashion.}
%Thereby, the system can asymptotically achieve reliable communication above this threshold, but cannot do so below it.}
Despite the accurate characterization of the convergence performance, the DE algorithm must track the entire PDF of LLRs and thus suffers from high computational complexity. By contrast, EXIT algorithm takes the MI into account and produces a visualized extrinsic-MI-convergence chart to determine the decoding thresholds under iterative decoding \cite{1194444,1291808}. %\footnote{The decoding threshold is defined as the minimum SNR that guarantee a reliable transmission for an LDPC-system in an asymptotic fashion.}
Building upon the MI metric, the EXIT algorithm can greatly simplify the threshold calculation of LDPC-systems without deteriorating the accuracy. Moreover, it has strong robustness against the variation of channel conditions. The EXIT algorithm can be directly applied to a large variety of channels, modulations and decoders, with only slight modifications. Also, it is very convenient to use the EXIT algorithm to characterize the decoding trajectories of concatenated systems, including BICM and non-binary CM systems. Thanks to the above superiorities, EXIT algorithms have been used as the dominant theoretical tool for analysis and design of BICM systems \cite{6359874,8338131,1413232,8186234,6241383,7935441,7801862,8611290,6952165,5960810}.

However, the standard EXIT algorithm, which yields two-dimensional extrinsic-MI-convergence charts and calculates the decoding thresholds for conventional LDPC codes, is not applicable to the PLDPC codes in BICM systems. More precisely, the PLDPC codes may contain some degree-$1$ variable nodes, punctured variables nodes, and parallel edges in the corresponding base matrices, which are generally not allowed in conventional LDPC codes \cite{5174517}. Moreover, some PLDPC codes possessing different base-matrix representations may share the same degree-distribution pair \cite{liva2006design}. The standard EXIT algorithm was developed based on the degree-distribution pairs and hence does not have the capability to deal with the above issues. To address this issue, a multi-dimensional EXIT algorithm, called {\em PEXIT algorithm}, was developed to analyze the statistical convergence property of PLDPC codes \cite{liva2006design,9424609}. In particular, the PEXIT algorithm tracks the convergence behavior of extrinsic MIs for each variable node and check node in the protograph, but not their average MIs specified by the degree-distribution pair. In spite of a lack of visualized MI-convergence chart, the PEXIT algorithm can yield accurate decoding thresholds by dealing with a small-size protographs \cite{8740906,8019817}. In parallel with the asymptotic PEXIT algorithms, some finite-length PEXIT algorithms were proposed for estimating the finite-length performance of PLDPC codes \cite{8269289,6253209}. Interested readers are referred to \cite{8740906,8883091} for the detailed steps of the PEXIT algorithm.
%in certain scenarios

{\em Remark:} The decoding threshold of a PLDPC code in a BICM-ID system is determined by the types of code, constellation as well as the interleaver used. In this sense, the PEXIT algorithm can be effectively used to guide the design of the BICM-ID systems with these three components.

{\bf \textit{Example 4:}} Table~\ref{tab:III} presents the decoding thresholds of the regular-$(3, 6)$ PLDPC code derived by the PEXIT algorithm in BICM systems with four different constellations over an AWGN channel. As seen, the Gray labeling exhibits the lowest decoding threshold among all the four labelings, irrespective of the modulation order and receiver type. However, for a fixed modulation order, the threshold difference between the ID and NI for the Gray labeling is much smaller than those for the other three labelings, which indicates that the former can only achieve a trivial iterative gain. It was demonstrated in \cite{706217,8740906,6809890} that the Gray labeling is best suited to BICM-NI, but is not well suited to BICM-ID, in various transmission environments.
%Table 3
\begin{table}[t]\scriptsize
\caption{Decoding thresholds $(E_b/N_0)_{\rm th}~{\rm(dB)}$ of the regular-$(3, 6)$ PLDPC code in BICM systems with four different labeling schemes over an AWGN channel.%, where the modulations used are $8$PSK and $16$QAM.
The maximum numbers of global iterations in the NI and ID cases are set as $t_{\rm GL,max}=1$ and $t_{\rm GL,max}=8$, respectively, while the maximum number of the local iterations is set as $t_{\rm BP,max}=25$.}\vspace{-1mm}
\centering
\begin{tabular}{|c|c|c|c|c|}
\hline
\diagbox[width=11em,height=2em,trim=l]{\hspace{-1mm}Modulation Type}{Constellation} & Gray & Anti-Gray & SP & MSEW \\ \hline
NI: $8$PSK & $2.587$ & $6.195$ & $5.013$ & $6.188$ \\ \hline
ID: $8$PSK & $2.323$ & $4.307$ & $3.049$ & $4.429$ \\ \hline
NI: $16$QAM & $3.415$ & $5.927$ & $6.687$ & $6.514$ \\ \hline
ID: $16$QAM & $3.201$ & $4.385$ & $4.724$ & $5.027$ \\ \hline
\end{tabular}
\label{tab:III}
\vspace{-2mm}
\end{table}%

\subsubsection{Asymptotic-Weight-Enumerator Analysis}\label{sect:IV-B}

According to \cite{ryan2009channel,7112076,8269289,5174517}, %. To tackle this intractable problem,
the AWE can effectively estimate the minimum Hamming weight (or Hamming distance) and its distribution of a PLDPC code ensemble, which are of great usefulness in predicting the asymptotic performance of a PLDPC code in the high-SNR region. % \cite{7112076,8269289,5174517}. %The AWE can be also used to yield an upper bound on the decoding threshold under maximum-likelihood (ML) decoding algorithm.
Moreover, PLDPC code ensembles, which have the desirable minimum-distance property, possess relatively fewer pseudocodewords and trapping sets compared with the ensembles that do not have such a property \cite{7353121,6262475}.
%\footnote{According to \cite{7353121,6262475}, the pseudocodewords and trapping sets are very harmful for the convergence of PB-LDPC codes under iterative decoding.}
As illustrated in \cite{7112076,1605713}, the PLDPC codes drawn from the ensembles having minimum distance and growing linearly with the codeword length (i.e., the linear-minimum-distance-growth property) always possess low error floors under BP decoding. Many theoretical and experimental evidences have suggested a performance tradeoff between the low-SNR region and high-SNR region for a PLDPC code \cite{7152893,7339427}. %In this subsection, we concisely portray the AWE analysis of PB-LDPC codes so as to formulate an effective solution for preserving their linear-minimum-distance-growth property in BICM systems.

\vspace{0.5mm}
\noindent{\bf (1) AWE Analysis for PLDPC Codes:} Consider a PLDPC-BICM system with the same parameters as in Section~\ref{sect:III-B}. The normalized logarithmic AWE function of the PLDPC code ensemble is defined as\vspace{-0.2cm}
%$r(\delta) =\lim_{N_{\rm T} \to \infty} \sup(\ln ({\cal A}_{\omega_{\rm HAM}})/N_{\rm T})$,
\begin{equation}\vspace{-0.2cm}
r(\delta) =\lim_{N_{\rm T} \to \infty} \sup(\ln ({\cal A}_{\omega_{\rm HAM}})/N_{\rm T}),
\label{eq:new-eq-5}
\end{equation}where $\sup(\cdot)$ is the supremum operation, $\delta$ is the normalized weight, ${\cal A}_{\omega_{\rm HAM}}$ is the ensemble weight enumerator, $\omega_{\rm HAM} = \delta N_{\rm T}$ is the Hamming weight (or distance), and $N_{\rm T}$ is the transmitted codeword length. One can easily examine the AWE function to identify the existence of the property of linear minimum-distance growth and to derive the {\em minimum-Hamming-distance growth rate} (MHDGR). Assume that the AWE begins with the first zero-crossing at $\delta =0$ and has the second zero-crossing at $\delta =\delta_{\rm HAM} > 0$. If $r(\delta)<0$ for all $\delta \in (0, \delta_{\rm HAM})$, then $\delta_{\rm HAM}$ is called the MHDGR of the code ensemble. Under this condition, the minimum Hamming distance grows linearly with the transmitted codeword length with an arbitrarily high probability in an asymptotic fashion. Hence, the PLDPC code ensembles with valid MHDGRs are superior to their counterparts that do not have this property in the high-SNR region. Furthermore, a larger MHDGR always leads to a better asymptotic performance in the high-SNR region \cite{6145509}.

\begin{figure}[tbp]
\center
\includegraphics[width=2.8in,height=2.1in]{{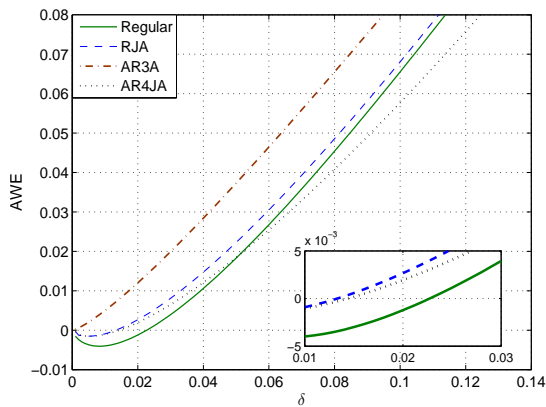}}
\vspace{-2.5mm}
\caption{AWE curves of the rate-$1/2$ regular-$(3, 6)$ PLDPC code, RJA code, AR3A code, and AR4JA code.}
\label{fig:Fig.9V}\vspace{-3mm}  %label for entire figure
\end{figure}

{\bf \textit{Example 5:}} Fig.~\ref{fig:Fig.9V} shows the AWE curves of the rate-$1/2$ regular PLDPC code, RJA code, AR$3$A code, and AR$4$JA code. It is apparent that the AR$3$A code does not have the property of linear minimum-distance growth, due to the lack of positive zero-crossing, while the other three types of PLDPC codes have such a property. In particular, the MHDGRs of the regular PLDPC code, RJA code and AR4JA code are $0.023, 0.013~{\rm and}~0.014$, respectively. This suggests that the AR$3$A code is more prone to error-floor behavior with respect to other three types of codes.
%As a further advance, Abu-Surra {\em et al.} have proposed a particular class of PLDPC codes with degree-$2$ variable nodes so as to achieve capacity-approaching decoding thresholds and to preserve linear-minimum-distance-growth property simultaneously \cite{5695100}.
%It was proved in \cite{5695100} that if there is no loop in the sub-graph constructed by the degree-$2$ variable nodes and their associated check nodes within a protograph, then the PLDPC code benefits from the linear-minimum-distance-growth property (e.g., the RJA and AR$4$JA codes in Fig.~\ref{fig:Fig.4V}).

\vspace{0.5mm}
\noindent{\bf (2) AWE Analysis for SC-PLDPC Codes:} Consider an SC-PLDPC-BICM system, in which the channel code is constructed by combining a group of disjoint protograph replicas together into a spatially-coupled chain. In contrast to the uncoupled PLDPC codes, the constraint length (i.e., $ \nu_{\rm SC}  = (\varsigma +1) Z n_{\rm P}$) of an SC-PLDPC code specifies the maximum number of variable nodes associated with the check nodes at any time instant $t$ in its corresponding derived graph. Precisely speaking, the constraint length for an SC-PLDPC code is equivalent to the codeword length for an uncoupled PLDPC code.
%In particular, the constraint length for an SC-PLDPC is defined as $ \nu_{\rm SC}  = Z(\varsigma +1) n_{\rm P}$.

As illustrated in \cite{6262475}, the minimum free distance $d_{\rm free}$ is more appropriate for evaluating the asymptotic performance of SC-PLDPC codes in the high-SNR region with respect to the minimum Hamming distance. Here, the minimum free distance of an SC-PLDPC code ensemble is defined as the minimum Hamming distance between any two individual codewords drawn from this code ensemble. An SC-PLDPC code ensemble is said to be asymptotically good in the high-SNR region if its minimum free distance grows linearly with the constraint length (i.e., $d_{\rm free} (L_{\rm SC}) = \delta_{\rm free} (L_{\rm SC}) \nu_{\rm SC}$) with an arbitrarily high probability, where $\delta_{\rm free} (L_{\rm SC})$ is defined as the {\em minimum free-distance growth rate} (MFDGR) for the code ensemble with a coupling length $L_{\rm SC} >\varsigma$. Benefiting from the convolutional effect, the SC-PLDPC codes usually possess better minimum-distance property than the corresponding uncoupled PLDPC codes. %In other words, the MFDGRs of SC-PLDPC codes are significantly larger than the MHDGRs of their corresponding uncoupled PLDPC codes.
In the following, both the minimum free distance and minimum Hamming distance are referred to as {\em minimum distance} unless ambiguity may arise.

It was pointed out in \cite{7152893} that the SC-PLDPC codes can retain the linear-minimum-distance-growth property if their uncoupled PLDPC codes have such a property.
The average minimum free distance of the TE-SC-PLDPC code ensemble is upper-bounded by the average minimum Hamming distance of its corresponding block-code ensemble, while that of the TB-SC-PLDPC code ensemble is lower-bounded by the average minimum Hamming distance of its corresponding block-code ensemble \cite{6262475}. With this property, the upper bound and lower bound of the MFDGRs for TE- and TB-SC-PLDPC codes can be respectively derived as \cite[eqs.\,(18)]{6262475} and \cite[eqs.\,(19)]{6262475}.
The above two bounds will converge to the same value as $L_{\rm SC}$ becomes large enough.
%More importantly, the minimum free distances of the TE-SC-PLDPC code ensemble and TB-SC-PLDPC code ensemble are respectively upper-bounded and lower-bounded by their corresponding average minimum Hamming distances.
%Moreover, one has
%%\begin{subequations}
%%\begin{equation}
%%\begin{array}{ll}
%$\delta_{\rm free}^{\rm TE} (L_{\rm SC})  \le \frac{L_{\rm SC} \delta_{\rm HAM}^{\rm TE} (L_{\rm SC})}  {\varsigma +1},~{\rm and}~
%\delta_{\rm free}^{\rm TB} (L_{\rm SC})  \ge \frac{L_{\rm SC} \delta_{\rm HAM}^{\rm TB} (L_{\rm SC})} {\varsigma +1}$,
%%\label{eq:MFDGR-TE-TB}
%%\end{array}
%%\end{equation}
%%%\end{subequations}
%where $\delta_{\rm free}^{\rm TE/TB} (L_{\rm SC})$ and $\delta_{\rm HAM}^{\rm TE/TB} (L_{\rm SC})$ are the MFDGR and MHDGR for the TE/TB-SC-PLDPC code, respectively \cite{6262475}. The above two bounds will converge to the same value as $L_{\rm SC}$ becomes large enough.

{\em Remark:} The minimum-distance metrics are only relevant to the type of code ensemble, but irrelevant to other components of PLDPC-BICM. Moreover, there exists a trade-off between the decoding threshold and minimum distance for a PLDPC code.

\begin{figure}[tbp]
\center
\includegraphics[width=2.8in,height=2.1in]{{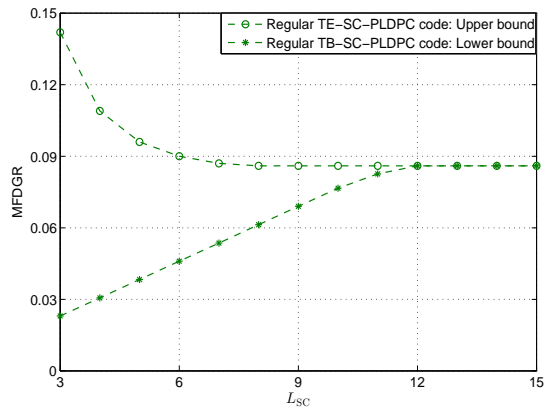}}
\vspace{-2mm}
\caption{MFDGR bounds of the regular-$(3, 6)$ SC-PLDPC codes with $\varsigma=2$. }
\label{fig:Fig.10V}
\vspace{-4mm}
\end{figure}

{\bf \textit{Example 6:}} Fig.~\ref{fig:Fig.10V} illustrates the MFDGR bounds of the rate-$1/2$ regular-$(3, 6)$ SC-PLDPC codes. %where the coupling width is $\varsigma=2$.
It can be observed that the upper bound and lower bound on MFDGRs of the regular SC-PLDPC code ensembles converge to the same value (i.e., $\delta_{\rm free} (L_{\rm SC}) = 0.086$) when the coupling length exceeds $12$, which is much larger than the MHDGR ($\delta_{\rm HAM} = 0.023$) of their uncoupled PLDPC code ensemble. Similar conclusion can be made for the RJA SC-PLDPC code ensembles \cite{7152893,8398231}.
%Likewise, the two bounds on MFDGRs of the RJA SC-PLDPC code ensembles simultaneously converge to $\delta_{\rm free} (L_{\rm SC}) = 0.056$ when $L_{\rm SC} \ge 9$, which is also much larger than the MHDGR ($\delta_{\rm HAM} = 0.013$) of their uncoupled counterpart.
%Thereby, the SC-PLDPC codes may achieve better high-SNR performance than their original PLDPC codes.
%Fig.10
%\begin{figure}[tbp]
%%\vspace{-0.5mm}
%\centering
%\subfigure[\hspace{-0.8cm}]{ %% label for first subfigure
%\includegraphics[width=3.0in,height=2.3in]{Fig.10(a).eps}}
%\subfigure[\hspace{-0.8cm}]{ %% label forsecond subfigure
%\includegraphics[width=3.0in,height=2.3in]{Fig.10(b).eps}}
%\vspace{-0.2cm}
%\caption{MFDGR bounds of the rate-$1/2$ (a) regular-$(3, 6)$ SC-PLDPC codes with $\varsigma=2$ and (b) irregular RJA SC-PLDPC codes with $\varsigma=1$. }
%\label{fig:Fig.10V}  %label for entire figure
%\vspace{-4mm}
%\end{figure}

\subsubsection{Harmonic Mean of Minimum Squared Euclidean Distance Analysis}\label{sect:IV-C}
%Apart from Hamming/free distance, Euclidean distance also plays a significant role in the overall performance of PB-LDPC-coded BICM systems.
%The MFDGR and MHDGR mentioned in Section~\ref{sect:IV-B} can only specify the performance of codes rather than the constellations (including the labeling function).

Aiming to analyze the constellation performance in BICM systems, Euclidean distance was introduced in \cite{669123}. %to derive the BER bounds of BICM systems. Actually, the BER induced by transmitted symbol errors is mainly determined by the minimum Euclidean distance between two adjacent signal points in the constellation.
In the past two decades, several constellation design methods were proposed to maximize the minimum squared Euclidean distance (MSED) \cite{1413232}. However, the MSED metric cannot ensure an accurate BER bound. Consequently, the harmonic mean of the MSED (i.e., HMMSED) was put forward to serve as a more relevant cost function for the BER bound of BICM systems \cite{924878,8611290,5960810}. Especially, the asymptotic performance of the signal constellations in BICM systems can be inherently interpreted by the HMMSED. Specifically, the Hamming/free distance and HMMSED are two dominant factors affecting the performance of BICM, where the former and the latter control the slope and horizontal gain of the asymptotic BER curve in the high-SNR region, respectively. In fact,
the HMMSED metrics can be utilized to optimize the asymptotic performance of signal constellations, which is independent of the code design.
To facilitate the design of signal constellations in PLDPC-BICM systems, the HMMSED analysis is discussed below.

Consider an $M$-ary modulation with a constellation $\chi$ and a labeling rule $\phi$. The HMMSED is expressed as \cite{924878}\vspace{-1.5mm}
\begin{equation}
d_{\rm h,A}^{2} (\chi,\phi) =  \left( \frac{1} {w 2^w} \sum_{\mu=1}^{w} \sum_{b=0}^{1} \sum_{x \in \chi_\mu^b} \frac{1} {|x-\tilde{x}_{\rm A}|^2} \right),
\label{eq:HMMED}\vspace{-1mm}
\end{equation}
where ${\rm A \in \{NI, ID \}},~w=\log_2 M$ is the number of labeling bits within a signal point (or a modulated symbol), $\chi_\mu^b$ is the constellation subset of signal points with the $\mu$-th labeling bit value $ b \in \{0, 1\}$, $x$ is a signal point belonging to $\chi_\mu^b$, and $\tilde{x}_{\rm A} \in \bar{\chi}_\mu^b$ represents the signal point closest to $x$ in the complementary set of $\chi_\mu^b$ in the case of BICM-NI, and also represents the signal point with the same labeling-bit value as those in $x$ except the $\mu$-th labeling bit in the case of BICM-ID with error-free feedback. In the following, $d_{\rm h,NI}^{2} (\chi,\phi)$ and $d_{\rm h,ID}^{2} (\chi,\phi)$ are referred to as {\em NI-HMMSED and ID-HMMSED}, respectively. Based on the two HMMSED metrics, one can promptly obtain the offset gain as\vspace{-0.1cm}
%$G_{\rm h} = 10 \log_{10} (d_{\rm h,ID}^{2} (\chi,\phi) / d_{\rm h,NI}^{2} (\chi,{\rm Gray}))\,{\rm (in~dB)}$,
\begin{equation}%\vspace{-0.1cm}
G_{\rm h} = 10 \log_{10} \frac{d_{\rm h,ID}^{2} (\chi,\phi} { d_{\rm h,NI}^{2} (\chi,{\rm Gray})}\,{\rm (in~dB)},
\label{eq:new-eq-6}
\end{equation}
which specifies the asymptotic iterative gain attained by an ideal BICM-ID with respect to the optimal Gray-labeled BICM-NI.

Actually, the ID-HMMSED is derived based on an ideal BICM-ID framework in which perfect {\it a priori} LLRs of the labeling bits are fed back to the demodulator. This assumption is unrealistic and may degrade the accuracy of performance evaluation for BICM-ID systems. From the design point of view, it is advisable to choose a constellation exhibiting a good balance between the NI-HMMSED and the ID-HMMSED, such that the BICM is able to achieve sufficiently good performance in both NI and ID scenarios after a few global iterations \cite{9210097}. Moreover, a powerful resistance to the feedback errors is particularly important for the constellation shaping in BICM-ID systems.

%{\em Remark:} The HMMSED metrics can be utilized to predict the asymptotic performance of signal constellations in BICM systems, which is independent of the code design.

{\bf \textit{Example 7:}} The NI-HMMSED $d_{\rm h,NI}^{2} (\chi,\phi)$, ID-HMMSED $d_{\rm h,ID}^{2} (\chi,\phi)$ and offset gain $G_{\rm h}\,{\rm (dB)}$ of four classic $16$QAM constellations are shown in Table~\ref{tab:IV}. As observed, the Gray labeling achieves the largest $d_{\rm h,NI}^{2} (\chi,\phi)$, but the smallest $d_{\rm h,ID}^{2} (\chi,\phi)$ and $G_{\rm h}$. This phenomenon reveals that the Gray labeling may not be suitable for BICM-ID systems. On the contrary, MSEW labeling suffers from the smallest $d_{\rm h,NI}^{2} (\chi,\phi)$, but benefits from the largest $d_{\rm h,ID}^{2} (\chi,\phi)$, which indicates that it may outperform other three labelings in BICM-ID scenario with perfect {\it a priori} information. However, the MSEW labeling may not achieve the best performance in the case with feedback errors. %Overall, the ID-HMMSED can only be utilized to predict the asymptotic performance of constellations in the ideal or quasi-ideal ID circumstance, while NI-HMMSED can be utilized to predict the asymptotic performance of constellations in generic NI circumstance.
%Table 4
\begin{table}[t]\scriptsize
\caption{NI-HMMSED $d_{\rm h,NI}^{2} (\chi,\phi)$, ID-HMMSED $d_{\rm h,ID}^{2} (\chi,\phi)$, and offset gain $G_{\rm h}~{\rm (dB)}$ of the $16$QAM constellations with the Gray labeling, anti-Gray labeling, SP labeling, and MSEW labeling.}\vspace{-1.5mm}
\centering
\begin{tabular}{|c|c|c|c|c|}
\hline
Constellation Type & Gray & Anti-Gray & SP & MSEW \\ \hline
$d_{\rm h,NI}^{2} (\chi,\phi)$ & $0.492$ & $0.400$ & $0.441$ & $0.400$ \\ \hline
$d_{\rm h,ID}^{2} (\chi,\phi)$ & $0.514$ & $0.993$ & $1.119$ & $2.364$ \\ \hline
$G_{\rm h}$ & $0.190$ & $3.050$ & $3.569$ & $6.817$ \\ \hline
\end{tabular}
\label{tab:IV}
\vspace{-3mm}
\end{table}%

\subsection{Summary}\label{sect:III-E}

In this section, the basic principles of PLDPC codes and SC-PLDPC codes are first reviewed. %Especially, we highlight their concepts, graphical and mathematical representations, as well as construction methods.
Then, three types of prevailing theoretical-analysis tools are introduced for the PLDPC-BICM, which can be used to develop system design and optimization methods. Specifically, the PEXIT algorithm, AWE analysis, and HMMSED analysis are discussed, which respectively yield three critical performance metrics, i.e., decoding threshold, minimum distance rate, and HMMSED, for the PLDPC-BICM. %Among the three types of performance metrics, the decoding threshold is the most popular metric, which can be employed to predict the asymptotic convergence performance of the overall BICM system involving channel code, constellation, and bit-mapper.

%In the following sections, we will focus on the deployment of the above two types of PB-LDPC codes, i.e., PLDPC codes and SC-PLDPC codes, for BICM systems and outline their representative design methodologies in order to provide desirable bandwidth-efficiency transmission solutions for modern and future communication and storage applications.

\section{Design of PLDPC-BICM over AWGN Channels}\label{sect:V}

This section summarizes the current research achievements in the design of PLDPC-BICM systems over AWGN channels, which can be used to model various communication scenarios, e.g., deep-space communication systems \cite{4383367,5174517}, satellite broadcasting systems \cite{8798970,esplugabinary}, optical communication systems \cite{7320948,7005396,hager2016analysis}. As discussed in Section~\ref{sect:II-A}, the BICM design developed over AWGN channels usually perform well over other memoryless channels if they have similar input-output MIs. Hence, the AWGN channel is considered as a proper surrogate channel for some complex channel models. %The proposed code-construction methods in such a scenario possibly provide desirable performance in practical communication applications.

\subsection{Design of PLDPC-BICM Systems}\label{sect:V-A}

\subsubsection{Code Construction}\label{sect:V-A-1} 
\begin{figure}[tbp]
%\vspace{-0.5cm}
\centering
\includegraphics[width=3.35in,height=1.75in]{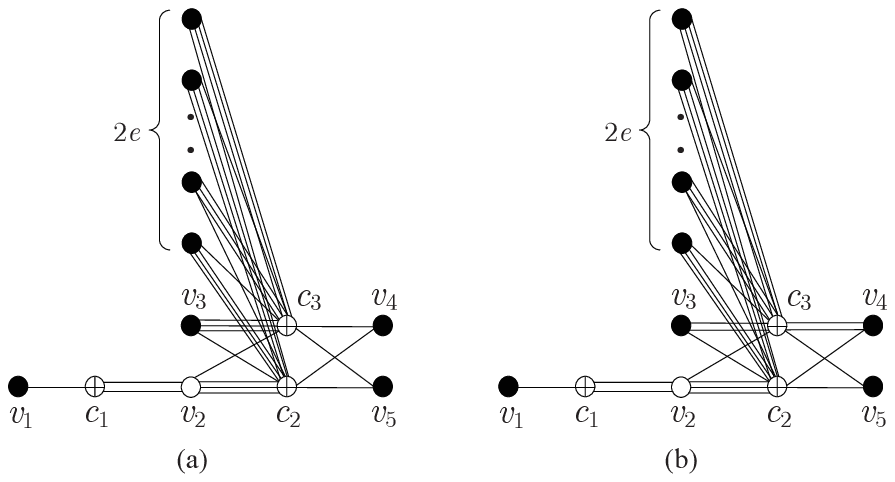}
\vspace{-3mm}
\caption{Structures of (a) AR$4$A code \cite{5174517} and (b) AR$4$JA code \cite{7320948}, with a code rate of $r_{\rm P}=(e+1)/(e+2)~(e=0,1,2,\ldots)$.}
\label{fig:Fig.11V}  \vspace{-3mm}
\end{figure}

In PLDPC-BICM systems, most related works exploited computer search methods to construct capacity-approaching PLDPC codes with linear-minimum-hamming-distance-growth property based on the PEXIT algorithm and AWE analysis.
The first attempt was made in \cite{1605713} to carry out a joint design of PLDPC codes and high-order modulation. Motivated by this pioneering work, the majority of coding efforts turned to investigating the interplay between the PLDPC codes and high-order modulations as well as the concatenation of these two types of techniques. For instance, a family of AR$4$A codes (see Fig.~\ref{fig:Fig.11V}(a)) were applied to the BICM-NI systems with various high-order modulations over AWGN channels \cite{5613828,7320948,7383250}. These studies concentrated on the optimization of bit-mapping schemes but not on PLDPC codes. In the following, several PLDPC-code constructions in BICM systems over different channels are discussed.
%Fig.11

\vspace{0.5mm}
\noindent{\bf (1) Code Design for PSK/QAM-aided BICM-NI:} To ensure the capacity-approaching performance of BICM-NI systems with $M$PSK/$M$QAM modulations, a family of RC PLDPC codes, called {\em AR4JA codes}, were constructed in \cite{1605713}, with the structure shown in Fig.~\ref{fig:Fig.11V}(b). %\footnote{For all the $M$PSK/$M$QAM-aided BICM systems mentioned in this paper, PSK modulations are adopted for $M \le 8$, while QAMs are used for $M \ge 16$ unless otherwise specified.}
Referring to this figure, the AR$4$JA codes cover a wide range of rate from $1/2$ to $1-\epsilon$, where $\epsilon$ is an arbitrarily small positive value. The higher-rate AR$4$JA codes are formulated by repeatedly adding $2e~(e=0,1,2,\ldots)$ degree-$4$ variable nodes into the rate-$1/2$ AR$4$JA code and then connecting them to the two highest-degree check nodes $c_2$ and $c_3$ (see Fig.~\ref{fig:Fig.11V}(b)). The AR$4$JA codes possess both desirable decoding thresholds and linear-minimum-Hamming-distance-growth property. Meanwhile, a novel bit-mapping scheme, i.e., {\em VDMM scheme} was constructed by combining the AR$4$JA code with $16$QAM modulation, which improves the spectral efficiency of PLDPC-based BICM-NI systems. The details of VDMM scheme will be further discussed in Section~\ref{sect:V-A-3}.

\vspace{0.5mm}
\noindent{\bf (2) Code Design for ASK-aided BICM-NI:} In parallel with the above research progress, a systematic study on the code construction and analysis of PLDPC-BICM-NI systems with probabilistic signal shaping was carried out in \cite{7339431,7282623}. In classic BICM-NI systems, the transmitted symbols are always assumed to follow a uniform distribution and the $w$ labeling bits within a given symbol are always treated as mutually independent, which leads to a non-trivial gap from the channel capacity. To overcome this weakness, a new symbol-distribution optimization scheme, called {\em probabilistic amplitude shaping (PAS)}, was proposed in \cite{7307154} to approach the capacity of BICM-NI with the use of rate-$(n_{\rm P}-1)/n_{\rm P}$ PLDPC codes and ASK modulations. Based on the PAS-aided BICM-NI framework, a PEXIT-based design scheme was developed in \cite{7339431,7282623} to construct outstanding PLDPC codes operating within $0.39 \sim 0.45~{\rm dB}$ to the channel capacities.

Specifically, consider a BICM-NI system with a rate-$(n_{\rm P}-1)/n_{\rm P}$ PLDPC code and an $M$-ary ASK modulation. To achieve the PAS goal, the protograph corresponding to the PLDPC code should include $n_{\rm P}=w$ variable nodes and $m_{\rm P}=w-1$ check nodes, where $w=\log_2 M$. The PLDPC code ${\bm \Lambda}=(v_1, v_2, \ldots, v_N)$ of length $N=Zn_{\rm P}$ is first mapped to an ASK-modulated symbol sequence $\bx=(x_1, x_2, \ldots, x_{N'})$ of length $N'$, where $N'=N/w=Z$, $x_k \triangleq (\hat{v}_{k,1},\hat{v}_{k,2},\ldots, \hat{v}_{k,w}) \in \chi$ is the $k$-th $(k=1,2,\ldots,N')$ modulated symbol, $\hat{v}_{k,\mu}$ is the $\mu$-th $(\mu=1,2,\ldots,w)$ labeling bit within the $k$-th symbol, and $\chi = \{ \pm 1, \pm 3, \ldots, \pm(2^w -1)\}$ is the constellation set. %For example, Fig.~\ref{fig:Fig.12V} shows the Gray-labeled $8$ASK constellation.
%%Fig.12
%\begin{figure}[tbp]
%%\vspace{-0.5cm}
%\centering
%\includegraphics[width=2.2in,height=0.6in]{Fig.12.eps}
%\vspace{-2.5mm}
%\caption{Constellation of the Gray-labeled $8$ASK modulation.}
%\label{fig:Fig.12V} \vspace{-3mm}
%\end{figure}

Totally, there are $Z(w-1)$ information bits and $Z$ parity bits in a PLDPC code of length $N=Z w$ in such a BICM-NI system. During the systematic encoding process, the $K=Z(w-1)$ independent and identically distributed (i.i.d.) information bits ${\bs}=(s_1, s_2,\ldots, s_K)$ are passed through a distribution matcher \cite{7339431}, so as to yield $Z$ groups of labeling bits $\{\hat{v}_{k,2},\hat{v}_{k,3},\ldots, \hat{v}_{k,w}\,|\, k=1,2,\ldots, Z \}$ with $A_{k} = \sum_{\mu = 2}^{w} \hat{v}_{k,\mu}{\bm \cdot} 2^{w-\mu}$ being the amplitude of the $k$-th group. Due to the effect of the distribution matcher, the $w-1$ resultant labeling bits in each group are no longer i.i.d., but follow a non-uniform discrete distribution, which are mutually dependent. Furthermore, the $Z$ parity bits ${\bm \Lambda}_{\rm P}=(v_{K+1}, v_{K+2},\ldots, v_{K+Z})$ can be promptly generated according to the checksum constraint ${\bm \Lambda} \bH^{\rm T} =(\bs, {\bm \Lambda}_{\rm P}) \bH^{\rm T} = {\bm 0}$, which controls the relationship between a systematic codeword ${\bm \Lambda} =(\bs, {\bm \Lambda}_{\rm P}) $ and its corresponding parity-check matrix $\bH^{\rm T}$. The $Z$ parity bits are appended to the information bits to constitute the overall systematic PLDPC code. They are consecutively employed to indicate the sign of the amplitudes of the $Z$ labeling-bit groups to form the overall transmitted symbols $\{x_k  \triangleq (\hat{v}_{k,1},\hat{v}_{k,2},\ldots, \hat{v}_{k,w}) \,|\, k=1,2,\ldots, Z \}$. Here, the first labeling bit of the $k$-th symbol $\hat{v}_{k,1} = {\rm sign}(x_k)$ is uniformly mapped from the $k$-th parity bits $v_{K+k}$. Thus, $x_k=+ A_k$ if $\hat{v}_{k,1} \triangleq v_{K+k}=0$, and $x_k=- A_k$ otherwise. In the end, the distribution of $M$ different signal points belonging to the Gray-labeled $M$ASK constellation is optimized, so as to maximize the channel input-output MI and approach the constellation-unconstrained capacity with Gaussian-distributed symbols. The above technique is referred to as PAS. Especially, the PAS technique concatenates a probabilistic-shaping-aided distribution matcher with a systematic PLDPC encoder at the transmitter, and exploits a bit-metric BP decoder at the receiver. %In such a scenario, the {\it a posteriori} LLR of the $\mu$-th labeling bit $\hat{v}_{k,\mu}$ within the $k$-th transmitted symbol $x_k$ output from the demodulator needs to be calculated, as follows:
%\begin{equation}
%L_{\rm APP,DEM} (k,\mu) \hspace{-0.5mm}=\hspace{-0.5mm}  \ln \hspace{-0.5mm}\left(\frac{\Pr (\hat{v}_{k,\mu} = 0)} {\Pr (\hat{v}_{k,\mu} = 1)} \hspace{-0.2mm}{\bm \cdot}\hspace{-0.2mm} \frac{\Pr (y_k | \hat{v}_{k,\mu} = 0)} {\Pr ( y_k | \hat{v}_{k,\mu} = 1)} \right)\hspace{-0.5mm},
%\label{eq:L-APP-DEM-PAS}
%\end{equation}
%where the first item is the {\it a priori} LLR of the $\mu$-th labeling bit $\hat{v}_{k,\mu}$, coming out of other labeling bits within the $k$-th symbol, and the second item is the channel initial LLR. %coming out of the channel.

To simplify the design of PLDPC codes in PAS-aided BICM systems, a surrogate channel was considered in \cite{7339431}. A channel is said to be {\em a proper surrogate channel} if the code optimized for such a channel cannot be further improved over the target BICM channel by varying the bit-mapping scheme. Through simulations, the binary-input AWGN channel was found to be proper surrogate for each bit-channel of an $M$ASK-aided BICM-NI channel with uniformly distributed input and PAS (non-uniform) input. Then, a novel PEXIT algorithm was developed for constructing the near-capacity PLDPC codes. %having near-capacity

{\bf \textit{Example 8:}} Utilizing the PEXIT-aided computer search method, the rate-$(w-1)/w$ PLDPC codes can be optimized for $M$ASK-aided BICM-NI systems with uniform input and PAS input under the Gray-labeling. The base matrices and thresholds of the optimized PLDPC codes for such frameworks with three different modulation orders (i.e., different values of $M$) are illustrated in \eqref{eq:B-MASK} and Table~\ref{tab:V}, respectively. As a baseline, the capacities and the capacity gaps are also included in the table.\vspace{-2mm}
\begin{eqnarray}
&{\bB}_{\rm 4ASK\textit{-}U} \hspace{-0.5mm}=\hspace{-0.5mm}\left[\begin{array}{cccccc}
2 & 1 & 1 & 2 & 1 & 4 \cr
1 & 1 & 1 & 2 & 2 & 5 \cr
1 & 0 & 2 & 1 & 0 & 6 \cr
\end{array}\right],\nonumber\\\vspace{3mm}
&\bB_{\rm 8ASK\textit{-}PAS} \hspace{-0.5mm}=\hspace{-0.5mm}
\left[\begin{array}{cccccc}
1 & 1 & 1 & 1 & 1 & 6 \cr
2 & 2 & 1 & 1 & 2 & 6 \cr
\end{array}\right],\nonumber\\\vspace{3mm}
&\bB_{\rm 64ASK\textit{-}PAS} = \left[\begin{array}{cccccccccccc}
2 & 2 & 2 & 1 & 2 & 2 & 6 & 2 & 2 & 0 & 6 & 6 \cr
1 & 1 & 1 & 2 & 1 & 1 & 6 & 1 & 0 & 2 & 6 & 6\cr
\end{array}\right].\vspace{-1mm}
\label{eq:B-MASK}
\end{eqnarray}
%\footnote{The capacity gap indicates the difference between the decoding threshold and the channel capacity.}
The rate-$1/2$ $4$ASK-U PLDPC code, rate-$2/3$ $8$ASK-PAS PLDPC code, and rate-$5/6$ $64$ASK-PAS PLDPC code are optimized for the $4$ASK modulation with uniform input, $8$ASK modulation with PAS input, and $64$ASK modulation with PAS input, respectively. As observed, all the three optimized PLDPC codes enjoy very desirable decoding thresholds, which are within $0.43~{\rm dB}$ from the channel capacities. The design method can be easily extended to the PLDPC codes with rates higher than $(w-1)/w$ for a given modulation order, such as $r_{\rm P}=3/4$ for $4$ASK \cite{7339431}. For simplicity, the result is omitted here.
%Table 5
\begin{table}[t]\scriptsize
\caption{Decoding thresholds $(E_b/N_0)_{\rm th}~{\rm (dB)}$ and capacity gaps $\Delta$ of the optimized PLDPC codes for three different $M$ASK-aided BICM-NI systems with uniform input and PAS input, where the Gray labeling is used.}
\centering\vspace{-1.5mm}
\begin{tabular}{|c|c|c|c|c|}
\hline
Modulation Scheme & $4$ASK-U & $8$ASK-PAS & $64$ASK-PAS  \\ \hline
$(E_b/N_0)_{\rm th}$ & $5.57$ & $7.77$ & $25.52$ \\ \hline
Capacity             & $5.27$ &	$7.34$ & $25.31$ \\ \hline
$\Delta$             & $0.30$ &	$0.43$ & $0.39$  \\ \hline
\end{tabular}
\label{tab:V}\vspace{-2.5mm}
\end{table}%
%\vspace{-0.5mm}

\vspace{0.5mm}
\noindent{\bf (3) Code Design for CPSK-aided BICM-ID:} In \cite{esplugabinary}, a new PEXIT function was proposed, tailored for the bilayer (i.e., $L=2$) root-PLDPC codes in the $M$CPSK-aided BICM-ID context. Then, a series of two-layer root-PLDPC codes were constructed for the BICM-ID systems to increase the transmission throughput and reduce the data-processing time for the next generation of global navigation satellite systems. Recall that the $M$CPSK modulation is an $M$-ary direct-sequence spread-spectrum (DSSS) modulation, which exploits a maximum-length pseudo-noise (PN) sequence with $\beta$ chips to carry $w=\log_2 M$ coded bits. Thus, a length-$N$ PLDPC code ${\bm \Lambda}=(v_1, v_2, \ldots, v_N)$ is first converted to a length-$N'$ non-binary codeword ${\bm \Lambda}_{\rm NB}=(v_{\rm NB,1}, v_{\rm NB,2}, \ldots, v_{{\rm NB},N'})$. %, where the $k$-th non-binary coded symbol is expressed by
%\begin{equation}
%v_{{\rm NB},k} =\sum\nolimits_{\mu=1}^w v_{w(k-1)+\mu} {\bm \cdot} 2^{w-\mu} \in \{ 0,1,\ldots, M-1\},
%\label{eq:binary-2-nonbinary-symbol}
%\end{equation}
%where $k=1,2,\ldots,N'~{\rm and}~\mu=1,2,\ldots,w$.
Later on, each non-binary symbol is represented by a circular shift of the pre-generated length-$\beta$ PN sequence ${\bm \varphi}_{\rm CPSK}$. In this setting, the $M$CPSK constellation set is constituted by $M$ different circular-shift versions of ${\bm \varphi}_{\rm CPSK}$. %The $q$-th element in an $M$CPSK constellation set should satisfy ${\bm \varphi}_{{\rm CPSK},q} (\tau) = {\bm \varphi}_{\rm CPSK} ({\rm mod}(\tau-\tau_q, \beta))$, where $q=0,2,\ldots, M-1, \tau =1,2,\ldots,\beta$, ${\bm \varphi}_{{\rm CPSK},q} (\tau)$ is the $\tau$-th component chip, $\tau_{q}$ is the circularly shifted length, and ${\rm mod}(a, b)$ is the modulo operation (i.e., $a~{\rm modulo}~b$).

As a type of maximum-distance separable (MDS) codes, root-PLDPC codes were initially proposed in \cite{6875246,6905810,8019817} to achieve full diversity and outage-limit-approaching performance over block-fading channels. It was demonstrated in \cite{8740906} that the structure of root-PLDPC codes perfectly matches the characteristics of block-fading channels.
%This type of codes can be also applied to BICM systems under other channel conditions.
Here, the root-PLDPC-code design for $M$CPSK-aided BICM systems over AWGN channels is discussed, without getting into details of such codes. The construction principle of root-PLDPC codes will be discussed in Section~\ref{sect:VI-B-1}, where the research progress relevant to the root-PLDPC-BICM over block-fading channels will be presented. %, since the structure of the root-PLDPC codes perfectly matches the characteristics of block-fading channels \cite{8740906}.
%Table 6
\begin{table*}[t]\scriptsize
\caption{Decoding thresholds $(E_b/N_0)_{\rm th}~{\rm (dB)}$ and capacity gaps $\Delta$ of the rate-$1/2$ optimized PLDPC codes, AR$3$A code, and AR$4$JA code in three different CSK-aided BICM-ID systems over an AWGN channel, with Gray labeling.}
\centering\vspace{-2mm}
\begin{tabular}{|c|c|c|c|c|c|c|c|}
\hline
\multirow{2}{*}{\diagbox[width=12em,height=2em,trim=l]{\hspace{-0.5mm}Modulation Type}{\hspace{1.5mm}Code Type}}
& \multicolumn{2}{|c|}{Optimized PLDPC} & \multicolumn{2}{|c|}{AR$3$A} & \multicolumn{2}{|c|}{AR$4$JA}  & \multirow{2}{*}{Capacity}\\
\cline{2-7}
& ${(E_b/N_0)}_{\rm th}$ & $\Delta$ & ${(E_b/N_0)}_{\rm th}$ & $\Delta$ & ${(E_b/N_0)}_{\rm th}$ & $\Delta$ & \\
\hline
GMSK & $-2.11$ & $0.65$ & $-0.45$ &	$2.31$	& $-0.20$ &	$2.56$ & $-2.76$ \\ \hline
QRC  & $1.36$  & $0.74$ & $3.36$ &	$2.74$	& $3.68$ &	$3.06$ & $0.62$ \\ \hline
OREC & $3.79$ & $1.54$ & $7.29$ &	$5.04$	& $7.92$ &	$5.67$ & $2.25$ \\ \hline
\end{tabular}
\label{tab:VI}
\vspace{-2mm}
\end{table*}%

{\bf \textit{Example 9:}} Aiming to obtain a desirable optimized bilayer root-PLDPC code for $4$CPSK-aided BICM-ID systems over an AWGN channel, a $4 \times 8$ base matrix is initialized, as\vspace{-1mm}
\begin{eqnarray}
\bB &=\left[\begin{array}{cccccccc}
1 & 0 & 0 & 0 & b_{1,5} & b_{1,6} & b_{1,7} & b_{1,8} \cr
0 & 1 & 0 & 0 & b_{2,5} & b_{2,6} & b_{2,7} & b_{2,8} \cr
b_{3,1} & b_{3,2} & b_{3,3} & b_{3,4} & 1 & 0 & 0 & 0 \cr
b_{4,1} & b_{4,2} & b_{4,3} & b_{4,4} & 0 & 1 & 0 & 0 \cr
\end{array}\right],
\label{eq:B-RP-N8-general}
\end{eqnarray}
where the value of the $(i,j)$-th element is assumed to be $b_{i,j} \in \{0,1,2,3 \}$ in order to keep relatively low computational complexity for code optimization and construction, with $i \in \{1,2,3,4 \}$ and $j \in \{1,2,3,4,5,6,7,8 \}$. Note that the structure of the base matrix \eqref{eq:B-RP-N8-general} strictly follows the definition of root-PLDPC codes given in Section~\ref{sect:VI-B-1}. After a PEXIT-aided search procedure, the base matrix corresponding to the optimized bilayer root-PLDPC code can be obtained, as\vspace{-1mm}
\begin{eqnarray}
\bB_{\rm OPT\textit{-}ROOT} &=\left[\begin{array}{cccccccc}
1 & 0 & 0 & 0 & 0 & 1 & 2 & 1 \cr
0 & 1 & 0 & 0 & 1 & 0 & 2 & 0 \cr
1 & 3 & 1 & 2 & 1 & 0 & 0 & 0 \cr
0 & 1 & 3 & 0 & 0 & 1 & 0 & 0 \cr
\end{array}\right].
\label{eq:B-RP-N8-optimized}
\end{eqnarray}
Both theoretical and simulation results demonstrated that the optimized bilayer root-PLDPC code accomplishes a remarkable gain of more than $0.5~{\rm dB}$ over the L$1$ civil LDPC codes adopted in the $4$CPSK-aided global positioning system \cite{esplugabinary}.

\vspace{0.5mm}
\noindent{\bf (4) Code Design for CPM-aided BICM-ID:} As another promising power- and bandwidth-efficiency modulation scheme, CPM has attracted enthusiastic interest in wireless communication systems due to its constant-envelope characteristic \cite{8444445}. Inspired by this appealing advantage, the joint design of the PLDPC codes and CPM modulations was investigated in \cite{6875180} by exploiting a modified PEXIT algorithm within the context of BICM-ID. Specifically, consider a CPM-aided BICM-ID system. A length-$N$ PLDPC code ${\bm \Lambda}=(v_1, v_2, \ldots, v_N)$ is interleaved and then processed by an $M$CPM modulator to generate a length-$N'$ $M$-ary symbol sequence $\bx=(x_1, x_2, \ldots, x_{N'})$, where $N'=N/w$ and $x_k \triangleq (\hat{v}_{k,1},\hat{v}_{k,2},\ldots, \hat{v}_{k,w})$ is the $k$-th~$(k=1,2,\ldots,N')$ modulated symbol; particularly, $x_k$ is taken value from the set $\chi = \{ \pm 1, \pm3, \ldots, \pm(2^w -1)\}$. According to a given symbol sequence $\bx$, a CPM signal transmitted over an AWGN channel can be generated as\vspace{-0.1cm}
%${\bm \varphi}_{\rm CPM} (\tau, \bx) = \sqrt{2E_{\rm s}/T} \cos(2 \pi f_0 \tau + \Upsilon(\tau, \bx) + \Upsilon_0)$,
\begin{equation}%\vspace{-0.1cm}
{\bm \varphi}_{\rm CPM} (\tau, \bx) = \sqrt{2E_{\rm s}/T} \cos(2 \pi f_0 \tau + \Upsilon(\tau, \bx) + \Upsilon_0),
\label{eq:new-eq-7}
\end{equation}where $E_{\rm s}$ is the average symbol energy, $T$ is the symbol duration, $f_0$ and $\Upsilon_0$ are the carrier frequency and initial phase shift, respectively, and $\Upsilon(\tau, \bx)$ is the information-carrying phase. %, $\theta$ is the modulation index, and $f(t)$ is the phase pulse determined by the pulse length $L_{\rm P}$.
Based on the CPM-aided BICM-ID framework, the extrinsic MI functions for the outer decoder and inner demodulator were derived, which can facilitate the PLDPC-code optimization. %Inspired by some appealing properties of the existing PLDPC codes and conventional LDPC codes,
Subsequently, a simple search method was proposed in \cite{6875180} to construct a series of capacity-approaching rate-$1/2$ PLDPC codes in this framework.
%%Fig.13
%\begin{figure*}[t]
%\centering
%\includegraphics[width=3.9in,height=1.16in]{Fig.13.eps}\vspace{-2mm}
%\caption{Protographs of the three rate-$1/2$ optimized PLDPC codes for BICM-ID systems with (a) GMSK, (b) QRC, and (c) OREC modulations.}
%\label{fig:Fig.13V}  %label for entire figure
%\end{figure*}

{\bf \textit{Example 10:}} Aiming to minimize the decoding threshold of a PLDPC code in the CPM-aided BICM-ID system, a $4 \times 8$ base matrix with three degree-$2$ variable nodes and three degree-$1$ variable nodes is initialized, as\vspace{-1mm}
\begin{eqnarray}
~~~~~\bB &=&
\left[\begin{array}{cccccccc}
1 & 1 & 0 & 0 & 0 & 0 & b_{1,7} & b_{1,8} \cr
0 & 1 & 1 & 0 & 1 & 0 & b_{2,7} & b_{2,8} \cr
0 & 0 & 1 & 1 & 0 & 1 & b_{3,7} & b_{3,8} \cr
0 & 0 & 0 & 1 & 0 & 0 & b_{4,7} & b_{4,8} \cr
\end{array}\right],
\label{eq:B-PLDPC-CPM}
\end{eqnarray}
where the seventh and eighth columns correspond to the variable nodes with degrees at least $3$. To limit the search space and lower the encoding complexity, the value of $b_{i,j}~(i \in \{1,2,3,4 \}, j \in \{7,8\})$ is assumed to satisfy $0 \le b_{i,j} \le 3$. Then, under three different CPM schemes, i.e., binary Gaussian minimum shift keying (GMSK, $M=2$), Gray-labeled quaternary raised-cosine (QRC, $M=4$), and Gray-labeled octal rectangular (OREC, $M=8$), the base matrices of optimized PLDPC codes can be obtained after a computer-based search \cite{6875180}, as \eqref{eq:B-CPM}.
%Fig.~\ref{fig:Fig.13V} shows the protographs of the three optimized PLDPC codes in \eqref{eq:B-CPM}.

For comparison, the decoding thresholds and capacity gaps of the rate-$1/2$ optimized PLDPC codes, AR3A code and AR4JA code in the CSK-aided BICM-ID systems are summarized in Table~\ref{tab:VI}. Obviously, the three optimized PLDPC codes have gaps of only $0.65~{\rm dB}, 0.74~{\rm dB},~{\rm and}~1.54~{\rm dB}$ to the channel capacities under GMSK, QRC, and OREC modulations, respectively, which are much smaller than those of AR$3$A and AR$4$JA codes. Moreover, the optimized PLDPC codes have decoding thresholds nearest to the capacities, while the AR$4$JA code has decoding thresholds farthest from the capacities.\vspace{-0.5mm}
\begin{eqnarray}
\hspace{5mm}\bB_{\rm GMSK} &=&
\left[\begin{array}{cccccccc}
1 & 1 & 0 & 0 & 0 & 0 & 1 & 2 \cr
0 & 1 & 1 & 0 & 1 & 0 & 2 & 1 \cr
0 & 0 & 1 & 1 & 0 & 1 & 1 & 1 \cr
0 & 0 & 0 & 1 & 0 & 0 & 1 & 1 \cr
\end{array}\right],\nonumber\\
\bB_{\rm ORC} &=&
\left[\begin{array}{cccccccc}
1 & 1 & 0 & 0 & 0 & 0 & 1 & 1 \cr
0 & 1 & 1 & 0 & 1 & 0 & 2 & 2 \cr
0 & 0 & 1 & 1 & 0 & 1 & 1 & 1 \cr
0 & 0 & 0 & 1 & 0 & 0 & 1 & 1 \cr
\end{array}\right],\nonumber\\
\bB_{\rm OREC} &=&
\left[\begin{array}{cccccccc}
1 & 1 & 0 & 0 & 0 & 0 & 1 & 1 \cr
0 & 1 & 1 & 0 & 1 & 0 & 3 & 3 \cr
0 & 0 & 1 & 1 & 0 & 1 & 1 & 1 \cr
0 & 0 & 0 & 1 & 0 & 0 & 1 & 1 \cr
\end{array}\right].
\label{eq:B-CPM}
\end{eqnarray}

{\em Remark:} The above works are restricted to the PLDPC-code design for regular-mapped BICM with different modulation schemes. By a glance at this issue for irregular-mapped BICM \cite{Zhao2020}, relevant studies are still in infancy.

\subsubsection{Constellation Shaping}\label{sect:V-A-2}

Since the inception of BICM, a great deal of research effort has been devoted to investigating constellation shaping in order to enhance the performance in both NI and ID scenarios. Of particular interest are the works that investigated the APSK constellation optimization for LDPC-BICM systems \cite{6064854,6515491,5598322,8878166,6241383}, proposing desirable constellations under different environments. In contrast to the BICM-NI scenario, the trade-off between the NI-HMMSED and ID-HMMSED has to be carefully taken into account when optimizing the constellations for BICM-ID. For instance, some innovative methods were proposed in \cite{6064854,5598322} to design APSK constellations having excellent HMMSEDs for achieving good performance in LDPC-BICM-ID systems.% (i.e., a few global iterations).

\begin{figure}[tbp]
\centering
\includegraphics[width=1.6in,height=1.53in]{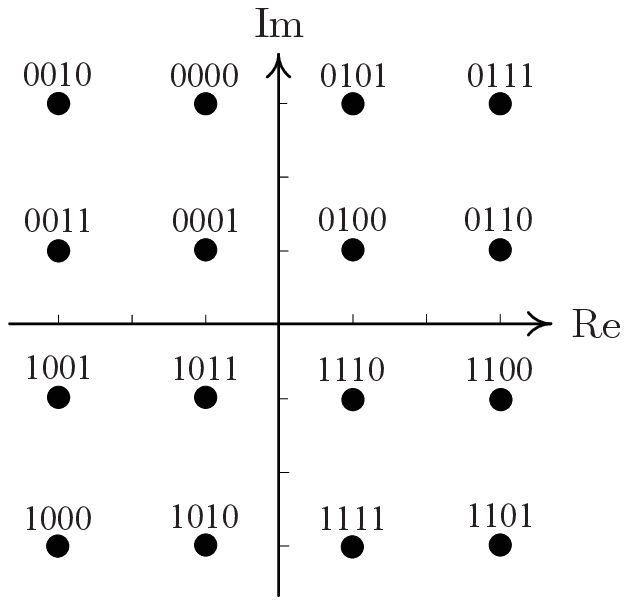}\vspace{-3mm}
\caption{Constellation of the adaptive Gray-labeled $16$QAM modulation.}\vspace{-2.5mm}
%{\color{blue}The coupling lengths of $(3,6)$ SC-P codes and RJA SC-P codes are assumed to be $12$ and $6$, respectively.
\label{fig:fig.14V}  \vspace{-2mm}
\end{figure}

Yet, the constellation shaping for PLDPC-BICM systems is relatively unexplored. To fill this gap, an adaptive $16$QAM constellation is proposed here for the PLDPC-BICM-ID systems by modifying the subset-partition rule on the Gray-labeled $16$QAM constellation. Fig.~\ref{fig:fig.14V} displays the structure of the proposed adaptive Gray-labeled $16$QAM constellation. Analysis confirms that the adaptive Gray-labeled constellation possesses the same NI-HMMSED (i.e., $d_{\rm h,NI}^2 (\chi, \phi) = 0.492 $) with respect to the Gray-labeled counterpart, with a relatively larger ID-HMMSED (i.e., $d_{\rm h,ID}^2 (\chi, \phi) = 0.615 $). This implies that the adaptive Gray-labeled constellation can obtain better performance than the Gray counterpart in the ID setting.

As a further progress, a generalized design method was proposed in \cite{Zhao2020} to construct an adaptive constellation based on an initial constellation with any labeling format (e.g., Gray and anti-Gray) and any modulation order. The designed adaptive constellation can be seamlessly combined with its corresponding initial constellation to formulate a powerful irregular mapping (IM) scheme, which can enhance the performance of the PLDPC-BICM-ID system over AWGN channels. Note that the principle of IM and its corresponding design guideline are ignored here, which will be discussed in the next subsection.

\subsubsection{Bit-Mapper (Interleaver) Design}\label{sect:V-A-3}

Given a well-constructed PLDPC code and a well-shaped constellation, the interface between these two components is also of great importance in determining the overall performance of BICM systems. The interface between a binary code and a high-order constellation is commonly called {\em bit mapper or bit interleaver}.

\vspace{0.5mm}
\noindent{\bf (1) Regular Mapping:} In $2005$, two VDMM schemes, i.e., {\em water-filling and reverse water-filling VDMMs}, were proposed in \cite{1605713} for Gray-labeled $16$QAM-modulated BICM-NI systems. Considering a given constellation, the principle of both VDMM schemes is that the coded bits within a codeword are assigned to different labeling-bit positions within a modulated symbol according to the variable-node degree distribution of the PLDPC code. In the water-filling VDMM, the variable nodes (i.e., coded bits) with the highest degrees are assigned to the labeling-bit positions with the highest protection degrees (i.e., the highest MI), while those with the lowest degrees are assigned to the labeling-bit positions with the lowest protection degrees. On the contrary, in the reverse water-filling VDMM, the mapping between the variable nodes and the labeling bits is assigned in a reverse order. Simulation results suggest that the water-filling VDMM scheme performs as good as the random bit-mapping (interleaving) scheme, while the reverse water-filling VDMM scheme performs worse than the random bit-mapping scheme by at least $0.1~{\rm dB}$. However, neither the water-filling VDMM nor the random mapping schemes can achieve optimal performance in PLDPC-BICM-NI systems. To address this issue, a PEXIT-assisted optimization strategy was developed to enhance the performance of VDMM in PLDPC-BICM-NI systems with $16$QAM modulation \cite{5613828}, where all possible permutations for the mapping between the variable nodes with different degrees in a protograph and the labeling bits with different protection degrees in an $M$-ary symbol are enumerated so as to achieve the lowest decoding threshold. The optimized VDMM attains a considerable performance gain over the conventional VDMM at the expense of a higher computational complexity.

There are two drawbacks for the VDMM schemes \cite{6133952,6777400}:
\begin{itemize}
\item
 The VDMM schemes are not effective if the number of variable nodes in a protograph is not equal to the number of labeling bits in a modulated symbol. %, i.e., $n_{\rm P} \neq w~({\rm where}~w = \log_2 M)$;
 \item
 The VDMM schemes are not effective if all the variable nodes have the same degree.
\end{itemize}

Hence, the conventional and optimized VDMM schemes can only work well for irregular PLDPC-BICM systems when, and only when $n_{\rm P} = w=\log_2 M$. This severely limits the application of such techniques to practical BICM systems. To break this limitation, a more feasible bit-mapping scheme, referred to as {\em TSLM scheme}, was developed in \cite{6133952}. The proposed TSLM scheme begins by permuting the original protograph with an initial lifting factor $z=w$, which is much smaller than the lifting factor used for producing a PLDPC code.

%Table 7
\begin{table}[t]\scriptsize
\caption{Decoding thresholds $(E_b/N_0)_{\rm th}~{\rm (dB)}$ and capacity gaps of the water-filling VDMM scheme, optimized VDMM scheme, and the TSLM scheme in the BICM-NI systems over an AWGN channel, where a rate-$1/2$ AR$4$JA code and Gray-mapped $16$QAM modulation are used.}\vspace{-1.5mm}
\centering
\begin{tabular}{|c|c|c|c|}
\hline
Bit-Mapping Scheme                & $(E_b/N_0)_{\rm th}$ & $\Delta$ & Capacity  \\ \hline
Water-filling VDMM \cite{1605713} & $2.853$ & $0.567$  & \multirow{3}{*}{ $2.286$} \\
\cline{1-3}
Optimized VDMM \cite{5613828}     & $2.714$ & $0.428$ & \\
\cline{1-3}
TSLM \cite{6133952}               & $2.812$ & $0.526$ & \\ \hline
%Capacity                          & $2.286$ & ${\rm N.A.}$ \\ \hline
\end{tabular}
\label{tab:VII}\vspace{-3mm}
\end{table}%

%\begin{tabular}{|c|c|c|c|c|c|c|c|}
%\hline
%\multirow{2}{*}{\diagbox[width=12em,height=2em,trim=l]{\hspace{-0.5mm}Modulation Type}{\hspace{1.5mm}Code Type}}
%& \multicolumn{2}{|c|}{Optimized PLDPC} & \multicolumn{2}{|c|}{AR$3$A} & \multicolumn{2}{|c|}{AR$4$JA}  & \multirow{2}{*}{Capacity}\\
%\cline{2-7}
%& ${(E_b/N_0)}_{\rm th}$ & $\Delta$ & ${(E_b/N_0)}_{\rm th}$ & $\Delta$ & ${(E_b/N_0)}_{\rm th}$ & $\Delta$ & \\
%\hline
%GMSK & $-2.11$ & $0.65$ & $-0.45$ &	$2.31$	& $-0.20$ &	$2.56$ & $-2.76$ \\ \hline
%QRC  & $1.36$  & $0.74$ & $3.36$ &	$2.74$	& $3.68$ &	$3.06$ & $0.62$ \\ \hline
%OREC & $3.79$ & $1.54$ & $7.29$ &	$5.04$	& $7.92$ &	$5.67$ & $2.25$ \\ \hline
%\end{tabular}
%\label{tab:VI}
%\vspace{-2mm}
%\end{table*}%

Accordingly, a $w m_{\rm P} \times w n_{\rm P}$ intermediate protograph can be formulated, with variable nodes distributed on $w$ different planes. Particularly, each plane corresponds to a replica of the protograph and comprises of $n_{\rm P}$ variable nodes. Now, all the $n_{\rm P}$ variable nodes of the $\mu$-th~$(\mu=1,2,\ldots, w)$ plane are mapped to $n_{\rm P}$ replicas of the $\mu$-th labeling bit within an $M$-ary modulated symbol, such that the $w n_{\rm P}$ variable nodes of the intermediate protograph can be mapped to $n_{\rm P}$ symbols. After that, the BICM framework is realized based on the intermediate protograph, which is convenient for further optimization. One can then expand the intermediate protograph to a PLDPC code with an expected codeword length (or symbol length). For instance, if the codeword length is $N= Z n_{\rm P}$, the secondary lifting factor for the expansion of the intermediate protograph becomes $Z'=\frac{N}{w n_{\rm P}}= \frac{Z}{w}$. Note that the proposed TSLM scheme is applicable to any protograph structure with any modulation order. Assuming that the PLDPC-BICM-NI systems use a rate-$\frac{1}{2}$ AR$4$JA code and Gray-mapped $16$QAM modulation, Table VII presents the decoding thresholds of the water-filling VDMM \cite{1605713}, optimized VDMM \cite{5613828}, and the TSLM schemes \cite{6133952}. As seen, the decoding threshold of the TSLM scheme is slightly smaller than that of the water-filling VDMM, but is slightly higher than that of the optimized VDMM. In fact, the optimized VDMM scheme resorts to a brute-force search, which requires much higher computational complexity than the TSLM scheme. It was shown in \cite{6133952} that the TSLM scheme allows the PLDPC-BICM-NI to attain close-to-capacity decoding thresholds in a wide range of code rates and modulation orders with relatively low computational complexity. %Benefiting from these merits, the TSLM scheme can be viewed as a competitive alternative for BICM systems.

Apart from the TSLM scheme, a generalized VDMM scheme was developed in \cite{6777400} for PLDPC-BICM-NI systems with $M$APSK modulation, which can also be applied to combine any protograph structure with any modulation order. Specifically, the $n_{\rm P}$ types of variable nodes in a PLDPC codeword are first re-ranked in a descending order according to their degrees, while the $w$ types of labeling bits in a symbol sequence are re-ranked in a descending order according to their protection degrees. In the sequel, the variable nodes are artificially divided into $w$ groups, which are successively mapped to the $w$ types of labeling bits. By this method, a water-filling-like procedure can be realized and thus desirable convergence and error performance can be achieved. As a further advance, a simplified optimization scheme for the generalized VDMM scheme was constructed to obtain optimal mapping performance with relatively lower complexity than the brute-force search-aided method \cite{5613828}. Also, the generalized VDMM scheme was applied to the PLDPC-RGB-LED-based BICM-VLC systems in \cite{7320948}.

In parallel with the VDMM-related scheme, another novel optimized bit-mapping scheme, referred to as {\em variable-node fractional-allocation mapping (VNFAM) scheme}, was proposed in \cite{hager2014improving} for BICM-NI fiber-optical communication systems.
%without optical inline dispersion compensation in.
In these systems, polarization-multiplexed (PM) $M$QAM signals are adopted for transmission. The proposed VNFAM scheme utilizes a modified PEXIT algorithm to search for an unrestricted matching between the variable nodes in a protograph and the labeling bits within a PM-$M$QAM modulated symbol. Results demonstrated that the PLDPC-BICM-NI with the VNFAM scheme can accomplish performance gains up to $0.25~{\rm dB}$ over its counterpart with the consecutive mapping scheme.

{\em Remark:} The generalized VDMM scheme is applicable only to irregular codes, while the TSLM and VNFAM schemes are applicable to both regular and irregular codes.

\vspace{0.5mm}
\noindent{\bf (2) Irregular Mapping:} In the traditional BICM frameworks, only one signal constellation is employed to modulate the PLDPC code, which is referred to as a {\em regular mapping}. With the development of BICM techniques, efforts have been made to explore more flexible mapping schemes in order to increase the degree of freedom of system design and to enhance the link adaption ability. As an alternative mapping scheme, IM that employs more than one constellation to modulate a codeword was proposed for BICM-ID systems in order to improve the convergence performance \cite{1576578}. It was verified that the deployment of IM in BICM-ID systems can have desirable capacity-approaching performance \cite{7589683}. Yet, how to devise excellent IM schemes with acceptable computational complexity remains to be a challenging issue for further investigation.

In $2011$, a novel IM scheme was proposed in \cite{5960810} by exploiting a modified adaptive binary switch algorithm (ABSA), which is able to achieve near-capacity performance in LDPC-BICM-ID systems with $M$QAM. In this IM scheme, the ABSA is exploited to search for a new constellation based on a given constellation (e.g., Gray- or quasi-Gray-labeled constellation). Then, both constellations are employed to modulate the LDPC code with a fixed mixing ratio. The mixing-ratio vector of the IM is defined as ${\bm \eta} = (\eta_1, \eta_2)$, where $\eta_\kappa = N_{\kappa}/N$ is the ratio between the number of coded bits mapped to the $\kappa$-th~$(\kappa=1,2)$ constellation and the codeword length.

As the first attempt to investigate the IM-PLDPC-BICM-ID over an AWGN channel, by mixing the adaptive Gray-labeled $16$QAM constellation (see Fig.~\ref{fig:fig.14V}) with its mother constellation (i.e., Gray-labeled $16$QAM constellation), a novel IM scheme, referred to as {\em protograph-based adaptive irregular mapping (PAIM) scheme}, was developed in \cite{Zhao2020}. It was shown that the PAIM scheme can significantly accelerate the decoding convergence of PLDPC codes in BICM-ID systems.
%Table 8
\begin{table}[t]\scriptsize
\caption{Decoding thresholds $(E_b/N_0)_{\rm th}~{\rm (dB)}$ of four different IM schemes ($\chi_1$: Gray-labeled constellation, $\chi_2$: another constellation) and Gray-labeled regular mapping scheme in the $16$QAM-aided BICM-ID systems over an AWGN channel, where the rate-$1/2$ regular-$(3, 6)$ PLDPC code is used.}\vspace{-1.5mm}
\centering
\begin{tabular}{|c|c|c|c|c|c|}
\hline
\diagbox[width=6.5em,height=2em,trim=l]{\hspace{-1.2mm}Code Type}{$\chi_2$}
& Anti-Gray & MSEW & ABSA \cite{5960810} & Adp-Gray  &  Gray \\ \hline
Regular    & $3.442$ & $4.053$ & $3.160$ & $3.108$ & $3.211$ \\ \hline
%RJA        & $3.286$ & $3.947$ & $3.004$ & $2.912$ & $3.126$ \\ \hline
%
%
%Regular & RJA \\ \hline
%Anti-Gray              & $3.442$ & $3.286$ \\ \hline
%MSEW                   & $4.053$ & $3.947$ \\ \hline
%ABSA \cite{5960810}    & $3.160$ & $3.004$ \\ \hline
%Adp-Gray               & $3.108$ & $2.912$ \\ \hline
%   Gray                & $3.211$ & $3.126$ \\ \hline
\end{tabular}
\label{tab:VIII}
\vspace{-2mm}
\end{table}%
%%Fig.15
%\begin{figure}[tbp]
%\centering
%\includegraphics[width=2.9in,height=2.2in]{{Fig.15a.eps}}
%%\subfigure[\hspace{-0.8cm}]{ %% label for first subfigure
%%\includegraphics[width=3.0in,height=2.3in]{Fig.15a.eps}}
%%\subfigure[\hspace{-0.8cm}]{ %% label forsecond subfigure
%%\includegraphics[width=3.0in,height=2.3in]{Fig.15b.eps}}
%\vspace{-0.2cm}
%\caption{BER curves of four different IM schemes and Gray-labeled regular-mapping scheme in the $16$QAM-aided BICM-ID systems over AWGN channels, where the rate-$1/2$ regular-$(3, 6)$ PLDPC code is used. The transmitted codeword length is assumed to be $N_{\rm T}=4800$.}
%\label{fig:Fig.15V}  %label for entire figure
%\vspace{-2mm}
%\end{figure}

{\bf \textit{Example 11:}} Assuming a Gray-labeled constellation (denoted by $\chi_1$) and a mixing-ratio vector ${\bm \eta} = (1/2, 1/2)$, the decoding thresholds of the IM schemes are now compared with four different realizations of the second component constellation (denoted by $\chi_2$) in the PLDPC-BICM-ID systems over an AWGN channel, where the channel code and modulation used are the rate-$1/2$ regular-$(3, 6)$ PLDPC code and $16$QAM, respectively. In the comparison, each IM scheme is composed of the constellations $\chi_1$ and $\chi_2$. The Gray-labeled regular mapping (i.e., $\chi_2$: Gray-labeled constellation) is also included as a benchmark. As seen from Table~\ref{tab:VIII}, %and Fig.~\ref{fig:Fig.15V},
the IM scheme employing both Gray- and adaptive-Gray-labeled constellations exhibits the best decoding threshold among the four schemes. Moreover, the IM scheme obviously outperforms the Gray-labeled regular-mapping scheme, which achieves optimal performance in the NI scenario. Hence, the Gray- and adaptive-Gray-labeled constellations constitute a promising IM scheme for the PLDPC-BICM-ID systems.

\subsection{Design of SC-PLDPC-BICM Systems} \label{sect:V-B}

\subsubsection{Code Construction} \label{sect:V-B-1}

During the past decade, SC-PLDPC codes have been extensively applied to BICM systems. With the advancement of the design methodologies for SC-PLDPC codes, their BICM relatives gradually stood out as a predominant bandwidth-efficient transmission solution for communication and storage systems. For example, based on the SC-PLDPC codes, a variety of BICM design paradigms were proposed for different constellation formats, modulation orders and code rates, with the aim to provide diverse requirements for practical applications and services \cite{8798970,7460483,8883091,9057491,8847350,6469365,6875180}. In the following, several representative works are discussed.

\vspace{0.5mm}
\noindent{\bf (1) Code Design for PSK/QAM-aided BICM-NI:} A so-called ``{\em universal}" design method was developed in \cite{6469365} for TE-SC-PLDPC codes, which not only achieves capacity-approaching performance under high-order modulations regardless of the labeling format but also has very low complexity. To facilitate the design and analysis, two convergence regions, i.e., {\em macro-convergence and micro-convergence regions}, are defined. The macro-convergence and micro-convergence regions are dominated by the degree distribution and spatial coupling effect of the TE-SC-PLDPC code, respectively. According to the theoretical analysis, some regular TE-SC-PLDPC codes with a proper edge-spreading rule can perform very well under different labeling formats, even perform better than the optimized LDPC codes. Assume that the TE-SC-PLDPC code is generated from a regular PLDPC code with a base matrix $\bB = \bB_{\rm S,1} +\bB_{\rm S,2}+,\ldots,+\bB_{\rm S, \varsigma+1}$, where $\bB_{\rm S,\mu}$ is the $\mu$-th sub-base matrix yielded by the ``matrix division'' operation. To trigger an asymmetric ``decoding wave" phenomenon from both the left-hand and the right-hand sides of the SC protograph, it is needed to introduce some irregularity to the $\varsigma + 1$ sub-base matrices.\footnote{The ``decoding wave" phenomenon is that the variable nodes associated with the lower-degree check nodes at both ends of the coupled chain converge faster than the variable nodes in the middle under BP decoding, leading to a wave-like decoding phenomenon \cite{7152893}.} The above simple operation can significantly accelerate the MI convergence of the TE-SC-PLDPC codes in the BICM-ID systems. Analytical and simulation results demonstrated that the {\em asymmetric} regular TE-SC-PLDPC codes can attain capacity-approaching decoding thresholds in the BICM-ID systems with various labelings. More importantly, the asymmetric TE-SC-PLDPC codes are superior to the conventional irregular LDPC codes optimized for the BICM-ID systems.
%Table 9
\begin{table}[t]\scriptsize
\caption{Decoding thresholds $(E_b/N_0)_{\rm th}~{\rm (dB)}$ of the rate-$3/4$ conventional regular TE-SC-PLDPC code, asymmetric regular TE-SC-PLDPC code, and three optimized irregular LDPC codes in the $16$QAM-aided BICM-ID systems over an AWGN channel, where the Gray, SP, and M$16$a labelings are considered. The capacity limit is $4.528~{\rm dB}$, and the parameters used for the TE-SC-PLDPC codes are $\varsigma=1$ and $L_{\rm SC}=100$.}\vspace{-1mm}
\centering
\begin{tabular}{|c|c|c|c|}
\hline
\diagbox[width=15.5em,height=2em,trim=l]{Code Type}{Constellation}  & Gray & SP & M$16$a \cite{schreckenbach2007iterative} \\ \hline
Conv. regular TE-SC-PLDPC            & $4.695$ & $5.001$ & $5.260$ \\ \hline
Asym. regular TE-SC-PLDPC            & $4.712$ & $4.577$ & $4.745$ \\ \hline
Opt. irregular LDPC-A (for Gray)     & $4.600$ & $6.217$ & $6.344$\\ \hline
Opt. irregular LDPC-B (for SP)       & $ > 7 $ & $4.739$ & $4.983$ \\ \hline
Opt. irregular LDPC-C (for M16a)     & $ > 7 $ & $4.729$ & $4.959$ \\ \hline
\end{tabular}
\label{tab:IX}
\vspace{-2mm}
\end{table}%

{\bf \textit{Example 12:}} Based on a rate-$3/4~(4, 16)$-regular protograph with base matrix $\bB=[4~4~4~4]$, one can construct different asymmetric regular TE-SC-PLDPC codes with coupling width $\varsigma=1$. For one realization, the two sub-base matrices split from the original base matrix $\bB$ are expressed as $\bB_1=[3 ~2~3~2]$ and $\bB_2=[1~2~1~2]$, respectively. According to its convolutional structure, the resultant asymmetric regular TE-SC-PLDPC code can be viewed as a regular code if the coupling length is sufficiently large. As observed, the structures of the sub-base matrices for the asymmetric regular TE-SC-PLDPC code are quite different from those of the conventional regular TE-SC-PLDPC code (i.e., as $\bB_1=[2~2~2~2]$ and $ \bB_2=[2~2~2~2]$), which lead to different decoding-wave behaviors. Table IX compares the decoding thresholds of the two TE-SC-PLDPC codes in the $16$QAM-aided BICM-ID systems with three different labeling schemes, where $L_{\rm SC}=100$. As benchmarks, three irregular LDPC codes are considered, which are particularly optimized for the Gray, SP, and M$16$a labelings. One can observe that the conventional regular TE-SC-PLDPC code has good performance only for Gray labeling, while the asymmetric regular counterparts have excellent performance for all the three labelings. On the other hand, the irregular LDPC-A, LDPC-B, and LDPC-C codes are unable to exhibit universally good performance for all the three labelings. Thereby, the asymmetric regular TE-SC-PLDPC code can be treated as a universally desirable candidate for BICM-ID systems, regardless of the labeling scheme. In addition, the decoding-wave analysis in \cite{6469365} also verified the merit of the asymmetric regular TE-SC-PLDPC code.

Apart from the TE-SC-PLDPC code design, an efficient design method was developed in \cite{9057491} for the TB-SC-PLDPC codes in $M$PSK/$M$QAM-aided BICM-NI systems with shuffled BP decoding. An appealing feature of the TB-SC-PLDPC codes is that the non-negligible rate loss incurred by the termination effect can be avoided, especially in the finite-coupling-length scenario. Different from most existing design methods, the proposed method constructs a novel type of TB-SC-PLDPC codes from two mother PLDPC codes to inherit the performance advantages of both codes. This type of TB-SC-PLDPC codes is referred to as {\em TB-SC double-PLDPC (DPLDPC) codes}. Assume that two PLDPC codes ${\cal A}$ and ${\cal B}$ correspond to the same-size $m_{\rm P} \times n_{\rm P}$ base matrices $\bB_{\cal A}$ and $\bB_{\cal B}$, respectively. The proposed code-design method includes a hybrid edge-spreading scheme and an EXIT-based degree-distribution optimization scheme. More precisely, the hybrid edge-spreading scheme is composed of a uniform edge-spreading rule and a replicative edge-spreading rule. The uniform edge-spreading rule is exploited to divide the base matrix of code ${\cal A}$ into two sub-base matrices $\bB_{{\cal A},1}$ and $\bB_{{\cal A},2}$ (i.e., coupling width $\varsigma=1$), while the replicative edge-spreading rule is used to divide the base matrix of code ${\cal B}$ into another set of two sub-base matrices $\bB_{{\cal B},1}$ and $\bB_{{\cal B},2}$. To guarantee the TB-SC-DPLDPC codes constructed from the above four sub-base matrices can combine both degree distributions of the two mother PLDPC codes ${\cal A}$ and ${\cal B}$, $\bB_{{\cal B},1}$ must equal $\bB_{{\cal A},1}$ (i.e., $\bB_{{\cal B},1}=\bB_{{\cal A},1}$) in the replicative edge spreading. Using this hybrid edge-spreading scheme, the TB-SC-DPLDPC codes can inherit the superiorities of the two mother PLDPC codes. To achieve an additional performance gain, one can optimize the structure of the sub-base matrices based on the PEXIT algorithm to formulate an improved TB-SC-DPLDPC (I-TB-SC-DPLDPC) code. The I-TB-SC-DPLDPC code benefits from both the hybrid edge-spreading and degree-distribution optimization, compared to the two mother PLDPC codes.

{\bf \textit{Example 13:}} Based on the rate-$1/2$ AR$4$JA code (code ${\cal A}$) and AR$3$A code (code ${\cal B}$) in {\em Example 2}, one can readily divide the base matrix $\bB_{\rm AR4JA}$ into two sub-base matrices $\bB_{\rm AR4JA,1}$ and $\bB_{\rm AR4JA,2}$ via the uniform edge-spreading rule, and divide the base matrix $\bB_{\rm AR3A}$ into another two sub-base matrices $\bB_{\rm AR3A,1}$ and $\bB_{\rm AR3A,2}$ via the replicative edge-spreading rule. Assuming the coupling length $L_{\rm SC}=4$, the base matrix of TB-SC-DPLDPC code is formulated as \cite{9057491}\vspace{-4mm}

{\small
\begin{eqnarray}
\hspace{-1.5mm}\bB_{\text{TB-SC-DP}} \hspace{-0.5mm}= \hspace{-1mm} \left[\hspace{-0.5mm}
\begin{array}{cccc}
  \bB_{\rm AR4JA,1}  & {\bm 0}           & {\bm 0}           & \bB_{\rm AR3A,2} \\
  \bB_{\rm AR4JA,2}  & \bB_{\rm AR3A,1}  & {\bm 0}           & {\bm 0}            \\
  {\bm 0}            & \bB_{\rm AR3A,2}  & \bB_{\rm AR4JA,1} & {\bm 0}            \\
  {\bm 0}            & {\bm 0}           & \bB_{\rm AR4JA,2}  & \bB_{\rm AR3A,1}  \\
  \end{array}\hspace{-0.5mm}\right]\hspace{-1mm},%\hspace{-1mm}
  \label{eq:B-SC-DP}
\end{eqnarray}}where $\bB_{\rm AR4JA,2}$ is derived in \cite[eq.\,(9)]{9057491}.
%\begin{eqnarray}\vspace{-2mm}
%\bB_{\rm AR4JA,2} =\left[
%\begin{array}{ccccc}
%1&1&0&0&0\\
%0&2&1&0&0\\
%0&0&1&1&1
%\end{array} \right].\nonumber
%\label{eq:B-AR4JA-sub2}
%\end{eqnarray}
Moreover, one has $\bB_{\rm AR3A,1}=\bB_{\rm AR4JA,1}= \bB_{\rm AR4JA}-\bB_{\rm AR4JA,2}$ and $\bB_{\rm AR3A,2}= \bB_{\rm AR3A}-\bB_{\rm AR3A,1}$. In particular, the variable nodes corresponding to the second columns of all the four sub-base matrices are punctured.

As is well known, the degree of the punctured variable nodes has a significant effect on the performance of the PLDPC codes. One can further optimize the degree of the punctured variable node in $\bB_{\rm AR4JA,2}$ so as to improve the decoding threshold of the TB-SC-DPLDPC code. To keep the encoding complexity low, the value of each element is assumed to be smaller than or equal to $3$. After a simple search, the improved version of $\bB_{\rm AR4JA,2}$ is obtained as \eqref{eq:B-I-AR4JA-sub2}.
Hence, the base matrix of the {\em improved TB-SC-DPLDPC} (I-TB-SC-DPLDPC) code is obtained by replacing $\bB_{\rm AR4JA,2}$ in \eqref{eq:B-SC-DP} with $\bB_{\rm I\textit{-}AR4JA,2}$.\vspace{-1mm}
\begin{eqnarray}
\bB_{\rm I\textit{-}AR4JA,2}  = \left[
\begin{array}{ccccc}
1&1&0&0&0\\
0&1&1&0&0\\
0&1&1&1&1
 \end{array} \right]
  \label{eq:B-I-AR4JA-sub2}
\end{eqnarray}

\begin{figure}[tbp]
\centering
\includegraphics[width=2.8in,height=2.1in]{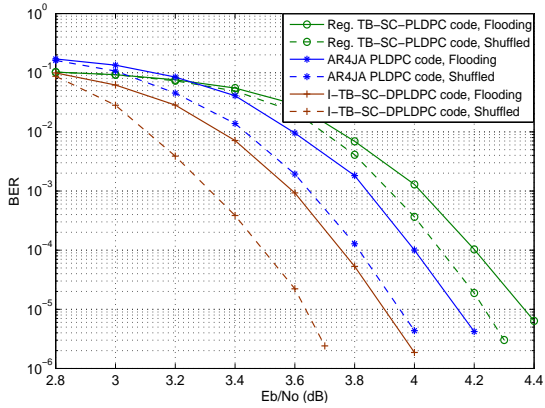}
\vspace{-0.15cm}
\caption{BER curves of the rate-$1/2$ I-TB-SC-DPLDPC, regular-$(3, 6)$ TB-SC-PLDPC, and AR$4$JA codes in a Gray-mapped $16$QAM-aided BICM-NI system over AWGN channels with $\varsigma=1, L_{\rm SC}=4,N_{\rm T}= 3840,~{\rm and}~t_{\rm BP,max}=25$.}
\label{fig:Fig.16V}  %label for entire figure
\vspace{-3.5mm}
\end{figure}

As illustrated in \cite{9057491}, the I-TB-SC-DPLDPC code possesses better convergence performance than the TB-SC-DPLDPC code, which further has better performance than the two mother PLDPC codes. Fig.~\ref{fig:Fig.16V} compares the BER performance of the I-TB-SC-DPLDPC code, regular TB-SC-PLDPC code, and AR$4$JA code in a Gray-mapped $16$QAM-aided BICM-NI system over an AWGN channel. As shown, % the shuffled schedule achieves better performance than the flooding schedule. Moreover,
the I-TB-SC-DPLDPC code is superior to the regular TB-SC-PLDPC code and the AR4JA code under both schedules, which commendably verifies the merit of the proposed design.
Specifically, at a BER of $10^{-5}$, the I-TB-SC-DPLDPC code has additional gains of about $0.3$ and $0.5$ dB over the AR4JA code and regular TB-SC-PLDPC code, respectively, under both BP decoding schedules.
%For example, at a BER of $10^{-5}$, the I-TB-SC-DPLDPC code attains a $0.3$-dB gain over the AR4JA code, which obtains an additional $0.3$-dB gain over the regular TB-SC-PLDPC code.
%Fig.16

Aiming to realize more efficient implementation of SC-PLDPC codes without loss of their performance superiority, the QC structure was taken into account in developing new code-construction methods. On the one hand, a couple of time-invariant code-construction methods were proposed to further reduce the encoding and decoding complexity of the traditional SC-PLDPC codes in the past five years \cite{8114265}. On the other hand, circulant-based PEG (i.e., circulant-based ``lifting") algorithm was seamlessly combined with the protographs in order to generate QC-PLDPC codes, which not only enable the linear encoding complexity with the aid of simple shift registers, but also produce similar error performance as the unstructured PLDPC codes \cite{8291725,7592443}. Inspired by the above advantages, an in-depth study was carried out in \cite{8798970} on the QC SC-PLDPC codes in BICM-ID systems, which achieves both high performance and efficient implementation in practical broadcasting applications. Specifically, a novel construction method was developed for the finite-coupling-length QC-TE-SC-PLDPC codes in BICM-ID systems with QPSK modulation over AWGN channels. With this method, a base-matrix optimization scheme was developed for the variable-node degree distribution and edge-spreading pattern of the original PLDPC codes to improve the decoding thresholds and minimum distance. Then, a two-step lifting algorithm was designed, which includes a pre-lifting procedure and a post-lifting procedure, to construct the QC structure and to improve the finite-length performance of the resultant QC-TE-SC-PLDPC codes. Results demonstrated that the proposed QC-TE-SC-PLDPC codes are capable of exhibiting excellent error performance in the BICM-ID context. However, the design method was carried with QPSK modulation only.%, and their feasibility was compared to other modulation orders.

\vspace{0.5mm}
\noindent{\bf (2) Code Design for ASK-aided BICM-NI:} Although the TB-SC-PLDPC codes can overcome the rate-loss issue of the TE-SC-PLDPC codes, the former is incapable of triggering the decoding wave without a termination operation \cite{8281448}. In \cite{7460483}, the TB-SC-LDPC-BICM-NI systems with $4$ASK modulation were investigated, attempting to resolve the ``decoding wave" problem with minimum code-rate loss over AWGN channels. Specifically, a random shortening technique was proposed for the TB-SC-LDPC codes to trigger a wave-like decoding behavior and to approach the performance of the TE-SC-LDPC codes with relatively smaller rate loss in the $4$ASK-aided BICM-NI systems. In that technique, the codeword is shortened by setting some coded bits to zero. Through such an operation, some {\it a priori} knowledge can be used to initiate the wave-like convergence under BP decoding. This technique is also applicable to the TB-SC-PLDPC codes, which may achieve similar performance in such a BICM framework. Although the shortening technique can improve the convergence performance of TB-SC-LDPC-BICM systems, it slightly decreases the code rate. %More importantly, no iterative receiver is considered in this work.

Following the works of \cite{7339431,7282623}, in \cite{8847350} the ASK-aided TE-SC-PLDPC-BICM-NI systems with PAS were studied, regarding especially the quantized BP decoder design. Specifically, a quaternary BP decoding algorithm was proposed for the rate-$(w-1)/w$ TE-SC-PLDPC codes in PAS-aided BICM-NI system with $M$ASK~($M \ge 4$) modulation, where $w=\log_2 M$. To ease the analysis of the proposed quantized decoding algorithm, a modified DE algorithm was developed. Analyses and simulations revealed that the proposed decoding algorithm ensures the TE-SC-PLDPC codes to achieve desirable decoding thresholds and error performance over $M$ASK-aided BICM-NI with PAS. Hence, the PAS-aided TE-SC-PLDPC-BICM possesses great potential to be a high-throughput transmission solution for optical communications.
%More importantly, it has been verified that the TE-SC-PLDPC codes and quantized BP decoding algorithms can also yield excellent finite-length error performance over the $M$ASK-aided BICM-NI AWGN channels with PAS.
Note that the above work considered only regular TE-SC-PLDPC codes, but not the irregular TE-SC-PLDPC-BICM-NI systems. Differing from \cite{7339431}, it did not study the code optimization, but only the decoder design.

\vspace{0.5mm}
\noindent{\bf (3) Code Design for CPM-aided BICM-ID:} In \cite{7249024}, a new type of SC-PLDPC codes, called {\em DT-SC-PLDPC codes}, was developed for CPM-aided BICM-ID systems. The DT-SC-PLDPC code was constructed in an analogous way to that of a TB-SC-PLDPC code, in order to avoid rate loss. Yet, in a DT-SC-PLDPC code, one must directly cut off all the $m_{\rm P} \varsigma$ check nodes at the last $\varsigma$ time instants and their associated edges from the SC protograph, instead of combining them with those at the first $\varsigma$ time instants. With this operation, the base matrix of a DT-SC-PLDPC code is obtained as
\cite[eq.\,(10)]{7249024},
%\eqref{eq:DT-base},
which includes $m_{\rm P} L_{\rm SC}$ rows and $n_{\rm P} L_{\rm SC}$ columns, leading to a code rate of $r_{\rm DT\text{-}SC} =1- m_{\rm P} /n_{\rm P} =r_{\rm P} $. The DT-SC-PLDPC codes can preserve most convolutional benefits of the conventional TE-SC-PLDPC codes since the structure in the left-hand side of the protograph is unchanged. It was verified that the DT-SC-PLDPC code can retain the rate advantage of its original protograph code by slightly degrading the decoding threshold.%\vspace{-4mm}

To boost the performance of CPM-aided PLDPC-BICM-ID system in \cite{6875180}, two regular DT-SC-PLDPC codes with $\varsigma =1, 2$ were constructed  in \cite{7249024} based on the rate-$1/2$ regular-$(3, 6)$ PLDPC code. As benchmarks, a series of regular TE-SC-PLDPC codes with $\varsigma =1, 2$ were formulated based on the rate-$1/2$ regular-$(3, 6)$ PLDPC code. It was demonstrated that, with a relatively large coupling length (i.e., $L_{\rm SC}=50$) in the GMSK-aided BICM-ID systems, the regular DT-SC-PLDPC code with $\varsigma =1$ can achieve an identical decoding threshold to the corresponding TE-SC-PLDPC code, which is much smaller than that of their original PLDPC code. However, the regular DT-SC-PLDPC code with $\varsigma =2$ suffers from an obvious threshold degradation, compared with the TE-SC-PLDPC code.
%In addition, both the regular DT-SC-PLDPC code and TE-SC-PLDPC code
%the both TE-SC-PLDPC codes with $\varsigma =2$ achieve small gains over their counterparts with $\varsigma =1$ at the prize of a slight rate loss, which accomplishes remarkable gains over the original PLDPC codes.
%Simulations have also been performed to verify the performance advantages of the DT-SC-PLDPC codes and the TE-SC-PLDPC codes as compared with their original photograph codes.
Due to the performance and rate advantages, the regular DT-SC-PLDPC code with $\varsigma =1$ may serve as a preferable candidate over the TE counterpart in $M$CPM-aided BICM-ID systems.

%\vspace{0.5mm}
%\noindent{\bf (4) Code design for ASK-aided BICM-NI Design:}
%Following the work of \cite{7339431,7282623} , the authors in \cite{8847350} have endeavored to investigate the SC-PLDPC-coded BICM-NI systems with PAS and paid special attention to the quantized BP decoder design. To be specific, a quaternary BP decoding algorithm has been proposed and applied to the rate-$(w-1)/w$ TE-SC-PLDPC codes in PAS-aided BICM-NI system with MASK modulation, where $w=\log_2 M$. To facilitate the analysis of the proposed quantized decoding algorithm and its corresponding BICM-NI system, a modified DE algorithm has been proposed in the same work. Both analytical and simulated results have suggested that the proposed quaternary BP decoding algorithm make the SC-PLDPC-coded BICM-NI system achieving very desirable performance, and thus possesses great potential to be a high-throughput FEC solution for optical communications. More importantly, it has been verified that the SC-PLDPC codes and quantized BP decoding algorithms optimized through the asymptotic DE technique can also yield excellent finite-length error performance over AWGN channels. Note that the above work have only considered regular TE- and DT-SC-PLDPC codes, but have not discussed the applicability and merit of the quantized decoding algorithm for irregular SC-PLDPC-coded BICM-NI systems with PAS. Differing from \cite{7339431}, this work has not touched upon the code-optimization aspect, but has only discussed the code-application aspect.
%
%
%

\subsubsection{Constellation Shaping}\label{sect:V-B-2}

Constellation shaping was commonly exploited to boost either the channel capacity of BICM-NI or the iterative performance of BICM-ID. A constellation is universally good for both BICM-NI and ID systems if it can simultaneously realize the above-mentioned two objectives. In spite of a surge of publications on constellation design of BICM systems appeared in the past two decades \cite{8053803,6515491,8186234,6241383,5199551,6204017}, the investigation tailored for SC-PLDPC codes remains to be explored. As a supplement to the existing constellations, a novel type of constellations was proposed in \cite{8883091} for the TB-SC-PLDPC-BICM-ID systems over AWGN channels, which can achieve both desirable BICM-NI capacity and BICM-ID performance. Based on this work, the TB-SC-PLDPC-hierarchical modulated (HM) BICM-ID systems were further studied and a type of structural quadrant (SQ) constellations was devised for such systems \cite{9210097}. The design principles for the current constellation shaping schemes in the above transmission scenarios are further discussed below.

\vspace{0.5mm}
\noindent{\bf (1) Traditional Modulation:} It was proved, in a variety of research works, that the Gray constellation as an optimal constellation scheme in BICM-NI systems is generally unsuitable for BICM-ID systems due to its trivial iterative gain. Aiming at improving the performance of SC-PLDPC-BICM-ID systems over AWGN channels, a two-step design method was proposed in \cite{8883091} to construct a novel type of constellations, called {\em labeling-bit-partial-match} (LBPM) {\it constellations}. The detailed procedure of the LBPM scheme is outlined as follows.

{\bf Step 1:} For a given $M$-ary PSK/QAM modulation, the $k$-th signal point $x_k$ belonging to a constellation is denoted as $x_k \triangleq \{\hat{v}_{k,1}, \hat{v}_{k,2},\ldots, \hat{v}_{k,w}\}$, where $\hat{v}_{k,\mu}$ is the $\mu$-th labeling bit within the $k$-th modulated symbol, $w=\log_2M, k=1,2,\ldots,N', N'=N/w$, and $\mu=1,2,\ldots, w$. Especially, an $M$-ary constellation can be viewed as $w$ parallel independent and binary input memoryless sub-channels. Then, the labeling bits with the maximum average MI are referred to as {\em high protection-degree labeling bits} and the remaining labeling bits are referred to as {\em low protection-degree labeling bits} \cite{8883091}. By analyzing the labeling-bit distribution, the number of high protection-degree labeling bits can be calculated as $w' \in (0, w)$. Hence, the high protection-degree labeling bits are set as the first $w'$ positions (i.e., $\{\hat{v}_{k,1}, \hat{v}_{k,2},\ldots, \hat{v}_{k,w'}\}$) within the modulated symbol, to attain excellent performance in the non-ID context.

{\bf Step 2:} The remaining $w-w'$ low protection-degree labeling bits are processed sequentially and then placed in the last $w-w'$ positions within the modulated symbol. In particular, one should reasonably set each low protection-degree labeling bit to maximize the Hamming distance between the adjacent signal points in the constructed constellation. This operation is of great importance to boost the iterative gain of the constellation.

Note that the first and second steps of the LBPM constellation are executed to improve the performance in the NI and ID scenarios, respectively. Based on the design method, the LBPM constellations can be easily constructed.

\begin{figure}[tbp]\vspace{-2mm}
\centering
\includegraphics[width=3.2in,height=1.53in]{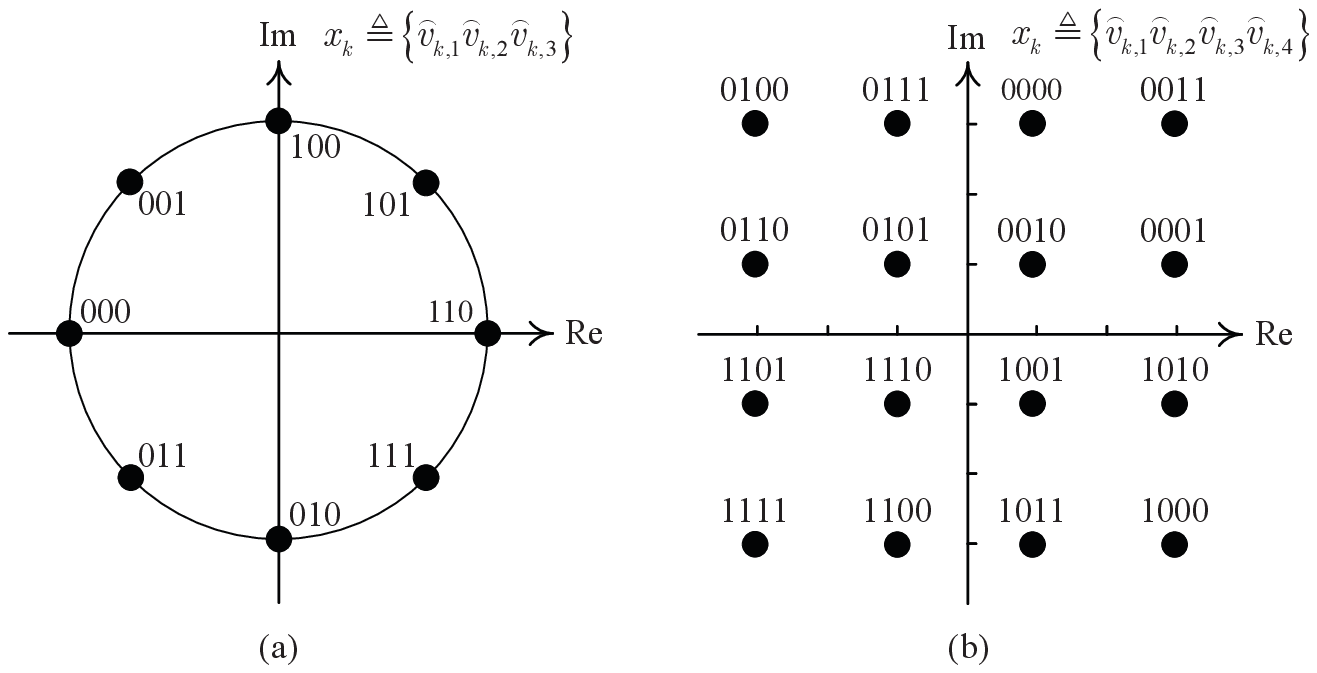}\vspace{-3mm}
\caption{Constellations of the LBPM schemes: (a) $8$PSK and (b) $16$QAM.}
\label{fig:fig.A1}\vspace{-2mm} %label for entire figure
\end{figure}
\begin{table}[tp]\scriptsize
\center
\caption{Decoding thresholds $(E_b/N_0)_{\rm th}~(\rm dB)$ of the regular-$(3,6)$ TB-SC-PLDPC code in the $8$PSK/$16$QAM-aided BICM-ID systems with four different constellations over AWGN channels. The parameters are $r_{\rm P}=1/2, \varsigma =2, L_{\rm SC}=12, t_{\rm GL,max}=8,~{\rm and}~t_{\rm BP,max}=25$.}\vspace{-0.5mm}
\begin{tabular}{|c||c|c|c|c|}
\hline
$8$PSK                     &  SP          & MSEW           &  anti-Gray    & LBPM\\
\hline
$(E_b/N_0)_{\rm th}$     & $3.049$      & $4.429$        & $4.307$       & $2.480$  \\
\hline\hline
$16$QAM                    &  SP          & MSEW           &  anti-Gray    & LBPM\\
\hline
$(E_b/N_0)_{\rm th}$     & $4.724$      & $5.027$        & $4.385$       & $3.626$   \\
\hline
\end{tabular}\label{tab:tab1}\vspace{-3.5mm}
\end{table}

{\bf \em Example 14:} The LBPM constellations for the $8$PSK and $16$QAM are shown in Fig.~\ref{fig:fig.A1}. In order to verify the superiority of the two LBPM constellations, the decoding thresholds of the regular-$(3, 6)$ TB-SC-PLDPC code in BICM-ID systems are calculated with results shown in Table~\ref{tab:tab1}, where the $8$PSK and $16$QAM modulations are considered. %The coupling length for the $(3, 6)$ TB-SC-PLDPC code is set to be $12$, while the code rate and codeword length are set to be $1/2$ and $4800$, respectively.
As benchmarks, three state-of-the-art constellations, i.e., anti-Gray-labeled, SP-labeled, and MSEW-labeled constellations, are included in the same table. It is observed that the $(3, 6)$ TB-SC-PLDPC code with the LBPM constellation exhibits smaller decoding threshold than those of the other three constellations.

\vspace{0.5mm}
\noindent{\bf (2) Hierarchical Modulation:} Aiming to satisfy the diverse quality-of-service requirement in emerging wireless-communication applications, such as IoT, the HM has been extensively studied. Furthermore, the HM-BICM can retain the spectral-efficiency benefit of BICM and can also realize UEP for different transmitted data streams. Inspired by these appealing characteristics, a large volume of research works related to the HM-BICM have been carried out \cite{9210097,6809890}.
%Of particular interest is the work of \cite{9210097}, in which the authors have delved into the SC-PLDPC-coded HM-BICM-ID systems. %The block diagram of an SC-PLDPC-coded HM-BICM-ID system is shown in Fig.~\ref{fig:HM-SM}.
%Based on such a system architecture, we elaborate the design principle of the HM constellation in \cite{9210097}.
%\begin{figure}[t]
%\center
%\includegraphics[width=3.5in,height=2.77in]{{HM.eps}}\vspace{-2mm}
%\caption{Block diagram of an SC-PLDPC-coded HM-BICM-ID system.}
%\label{fig:HM-SM}\vspace{-1.5mm}
%\end{figure}

Distinguished from the traditional BICM systems, HM-BICM systems can be considered as layered systems, with constellations decomposed into several different layers in order to deal with different data streams. Thus, the existing constellation design methods may not work well for HM-BICM systems. Owing to this issue, several constellations, e.g., MSED \cite{6809890}, MSED-B \cite{6809890}, M3 \cite{7807354}, and hierarchical bandwidth modulation (HBM) constellations \cite{8761547}, were proposed in the context of HM. There is still room for further improvement of the existing HM constellations because they do not substantially consider the effect of the ID framework.
%For an $M$-ary HM constellation, the $w$ labeling bits within a signal point is uniformly divided into $w/2$ groups in a sequential order, and thus each group is composed of $2$ labeling bits. Then, the group-$1$, group-$2, \dots,$ and group-$(w/2)$ labeling bits correspond to different layers and have different protection priorities.
Recently, a two-step design method was proposed \cite{9210097} to construct a novel type of constellations, i.e., SQ constellations, which are applicable to any $M$-ary HM-BICM system. Analytical and simulation results demonstrate that the proposed SQ constellation outperforms the existing MSED, MSED-B, M$3$, and HBM constellations in both HM-BICM and HM-BICM-ID systems over AWGN channels.

\subsubsection{Bit-Mapper (Interleaver) Design}\label{sect:V-B-3}

%Bit-mapping design allows the system to explore the best interface between the SC-PLDPC codes and signal constellations and to accomplish further performance improvement.
Before 2015, there seemed to have no research work addressing the bit-mapping design for SC-PLDPC codes. Then, a couple of efficient bit-mapping schemes were developed for SC-PLDPC-BICM systems \cite{7005396,8883091}. The design frameworks in \cite{7005396} and \cite{8883091} have great potential for bandwidth-efficient fiber-optical systems and wireless-communication systems, respectively.

\vspace{0.5mm}
\noindent{\bf (1) Regular Mapping:} In \cite{7005396}, a systematic study was proposed on the bit-mapping optimization for SC-PLDPC-BICM-NI systems over AWGN channels. In these systems, both TE- and TB-SC-PLDPC codes are considered as channel codes, while the Gray-labeled PM-$M$QAM is considered as a modulation scheme. Two different optimized bit-mapping schemes were proposed for TE- and TB-SC-PLDPC codes based on the UEP property offered by the different labeling bits within a PM-$M$QAM-modulated symbol. More precisely, the fast convergence of SC-PLDPC codes in these BICM systems is attributed to the optimized bit-mapping schemes that proportionally assign the coded bits located in different positions of the coupled chain to the labeling bits with different protection degrees.

On the one hand, for a TE-SC-PLDPC code, the lower-degree check nodes at both ends of the TE-SC protograph must pass more reliable messages to their associated variable nodes, which significantly accelerate the decoding convergence of such variable nodes. In this sense, the variable nodes at the beginning and the end of the TE-SC protograph should converge faster than the variable nodes in the middle. Consequently, the decoding wave can be triggered by default from both ends to the middle of the coupled chain due to the inherent termination boundary. It was illustrated in \cite{7005396} that the decoding-wave phenomenon can be dramatically strengthened by assigning the variable nodes at the beginning and end of the TE-SC protograph to the labeling bits with low protection degrees within a modulated symbol, while assigning the variable nodes in the middle area to the labeling bits with high protection degrees. This technique is referred to as {\em optimized TE bit-mapping scheme} for the BICM-NI systems. Note that in the optimized TE bit-mapping scheme, the proportion of VNs at the beginning and the end of the coupled chain that are assigned to the low protection-degree labeling bits must be optimized so as to guarantee the excellent performance, which induces an additional computational overhead.

On the other hand, for a TB-SC-PLDPC code, the variable nodes in the TB-SC protograph (i.e., coupled chain) possess the same convergence behavior as those in its corresponding original protograph. In particular, all the variable nodes of a regular TB-SC-PLDPC code must have identical convergence performance regardless of the coupling length. Inspired by the inherent convergence feature of the TE-SC-PLDPC codes, the UEP offered by the $w$ labeling bits within an $M$-ary symbol can be utilized to create an artificial termination boundary, which is of great significance to triggering the wave-like decoding convergence of the TB-SC-PLDPC code. Assume that the BP (or the windowed BP) decoding algorithm starts from the first time instant of the coupled chain. In the {\em optimized TB bit-mapping scheme} \cite{7005396}, the coded bits at the beginning of the TB-SC protograph are mostly assigned to the high protection-degree labeling bits within a symbol. Accordingly, the remaining coded bits in other areas (i.e., in the middle and at the end) of the coupled chain can be decoded in a more efficient way by using reliable soft information of such coded bits at the beginning of the coupled chain. As a result, a locally improved decoding convergence is ensured using the above mapping scheme, which is able to initiate the wave-like decoding behavior.

%\begin{figure*}[tbp]
%\center\vspace{-2mm}
%\includegraphics[width=4.5in,height=2.45in]{{Fig.Interleaver-v1.eps}}\vspace{-2.5mm}
%\caption{Principle of the VNMM scheme for an $M$-ary modulation in a TB-SC-PLDPC-coded BICM-ID system.}
%\label{fig:Interleaver}\vspace{-1.5mm}
%\end{figure*}
However, it was found that allocating the coded bits at the beginning of the coupled chain to the labeling bits with relatively high protection degree \cite{7005396} is not able to completely trigger the decoding wave. In fact, coded bits at both the beginning and the end of the coupled chain play the same role on trigging the wave-like decoding. As an amendment to the work of \cite{7005396}, the effect of spatial-coupled structure of the TB-SC-PLDPC codes on the decoding convergence of the BICM systems was re-examined in \cite{8883091}. To further enhance the system performance, a new bit-mapping scheme, referred to as {\em VNMM scheme}, was designed. The principle of the VNMM scheme is illustrated as follows.

For a given $M$-ary PSK/QAM LBPM constellation (see Section~\ref{sect:V-B-2}), the input-output MI should be first analyzed so as to determine the protection-degree distribution of $w~(w=\log_2 M \ge 3)$ different labeling bits within a modulated symbol.
%\footnote{The protection degree of a labeling bit is considered as the reliability of its corresponding sub-channel \cite{8883091}.}
Then, the two high protection-degree labeling bits can be found, while the remaining $w-2$ labeling bits are referred to as the low protection-degree labeling bits. In the bit-to-symbol mapping process, an entire codeword is first divided into $w$ blocks in a sequential order. Then, $w$ coded bits are attracted from the $w$ different blocks to generate an $M$-ary modulated symbol. Specifically, the two coded bits, which are separately extracted from the first block and the last block, will be mapped to the two high protection-degree labeling bits within the symbol. On the other hand, the remaining $w-2$ coded bits, which are extracted from the remaining $w-2$ blocks, will be mapped to the $w-2$ low protection-degree labeling bits in a sequential order.%\vspace{1mm}

{\em Remark:} The optimized TB bit-mapping scheme in \cite{7005396} and the VNMM scheme in \cite{8883091} are effective for both regular and irregular codes, while the VDMM-based schemes in \cite{1605713,5613828,6777400} are effective only for the irregular codes.

{\bf \em Example 15:} Consider a TB-SC-PLDPC-BICM system using an LBPM-labeled $16$QAM, where $w=4$. To realize the VNMM scheme, the entire codeword is divided into $w$ blocks in a sequential order, where each block includes $N'=\frac{N_{\rm SC}}{w}$ coded bits. According to the LBPM-labeled $16$QAM constellation in Fig.~\ref{fig:fig.A1}(b), the input-output MIs of four labeling bits satisfy $I(\hat{v}_{k,1}; y_k) =I(\hat{v}_{k,2}; y_k) > I(\hat{v}_{k,4}; y_k) > I(\hat{v}_{k,3}; y_k)$. Thereby, the high protection-degree labeling bits are $\hat{v}_{k,1}$ and $\hat{v}_{k,2}$. %In particular, $\hat{v}_{k,1}$ and $\hat{v}_{k,2}$ are the high protection-degree labeling bits, while $\hat{v}_{k,3}$ and $\hat{v}_{k,4}$ are the low protection-degree labeling bits.

In the mapping procedure, the four coded bits involved in each symbol can be extracted from the four different sub-blocks separately. Specifically, the coded bits in the first and the last blocks must be mapped to the labeling bits $\hat{v}_{k,1}$ and $\hat{v}_{k,2}$, respectively. Meanwhile, the coded bits in the second and the third blocks must be mapped to the remaining two labeling bits $\hat{v}_{k,3}$ and $\hat{v}_{k,4}$, respectively. By using the VNMM scheme, the wave-like convergence of the TB-SC-PLPDC code can be initiated similarly to that of the TE-SC-PLDPC code without any rate loss. In this BICM system, the variable nodes at both the beginning and the end of the codeword are protected with a higher priority, so as to accelerate their MI/LLR convergence.
%{\bf \em Example 18:} By exploiting the EXIT algorithm, one can find that the decoding threshold of the TB-SC-PLDPC-coded BICM-ID systems can be further improved with the use of VNMM scheme. In particular, under the LBPM-labeled $8$PSK modulation, the regular-$(3, 6)$ TB-SC-PLDPC code can achieve a threshold gain of about $0.32$ dB with the aid of the VNMM scheme.
%Fig.~\ref{fig:side:wave} also illustrates the decoding waves of the VNMM-aided regular-$(3,6)$ TB-SC-PLDPC code in an LBPM-labeled BICM-ID system. Referring to this figure, the VNMM successfully triggers the wave-like convergence from both ends to the middle area of the codeword. In addition, the a-posteriori MIs in all the variable-node positions (i.e., time instants) converge faster as the number of global iterations increases (i.e., $t_{\rm GL}=1,2,\ldots,8$).
\begin{figure}[t]
\centering\vspace{-3mm}
%\subfigure[\hspace{-0.5cm}]{ %% label for first subfigure
%\includegraphics[width=2.8in,height=2.1in]{{PSK-BER.eps}}\vspace{-2.5mm}
\subfigure[\hspace{-0.5cm}]{ %% label forsecond subfigure
\includegraphics[width=2.8in,height=2.1in]{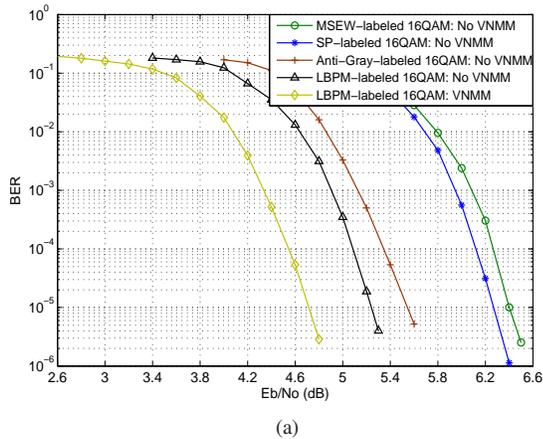}}\vspace{-2.5mm}
\caption{BER curves of the regular-$(3, 6)$ TB-SC-PLDPC code in the LBPM-labeled $16$QAM-aided BICM-ID systems with $r_{\rm SC}=1/2, \varsigma =2, L_{\rm SC}=12, N_{\rm T}=4800, t_{\rm GL,max}=8,\,{\rm and}\,t_{\rm BP,max}=25$.}
\label{fig:BER curves}\vspace{-2mm}
\end{figure}

%For further elaboration,
{\bf \em Example 16:} Fig.~\ref{fig:BER curves} depicts the BER curves of the regular-$(3, 6)$ TB-SC-PLDPC code in the $16$QAM-aided BICM-ID systems with different labelings and bit-mapping schemes.
%The results demonstrate that the LBPM labeling achieves better performance than other three counterparts.
The results demonstrate that the LBPM labeling achieves gains of about $0.3$, $1.0$ and $1.2$ dB over the anti-Gray, SP and MSEW labelings, respectively, at a BER of $10^{-5}$.
Moreover, the TB-SC-PLDPC-BICM-ID systems with the LBPM labeling obtain an additional performance gain of about $0.5$ dB by using the VNMM scheme. %Furthermore, the feasibility of the LBPM constellations and VNMM scheme was verified by the higher modulation orders such as $64$QAM.

\vspace{0.5mm}
\noindent{\bf (2) Hybrid Mapping:} In order to trigger the decoding-wave phenomenon of TB-SC-PLDPC codes in $M$QAM-aided BICM-ID systems, a {\em hybrid mapping scheme} was proposed in \cite{7593089}. This scheme adopts an SP-labeled constellation and another optimized constellation simultaneously to deal with a given codeword. In this scheme, each codeword is divided into $L_{\rm SC}$ blocks according to the coupling length $L_{\rm SC}$. Then, the SP-labeled constellation mapper is employed for the first $T_{\rm SC}~(T_{\rm SC}<L_{\rm SC})$ blocks, while another constellation mapper is adopted for the remaining $L_{\rm SC} - T_{\rm SC}$ blocks. In this way, for a sufficiently large $T_{\rm SC}$, the BP algorithm can locally decode the more reliable code bits, which can be treated as an effective termination to trigger the wave-like decoding convergence. However, the performance of the hybrid mapping scheme is determined by the parameter $T_{\rm SC}$. For any given codeword, it is necessary to search for the optimal value of $T_{\rm SC}$, which leads to an additional computational overhead. Especially, if the optimized constellation is implemented by the Gray-labeled constellation, combining the TB-SC-PLDPC codes with the hybrid mapping scheme can achieve desirable performance gains over the TE-SC-PLDPC counterparts in BICM-ID scenarios, without encountering any rate loss. Owing to these advantages, the hybrid mapping scheme turned out to be a promising alternative to the regular mapping scheme for BICM systems. It should be noted that the hybrid mapping scheme is exploited only for the $16$QAM-aided BICM-ID system in \cite{7593089}. %Possible extensions of this mapping scheme to other modulation orders (e.g., the $64$QAM) have not been adequately investigated.

{\em Remarks:} The hybrid mapping scheme has a similar principle to the IM scheme \cite{5960810,7589683,Zhao2020}, because both schemes employ two different constellations to modulate a single codeword.

\vspace{-1.5mm}
\subsection{Summary}\label{sect:V-C}\vspace{-0.3mm}

In this section, some design paradigms for PLDPC-BICM systems over AWGN channels are discussed. First, an overview of PLDPC-code constructions, constellation and bit-mapper designs for BICM systems is presented. Furthermore, the state-of-the-art progress of the SC-PLDPC-BICM system design is reported. For the SC-PLDPC-BICM systems, sophisticated channel-coding and bit-mapping schemes are developed to trigger the wave-like decoding phenomenon, which help improve the ID performance.%\vspace{-1mm}

\section{Design of PLDPC-BICM over Fading Channels}\label{sect:VI}\vspace{-0.7mm}

This section presents the recent achievements related to designing PLDPC-BICM systems over both ergodic fast-fading channels and non-ergodic block-fading (i.e., slow-fading) channels. The design techniques can be viewed as promising alternatives to realize high-spectral-efficiency transmission in a variety of wireless-communication applications.\vspace{-2mm}
%Table 13
\begin{table*}[t]\scriptsize
\caption{Decoding thresholds $(E_b/N_0)_{\rm th}~{\rm (dB)}$ and capacity gaps $\Delta$ of the rate-$1/2$ AR$4$JA code and EAR$4$JA code in the BICM-NI systems over a Nakagami fast-fading channel with fading depth $m=1$. The Gray-labeled QPSK, $8$PSK and $16$QAM modulations are assumed.}
\centering
\begin{tabular}{|c|c|c|c|c|c|c|c|c|c|}
\hline
\multirow{2}{*}{\diagbox[width=12em,height=2em,trim=l]{Code Type}{\hspace{2mm}Modulation Type}}
& \multicolumn{3}{|c|}{QPSK} & \multicolumn{3}{|c|}{$8$PSK} & \multicolumn{3}{|c|}{$16$QAM}  \\
\cline{2-10}
& ${(E_b/N_0)}_{\rm th}$ & $\Delta$ & Capacity & ${(E_b/N_0)}_{\rm th}$ & $\Delta$ & Capacity  & ${(E_b/N_0)}_{\rm th}$ & $\Delta$  & Capacity  \\
\hline
AR$4$JA & $2.326$ & $0.474$ & \multirow{2}{*}{1.852} & $4.019$	& $0.631$ & \multirow{2}{*}{3.388} & $4.783$ & $0.654$ & \multirow{2}{*}{4.129}  \\
\cline{1-3} \cline{5-6} \cline{8-9} EAR$4$JA & $2.158$ & $0.306$ &                       & $3.722$	 & $0.334$ &                        & $4.556$ & $0.427$ &  \\
\hline
%& \multirow{2}{*}{1.852}
\end{tabular}
\label{tab:XIII}
\vspace{-3mm}
\end{table*}%

\subsection{System Design for Ergodic Fast-Fading Channels}\label{sect:VI-A}

Recall the discussion in Section~\ref{sect:II-A-4}, the PLDPC codes that perform well over an AWGN channel can also perform well over other ergodic channel models \cite{7112076,8740906}. For this reason, the PLDPC codes optimized for BICM systems over AWGN channels can be directly applied to ergodic fast-fading channels \cite{6133952,8999517,8798970}. It has been shown that the AWGN-optimized PLDPC codes do exhibit excellent performance over fast-fading channels. %, which can meet the reliability requirement for wireless communication applications.
Hence, there are a few works devoted to the PLDPC-code design for BICM systems in such scenarios. More research effort was devoted to developing bit-mapping schemes for SC-PLDPC-BICM systems \cite{924878,1532194,6133952,8999517,8611290,8798970}.

\subsubsection{Design of PLDPC-BICM Systems}\label{sect:VI-A-1}
%Here, we succinctly introduce the research progress of PLDPC-coded BICM system over ergodic fast-fading channels reported in the existing literature.

%Since no related work discussed constellation shaping and bit-mapper design in PLDPC-BICM system over ergodic fast-fading channels, only the progress of code construction is discussed here.

It was proved that the AR$4$JA code possesses both excellent decoding threshold and MHDGR, and thus performs well in both low- and high-SNR regions over AWGN channels \cite{5174517}. Based on the structure of the rate-$1/2$ AR$4$JA code in Fig.~\ref{fig:Fig.6V}(d), an {\em EAR4JA code} was constructed in \cite{6133952}, by adding two variable nodes and one check node into the protograph. The base matrix of the rate-$1/2$ EAR$4$JA code is shown in \eqref{eq:EAR4JA}, where the variable node corresponding to the second column is punctured. Using the PEXIT algorithm, one can easily calculate the decoding threshold of the EAR$4$JA code as $0.395~{\rm dB}$ over a BPSK-modulated AWGN channel, which has a gap of only $0.208~{\rm dB}$ to the channel capacity. However, no work discussed bit-mapper design for PLDPC-BICM over ergodic fast-fading channels.
\begin{eqnarray}
{\bB}_{\rm EAR4JA} &=\left[\begin{array}{ccccccc}
1 & 2 & 0 & 0 & 1 & 0 & 0\cr
0 & 3 & 1 & 1 & 1 & 0 & 1\cr
0 & 1 & 2 & 2 & 2 & 1 & 1\cr
0 & 2 & 0 & 0 & 0 & 2 & 0\cr
\end{array}\right]
\label{eq:EAR4JA}
\end{eqnarray}
%%fig.17
%\begin{figure}[tbp]
%\centering
%\includegraphics[width=2.15in,height=0.8in]{Fig.17.eps}\vspace{-2mm}
%\caption{Protograph of the rate-$1/2$ EAR$4$JA code.}
%\label{fig:Fig.17}\vspace{-3mm}  %label for entire figure
%\end{figure}

{\bf \em Example 17:} Table~\ref{tab:XIII} compares the decoding thresholds and capacity gaps of the rate-$1/2$ AR$4$JA code and EAR$4$JA code in the $M$PSK/$M$QAM-aided BICM-NI systems over a Nakagami fast-fading channel. As seen, AR$4$JA-coded BICM-NI systems possess decoding thresholds of about $0.47 \sim 0.66~{\rm dB}$ to the corresponding capacity limits. Moreover, the EAR$4$JA code that outperforms the AR$4$JA code over an AWGN channel also preserves it advantage over a fast-fading channel. For example, the EAR$4$JA code not only has gains of about $0.17 \sim 0.30~{\rm dB}$ over the AR$4$JA code, but also operates about $0.3 \sim 0.43~{\rm dB}$ away from the corresponding capacity limits. These demonstrate that there is no need to re-optimize the PLDPC codes over ergodic fast-fading channels.
%%Table 14
%\begin{table*}[t]\scriptsize
%\caption{Decoding thresholds $(E_b/N_0)_{\rm th}~{\rm (dB)}$ of the regular-$(3, 6)$ TE-SC-PLDPC codes and irregular RJA TE-SC-PLDPC codes with different coupling lengths in the Gray-labeled $16$QAM-aided BICM-NI systems with and without the SPMM scheme over a Nakagami fast-fading channel with fading depth $m=1$. The coupling width and lengths for the regular-$(3, 6)$ TE-SC-PLDPC codes are $\varsigma=2~{\rm and}~L_{\rm SC}=8, 10, 12$, while those for the irregular RJA TE-SC-PLDPC codes are $\varsigma=1~{\rm and}~L_{\rm SC}=4, 5, 6$.}
%\centering
%\begin{tabular}{|c|c|c|c|c|c|c|}
%\hline
%\multirow{2}{*}{code rate}
%& \multicolumn{3}{|c|}{Regular TE-SC-PLDPC} & \multicolumn{3}{|c|}{RJA TE-SC-PLDPC} \\
%\cline{2-7}
%& $L_{\rm SC}$ & With SPMM & Without SPMM & $L_{\rm SC}$ & With SPMM & Without SPMM \\
%\hline
%$3/8$ &	 $8$  & $4.506$ & $4.956$ & $4$ & $4.297$ &	$4.741$ \\ \hline
%$2/5$ &	 $10$ &	$4.748$ & $5.058$ &	$5$ & $4.438$ &	$4.891$\\ \hline
%$5/12$ & $12$ &	$4.921$ & $5.109$ &	$6$ & $4.608$ &	$5.058$\\ \hline
%\end{tabular}
%\label{tab:XIV}
%\vspace{-2mm}
%\end{table*}%

\subsubsection{Design of SC-PLDPC-BICM Systems}\label{sect:VI-A-1}

In the current literature, there did not seem to be much work on designing BICM systems deploying SC-PLDPC codes over fast-fading channels.

\noindent{\bf (1) Code Construction:} In recent progress, a new QC-SC-PLDPC-BICM scheme was proposed in \cite{8798970} using a conventional pseudo-random interleaver to support efficient and reliable transmission of QPSK-aided BICM-ID systems. In particular, a new construction method, which consists of a base-matrix optimization scheme and a two-step lifting scheme, was developed for the QC-TE-SC-PLDPC codes to achieve capacity-approaching performance over both AWGN channels and ergodic fast-fading channels. More precisely, the design of capacity-approaching QC-TE-SC-PLDPC codes proposed in \cite{8798970} was formulated over an AWGN channel. Subsequently, the designed codes were applied to ergodic fast-fading channels under the same framework. BER and word error rate (WER) simulations were performed, which demonstrated that the QC-TE-SC-PLDPC codes designed over the AWGN channel also exhibit desirable performance over the Nakagami fast-fading channel. It was further conjectured that the QC-TE-SC-PLDPC codes are able to universally achieve near-capacity performance in BICM-ID systems even for the case of higher-order modulation. As such, the QC-TE-SC-PLDPC-BICM-ID appears to be a good attractive transmission scheme for practical wireless systems due to its low implementation complexity. The details of the design, especially the base-matrix optimization scheme and two-step lifting scheme, has been discussed in Section~\ref{sect:V-B-1}.

\vspace{0.5mm}
\noindent{\bf (2) Bit-Mapper Design:}  According to \cite{7152893}, the coded bits in the middle area of the coupled chain are much more difficult to converge than those at the beginning and the end for a TE-SC-PLDPC code when the coupling length $L_{\rm SC}$ is a finite value. Motivated by this property, a position-based bit-mapping scheme was proposed for the TE-SC-PLDPC codes in BICM-NI systems \cite{7005396}. Yet, this scheme is not well suited for irregular TE-SC-PLDPC codes (e.g., the RJA TE-SC-PLDPC codes \cite{7152893} and RA TE-SC-PLDPC codes \cite{7339427}), in which some coded bits at the beginning and the end of the coupled chain may converge slower than those in the middle. This is one salient property of the irregular TE-SC-PLDPC codes distinguished from the regular TE-SC-PLDPC codes. To overcome the weakness of the work of \cite{7005396}, a spatial-position matched mapping (SPMM) scheme was constructed in \cite{8999517} for the TE-SC-PLDPC-BICM-NI system based on MI analysis. The SPMM can enhance the decoding-wave phenomenon and accelerate the convergence of TE-SC-PLDPC-BICM systems irrespective of the code structure.
%%Fig.18
%\begin{figure}[tbp]
%\centering
%\includegraphics[width=2.9in,height=2.2in]{Fig.18.eps}
%\vspace{-0.2cm}
%\caption{A-posteriori MIs of the variable nodes at different positions for the RJA TE-SC-PLDPC code in the Gray-labeled $16$QAM-aided BICM-NI system over a Nakagami fast-fading channel. The parameter are $m=1, r_{\rm SC}= 5/12,\varsigma=1, L_{\rm SC}=6,n_{\rm SC}= 24,m_{\rm SC}= 14,~{\rm and}~t_{\rm BP,max}=100$.}
%\label{fig:Fig.18V}  %label for entire figure
%\vspace{-4mm}
%\end{figure}

Now, consider a TE-SC-PLDPC-BICM-NI system with Gray-labeled $M$PSK/$M$QAM. The principle of the SPMM scheme can be summarized as follows. Assume that the length of the TE-SC-PLDPC code is $N=Z n_{\rm SC}=Z n_{\rm P} L_{\rm SC}$, where $Z$ is the lifting factor, $n_{\rm SC}$ and $n_{\rm P}$ are the numbers of the variable nodes of the TE-SC protograph and the uncoupled original protograph, respectively, $L_{\rm SC}$ is the coupling length, and the codeword length $N$ is a multiple of $w=\log_2 M~({\rm i.e.,}~N'= N/w)$. At the beginning, the entire codeword is divided into $n_{\rm SC}$ blocks in a sequential order, i.e., ${\cal V}_{\rm SC} = \{{\cal V}_{\rm B,1},{\cal V}_{\rm B,2},\ldots, {\cal V}_{{\rm B},n_{\rm SC}} \}$, where ${\cal V}_{{\rm B},j}$ is the $j$-th block that includes all the $Z$ variable nodes at the $j$-th position. Subsequently, the $n_{\rm SC}$ blocks ${\cal V}_{\rm B,1},{\cal V}_{\rm B,2},\ldots, {\cal V}_{{\rm B},n_{\rm SC}} $ are re-ranked to ${\cal V}_{\rm B,\iota(1)},{\cal V}_{\rm B,\iota(2)},\ldots, {\cal V}_{{\rm B},\iota(n_{\rm SC})} $ in a descending order of their {\it a posteriori} MIs output from the outer decoder. Furthermore, the $n_{\rm SC}$ re-ranked blocks are consecutively converted to $w$ intermediate blocks ${\cal V}_{\rm G,1},{\cal V}_{\rm G,2},\ldots, {\cal V}_{{\rm G},w} $, where ${\hspace{-0.5mm}\cal V}_{{\rm G},\mu}$  %= \{\hspace{-0.5mm} {\cal V}_{{\rm B},\iota((\mu-1)(n_{\rm SC}/w)+1)}, \hspace{-0.5mm}{\cal V}_{{\rm B},\iota((\mu-1)(n_{\rm SC}/w)+2)},\hspace{-0.3mm}\,\ldots\,\hspace{-0.3mm}, {\cal V}_{{\rm B},\iota(\mu(n_{\rm SC}/w))} \hspace{-0.5mm}\}$
contains $n_{\rm SC}/w$ blocks (i.e., $N'$ variable nodes), and $\mu=1,2,\ldots,w$. Finally, aiming to balance the convergence speeds of the MIs for all the variable nodes as much as possible in the bit-to-symbol mapping procedure, the $w$ intermediate blocks ${\cal V}_{\rm G,1},{\cal V}_{\rm G,2},\ldots, {\cal V}_{{\rm G},w} $ are consecutively mapped to the $w$ labeling bits within a modulated symbol in an ascending order of their protection degrees. Specifically, all the variable nodes in ${\cal V}_{{\rm G},\mu}$ are mapped to the labeling bits with the $\mu$-th lowest protection degree within the length-$N'$ $M$-ary modulated symbol sequence. Based on this mapping procedure, the variable nodes with relatively slower convergence speed can be protected with relatively higher protection degrees, and thus the decoding-wave effect of the TE-SC-PLDPC code can be further enhanced. As such, the SPMM scheme is a more feasible and precise method with respect to the bit-mapping optimization scheme in \cite{7005396}, because the former substantially uses the MI distribution of the coded bits to accelerate the decoding convergence.
%\begin{figure}[t]
%\centering%\vspace{-2mm}
%%\subfigure[\hspace{-0.5cm}]{ %% label for first subfigure
%\includegraphics[width=3.0in,height=2.3in]{Fig.19a.eps}
%%\subfigure[\hspace{-0.5cm}]{ %% label forsecond subfigure
%%\includegraphics[width=3.0in,height=2.3in]{Fig.19b.eps}}
%\caption{BER curves of the RJA TE-SC-PLDPC codes in the Gray-labeled 16QAM-aided BICM-NI systems with and without the SPMM scheme over a Nakagami fast-fading channel. The transmitted codeword length and maximum local iteration number are assumed as $N_{\rm T}=4800~{\rm and}~t_{\rm BP,max}=100$, respectively; while other parameters are the same as those in Table~\ref{tab:XIV}.}
%\label{fig:Fig.19V}\vspace{-2mm}
%\end{figure}
\begin{figure}[tbp]
\vspace{-2mm}
\center
\includegraphics[width=2.8in,height=2.1in]{{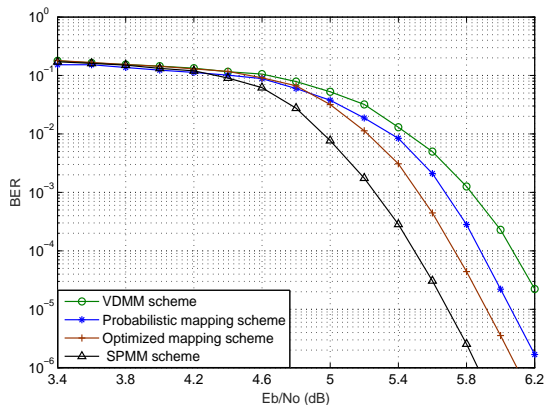}}
\vspace{-0.2cm}
\caption{BER curves of the RJA TE-SC-PLDPC code with four different bit-mapping schemes in the Gray-labeled $16$QAM-aided BICM-ID systems over a Nakagami fast-fading  channel with $m=1, r_{\rm SC}= 5/12,\varsigma=1, L_{\rm SC}=6,N_{\rm T}=4800,~{\rm and}~t_{\rm BP,max}=100$.}
\label{fig:Fig.20V} \vspace{-2.5mm}
\end{figure}

{\bf \em{Example 18:}} %We present the BER curves of the irregular RJA TE-SC-PLDPC codes in the Gray-labeled $16$QAM-aided BICM-NI systems with and without the SPMM scheme over a Nakagami fast-fading channel with $m=1$ in Fig.~\ref{fig:Fig.19V}. One can easily find that the RJA TE-SC-PLDPC codes can achieve much better error performance by using the SPMM scheme for all the three coupling lengths (i.e., code rates) in the BICM-NI systems. As a further advance,
Consider the BICM-NI systems deploying the rate-$5/12$ RJA TE-SC-PLDPC code and Gray-labeled $16$QAM modulation. Fig.~\ref{fig:Fig.20V} compares the performance of the SPMM scheme with the VDMM scheme \cite{1605713}, the probabilistic mapping scheme \cite{5199551}, and the optimized mapping scheme \cite{8170954}.
As observed from this figure, the SPMM needs an SNR of about $5.6$ dB to achieve a BER of $2 \times 10^{-5}$, while the optimized mapping, probabilistic mapping and VDMM need SNRs of about $5.85$ dB, $6.0$ dB and $6.2$ dB to do so, respectively.
The results verify the superiority of the SPMM scheme.

Finally, it is noted that there does not seem to be any work on the design of signal constellations for the SC-PLDPC-BICM systems over ergodic fast-fading channels.%, which therefore will not be discussed in this article.\vspace{-0.8mm}

\subsection{System Design for Non-Ergodic Block-Fading Channels} \label{sect:VI-B}

The block-fading channel has attracted considerable attention during the past three decades because it can be used to model many communication scenarios with slowly varying fading, such as orthogonal frequency division multiplexing systems, free-space optical systems, cooperative communication systems, and cellular networks \cite{7105928,6205592,6875246}. In this channel model, the fading gain remains constant on each symbol block but randomly varies on different blocks. As a type of non-ergodic channel, block-fading channel possesses some salient features quite different from those in ergodic AWGN and fast-fading channels. Of particular significance is that the block-fading channel is not information stable. For this reason, the information outage-probability limit, instead of the Shannon capacity limit, is utilized to characterize the fundamental performance limit of block-fading channels \cite{6905810,2000On}. %As a remedy, the information outage-probability limit was introduced in \cite{2000On} to serve as a key metric to accurately measure the theoretically lower limit on the WER for any CM system over a block-fading channel.
In recent years, PLDPC codes along with their BICM schemes have been recommended for such non-ergodic channels \cite{8740906,8019817,7080854,6875246,6905810}, which are obviously superior to the previously-proposed convolutional- and turbo-coded counterparts \cite{6952165,1564429,2000On}.

Assuming a BICM system deploying an $M$-ary modulation and a rate-$r$ PLDPC code, the achievable diversity order of a PLDPC code over a block-fading channel is bounded by %\vspace{-1mm}%the Singleton bound, i.e.,
%\begin{equation}
$d \leq m \left(1 + \left\lfloor L\left( 1- R_{\rm SE} / \log_2|\chi| \right) \right\rfloor \right)$  \cite{8740906},
%\label{eq:diversity-bound}%\vspace{-2mm}
%\end{equation}
where $L$ is the number of fading blocks, $m$ is the fading depth, $R_{\rm SE}=r \log_2 M = r w$ is the spectral efficiency, and $\chi$ is the signal constellation set. This bound implies an optimal trade-off between the spectral efficiency (or code rate) and the diversity order. A PLDPC code is said to have full diversity if and only if $d=mL$. As a result, the highest achievable spectral efficiency (resp. code rate) for a full-diversity PLDPC-BICM system over a BF channel is $R_{\rm SE, max} =\log_2 |\chi|/L = w/L$ (resp. $r_{\rm max} = \frac{1}{L}$). Accordingly, designing a rate-$\frac{1}{L}$ full-diversity PLDPC code and an appropriate $M$-ary modulation is indispensable in realizing high-efficiency transmissions over BICM block-fading channels \cite{6205592,1564429}.

\subsubsection{Design of PLDPC-BICM Systems} \label{sect:VI-B-1}

In the past five years, a flurry of research activities took place in devising PLDPC-BICM systems over block-fading channels. In \cite{8740906}, PLDPC-BICM systems were carefully investigated over $M$PSK/$M$QAM-modulated Nakagami block-fading channels from the perspectives of code construction, modulation-strategy design, and bit-mapper optimization. Besides, a number of works studied the $M$-ary chaos-based SS BICM systems, i.e., $M$DCSK-aided BICM systems, with the use of PLDPC codes over such channels \cite{7093149,7935433,7968491,7956256,8338131,8012533}. In the following, these PLDPC-BICM designs over block-fading channels will be discussed.%, conditioned on $M$PSK/$M$QAM and $M$DCSK modulations.

\vspace{0.5mm}
\noindent{\bf (1) PSK/QAM-aided PLDPC-BICM Design:} To achieve full diversity over block-fading channels, several types of powerful LDPC codes, including root-LDPC codes \cite{5550477,7489038} and root-PLDPC codes \cite{8019817,7080854,6905810}, were proposed. Although all these LDPC codes are able to attain outage-limit-approaching performance with the maximum code rate (i.e., $r_{\rm max}=1/L$) over block-fading channels, the conventional root-LDPC codes are randomly constructed and thus suffer from relatively high encoding and decoding complexity. On the contrary, the structured root-PLDPC codes emerged to be a more desirable choice for such environments, because they possess both advantages of full diversity and simple implementation. As a special type of check nodes, rootchecks are the key component in root-PLDPC codes, which guarantees full diversity for all the information-bearing coded bits. However, it is very challenging for the root-PLDPC codes to adapt to the high-order modulations without losing full diversity \cite{4558590,8740906,1564429}. Here, the principle of root-LDPC codes is briefly introduced, followed by some discussions on their feasibility to BICM-ID framework.

%$\langle{\rm i}\rangle$ Root-PLDPC-Code Construction: %Rootcheck is the key component to guarantee the full-diversity property of root-PLDPC codes (i.e., $d=L+1$).
Consider an $L$-layer root-PLDPC code transmitted over a block-fading channel. A type-$l~(l=1,2,\ldots,L)$ {\em rootcheck} is defined as a check node having one edge connecting to an information-bearing variable node transmitted on the $l$-th fading gain $\alpha_l$, and the remaining edges connecting to other variable nodes are transmitted on another fading gain $\alpha_{l'}~(l' \neq l)$. The maximum achievable code rate for a full-diversity PLDPC code in such a scenario is $r_{\rm max}=1/L$. To guarantee full diversity and maximum rate, an $L$-layer root-PLDPC code must include $L$ different types of rootchecks.
%Each information-bearing variable node in the $l$-th fading block must simultaneously connect to $L-1$ type-$l$ rootchecks via $L-1$ single edges, which separately span the remaining $L-1$ fading blocks.
Also, to achieve full diversity of the information-bearing variable nodes, each type-$l$ rootcheck set must be composed of $L-1$ different rootchecks, which separately span the remaining $L-1$ fading blocks \cite{8019817}.
%For brevity, the $L-1$ type-$l$ rootchecks are referred to as {\em type-$l$ rootcheck set}.

Based on the principle of rootcheck, one can easily construct an $L$-layer root-PLDPC code ${\bm \Lambda} = ({\cal V}_1,{\cal V}_2, \ldots,{\cal V}_{L})$ based on an $m_{\rm P}\times n_{\rm P}$ protograph, as follows.
%Consider an $m_{\rm P}\times n_{\rm P}$ protograph that corresponds to a rate-$1/L$ PLDPC code.
At the beginning, the $n_{\rm P}/L$ information bits are uniformly divided into $L$ subsets, in which the $l$-th subset is transmitted on the $l$-th fading gain $\alpha_l$. Meanwhile, the $(L-1)n_{\rm P}/L$ parity bits are uniformly divided into $L$ subsets, to effectively protect the information bits in each fading block. The $l$-th information-bit subset and $l$-th parity-bit subset constitute the $l$-th code block, and the $L$ different blocks constitute the overall variable-node set of the root-PLDPC code. To achieve full diversity, each information-bearing variable node in the $l$-th block must be simultaneously connected to all the $L-1$ rootchecks belonging to the type-$l$ rootcheck set with $L-1$ single edges. On the other hand, the parity-check variable nodes in the $l$-th block must be connected to other $L-1$ types of rootchecks. The $L$ different types of rootcheck sets constitute the overall check-node set of the root-PLDPC code. As a consequence, a full-diversity $L$-layer root-PLDPC code can be constructed by combining the variable-node set, check-node set, and their associated edges, into a single entity.
%\begin{figure}[tbp]
%%\vspace{-0.3cm}
%\center
%\includegraphics[width=3.45in,height=0.6in]{{New-fig-1.eps}}
%\vspace{-0.2cm}
%\caption{Base-matrix structure of an $L$-layer root-PLDPC code.}
%\label{fig:Fig.new-fig1} \vspace{-3mm}
%\end{figure}

According to the above construction procedure, the number of variable nodes in a root-PLDPC protograph must be $n_{\rm P} = L^2$ in order to ensure the lowest encoding complexity. In the root-PLDPC code, the information bits rather than the parity bits can realize full diversity, which is called {\em UEP}. The root-PLDPC codes are capable of achieving excellent performance because only the information bits are counted in performance measurement.
%The $(L-1)L \times L^2$ base matrix $\bB_{\rm RP}$ of an $L$-layer root-PLDPC code is shown in Fig.~\ref{fig:Fig.new-fig1},
%where ${\bm 1}$ and $\bB_{{\rm i}l,\mu}~(l= 1,2,\ldots, L; \mu =1,2,\ldots,L-1)$ are the all-ones matrix and random matrix of size $(L-1) \times 1$, respectively; ${\bm 0}$ and $\bB_{{\rm p}l,\mu}$ are the all-zeros matrix and random matrix of size $(L-1) \times (L-1)$, respectively; ${\cal V}_{{\rm i}l}$ and ${\cal V}_{{\rm p}l}$ are the information-bearing variable-node subset and parity-check variable-node subset in the $l$-th fading block, respectively; ${\cal V}_l= ({\cal V}_{{\rm i}l}, {\cal V}_{{\rm p}l})$ and ${\cal C}_l$ are the $l$-th variable-node set and the type-$l$ rootcheck set, respectively. Hence, the entire variable-node set and check-node set of the rate-$1/L$ root protograph can be denoted by ${\cal V}_{\rm RP} = ({\cal V}_1,{\cal V}_2, \ldots,{\cal V}_{L})$ and ${\cal C}_{\rm RP} = ({\cal C}_1,{\cal C}_2, \ldots,{\cal C}_{L})$, respectively.
Moreover, one can clearly distinguish the information bits from the parity bits in a root-PLDPC code \cite{7080854}. %Note that the above base matrix can be easily extended to the scenario with larger sizes, i.e., $z(L-1)L \times zL^2$ with $z \ge 2$.

{\bf \em Example 19:} According to the concepts of root-PLDPC codes, one can easily derive the base matrices of the rate-$1/2$ regular-$(3, 6)$ bilayer root-PLDPC code and the rate-$1/3$ regular-$(4,6)$ three-layer root-PLDPC code, as \eqref{eq:B-matrix-RP-23}.
To simplify the exposition, the regular-$(3, 6)$ bilayer root-PLDPC code and the regular-$(4,6)$ three-layer root-PLDPC code are called {\em RP-2 code} and {\it RP-3 code}, respectively. It was proved in \cite{8019817} that the above two types of root-PLDPC codes can achieve full diversity over BPSK-modulated block-fading channels with $L=2$ and $L=3$, respectively. Likewise, full-diversity irregular root-PLDPC codes can be constructed by varying the row/column weights of the base matrix while keeping the rootcheck structure unchanged.\vspace{-1mm}
\begin{eqnarray}
\hspace{-1mm} {\bB}_{\rm RP\textit{-}2} \hspace{-1mm}=\hspace{-1mm} \left[\hspace{-1mm}\begin{array}{lll}
1 & 0 ~|~ 2 & 3 \cr
2 & 3 ~|~ 1 & 0 \cr
\end{array}\hspace{-1mm}\right],\,%\nonumber\\
\bB_{\rm RP\textit{-}3} \hspace{-1mm}=\hspace{-1mm}
\left[\hspace{-1mm}\begin{array}{ccccccccccc}
1 & 0 & 0 & | & 1 & 2 & 2 & | & 0 & 0 & 0 \cr
1 & 0 & 0 & | & 0 & 0 & 0 & | & 1 & 2 & 2 \cr
1 & 2 & 2 & | & 1 & 0 & 0 & | & 0 & 0 & 0 \cr
0 & 0 & 0 & | & 1 & 0 & 0 & | & 1 & 2 & 2 \cr
1 & 2 & 2 & | & 0 & 0 & 0 & | & 1 & 0 & 0 \cr
0 & 0 & 0 & | & 1 & 2 & 2 & | & 1 & 0 & 0 \cr
\end{array}\hspace{-1mm}\right]\hspace{-1mm}.
\label{eq:B-matrix-RP-23}
\end{eqnarray}%\vspace{-5mm}

However, when the modulation order $M$ varies from $2$ to $2^w~(w \ge 2)$, the root-PLDPC codes are unable to achieve full diversity. To overcome this weakness, a novel modulation strategy was proposed in \cite{8740906} to determine the order of $M$PSK/$M$QAM, which guarantees full-diversity order and maximum spectral efficiency for the root-PLDPC-BICM systems.

Suppose that an $L$-layer root-PLDPC code is deployed in an $M$PSK/$M$QAM-aided BICM system over a block-fading channel.
%The protograph corresponding to the $L$-layer root-PLDPC code consists of $m_{\rm P}=(L-1)L$ check nodes and $n_{\rm P}=L^2$ variable nodes. Then, a $z m_{\rm P} \times z n_{\rm P}$ intermediate protograph can be constructed by lifting the original $ m_{\rm P} \times n_{\rm P}$  protograph, where $z$ is much small than the lifting factor $Z$ of the root-PLDPC code (i.e., $1 \le z \ll Z$). In this situation, the $z$ replicas of variable-node set ${\cal V}_l$ in the $l$-th fading block are generated to formulate an intermediate variable-node set ${\cal V}_{{\rm IP},l} = ({\cal V}_{l,1},{\cal V}_{l,2},\ldots, {\cal V}_{l,z})$, where ${\cal V}_{l,\mu}= {\cal V}_{{\rm i}l,\mu}\cup {\cal V}_{{\rm p}l,\mu}~(\mu=1,2,\ldots,z)$ is the $\mu$-th replica of ${\cal V}_l$ that includes a single information bit and $L-1$ parity bits. To preserve the full-diversity property of the root-PLDPC code in the BICM system, the relationship among the modulation order $M$, the number of fading blocks $L$, and the initial lifting factor $z$ must be carefully analyzed by taking the non-ergodicity into account. Actually, there are three possible cases for the modulation-order selection based on a given $z$ and a given $L$, i.e., Case $1~({w =zL})$, Case $2~({w >zL})$, and Case $3~({w <zL})$.
Then, as demonstrated in \cite{8019817}, the modulation order must satisfy $M=2^w=2^{zL}$ in order to enable the root-PLDPC-BICM system to achieve full diversity with both maximum code rate and maximum spectral efficiency, where $z$ is a positive integer. For simplicity, one may assume that $z=1$ when developing a bit-mapping scheme or related techniques. %However, the corresponding methodologies can be easily generalized to the cases of $z>1$.

In addition, a UEP-based bit-mapping scheme was developed in  \cite{8740906} to further accelerate the decoding convergence of root-PLDPC-BICM over block-fading channels. Consider the scenario that an $L$-layer root-PLDPC code is transmitted over a block-fading channel in an $M$PSK/$M$QAM-aided BICM system, where the protograph size is $(L-1)L \times L^2$, and $M=2^w=2^L$. %The root-PLDPC code corresponding to the protograph includes $n_{\rm P}=L^2$ variable nodes and $m_{\rm P}=(L-1)L$ check nodes.
Here, the $l$-th variable-node set transmitted on the $l$-th fading gain $\alpha_l$ is expressed as ${\cal V}_{l}= (v_{{\rm i}l,1}, v_{{\rm p}l,1},v_{{\rm p}l,2}\ldots, v_{{\rm p}l,L-1})$, which can be exactly mapped to an $M$-ary symbol $x_k \triangleq (\hat{v}_{k,1},\hat{v}_{k,2},\ldots, \hat{v}_{k,w})$ transmitted with the same fading gain, where $v_{{\rm i}l,l}, v_{{\rm p}l,\mu'},  \,{\rm and}\,\hat{v}_{k,\mu}$ are the single information-bearing variable node in ${\cal V}_l$, the $\mu'$-th parity-check variable node in ${\cal V}_l$, and the $\mu$-th labeling bit within $x_k$, respectively, $k=1,2,\ldots,N'$. In the UEP-based bit-mapping scheme,
the information-bearing variable node $v_{{\rm i}l,1}$ should be mapped to the lowest protection-degree labeling bit within the symbol, while the $L-1$ parity-check variable nodes $v_{{\rm p}l,1},v_{{\rm p}l,2}\ldots, v_{{\rm p}l,L-1}$ should be mapped to the remaining $w-1$ labeling bits within the symbol. The UEP-based bit-mapping scheme can realize the best combination between the root-PLDPC code and the $M$PSK/$M$QAM modulation because it exploits the UEP property of both high-order modulation and rootcheck structure. %Due to the space limitation, the detailed proof of the above proposition is omitted here.
As illustrated in \cite{8019817}, the anti-Gray labeling achieves a significant performance gain over the Gray-labeling in the root-PLDPC-BICM-ID system over block-fading channels, while the former is slightly inferior to the latter without ID. Hence, the anti-Gray labeling is a better choice than the Gray labeling in this scenario.

\begin{figure}[tbp]
%\vspace{-0.3cm}
\center
\includegraphics[width=2.8in,height=2.1in]{{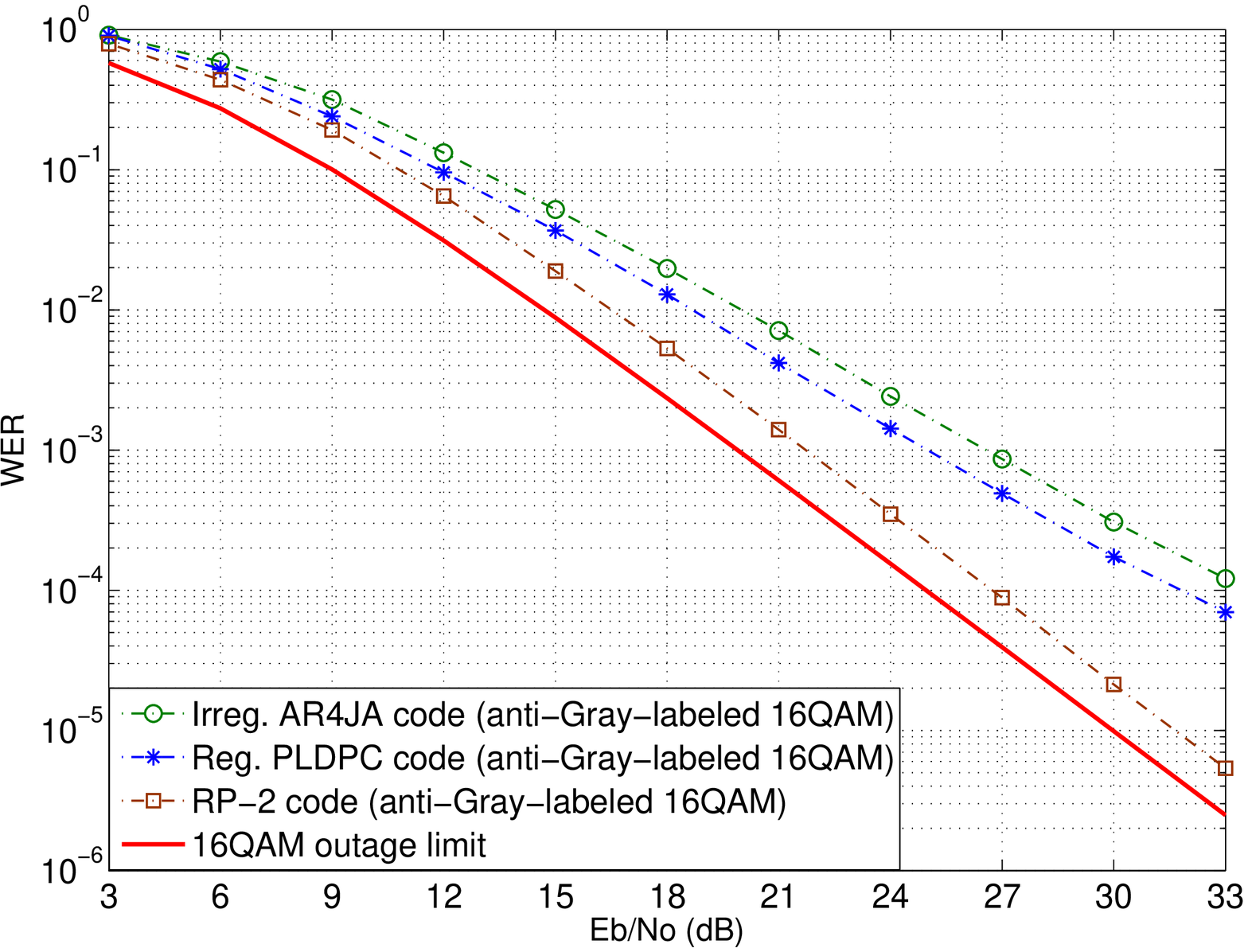}}
\vspace{-0.2cm}
\caption{WER curves of the root-PLDPC-BICM-ID system and two conventional PLDPC-BICM-ID systems over a Nakagami block-fading channel. The codes used are the bilayer RP-2 code, regular-$(3, 6)$ PLDPC code, and irregular AR$4$JA code, the modulation used is anti-Gray-labeled $16$QAM, respectively, and the parameters used are $m=1, L=2, r_{\rm P}=1/2, N_{\rm T}=1024, t_{\rm GL,max}=5, ~{\rm and}~t_{\rm BP,max}=50$.}
\label{fig:Fig.24V} \vspace{-3mm}
\end{figure}

{\bf \em Example 20:} Fig.~\ref{fig:Fig.24V} shows the WER curves of the bilayer RP-$2$ code, regular-$(3, 6)$ PLDPC code, and irregular AR$4$JA code in the BICM-ID systems over a Nakagami block-fading channel. Among the three types of PLDPC codes, only the RP-$2$ code can achieve full diversity and outage-limit-approaching performance.
%Finally, we test the performance of the root-PLDPC-BICM-ID systems over the Nakagami block-fading channels with different fading depths in Fig.~\ref{fig:Fig.25V}. As expected, the root-PLDPC-coded BICM-ID scheme preserves full diversity and accomplishes near-outage-limit performance as the fading depth varies.
Additionally, simulations on the RP-$3$ code over the BICM block-fading channels verified its excellent performance.

%\begin{figure}[tbp]
%%\vspace{-0.3cm}
%\center
%\includegraphics[width=3.0in,height=2.3in]{{Fig.25.eps}}
%\vspace{-0.2cm}
%\caption{WER results of the root-PLDPC-coded BICM-ID system over Nakagami BF channels with different fading depths. The code and modulation used are the bilayer RP-2 code and anti-Gray-labeled 16QAM, respectively, and the parameters used are $r_{\rm P}=1/2, N_{\rm T}=1024, L=2, t_{\rm GL,max}=5, ~{\rm and}~t_{\rm BP,max}=50$.}
%\label{fig:Fig.25V} \vspace{-3mm}
%\end{figure}

\vspace{0.5mm}
\noindent{\bf (2) DCSK-aided PLDPC-BICM Design:} DCSK modulation is a non-coherent SS-modulation scheme that benefits from excellent anti-fading robustness, powerful near-far resilience, and low implementation complexity. During the past two decades, a great deal of research effort has been devoted to optimizing the performance of DCSK as well as their high-order counterparts (i.e., $M$DCSK) and to facilitating their implementation \cite{9098915,8928564}. Based on the current research advancements, $M$DCSK has become an appealing modulation technique for low-power and low-complexity wireless applications, such as wireless personal area networks and wireless sensor networks. To realize high-reliability and high-throughput transmission in such short-range wireless-communication systems, the joint design of PLDPC codes and $M$DCSK was carefully investigated over block-fading channels \cite{7442517,8338131,7956256,8012533,7109922,9599361}.

More precisely, there are two types of $M$DCSK modulations, i.e., Walsh-code-based $M$DCSK and constellation-based $M$DCSK, which can construct $M$-ary chaotic signals with the $M$-order Walsh code and $M$-ary constellation, respectively \cite{7442517,7109922}. Both $M$DCSK modulations can be used to formulate robust PLDPC-BICM systems over block-fading channels.

%$\langle{\rm i}\rangle$ Walsh-Code-based $M$DCSK:
In the BICM-ID system with a Walsh-code-based $M$-DCSK modulation, a length-$N$ PLDPC codeword ${\bm \Lambda}=(v_1, v_2, \ldots, v_N)$ is first converted to a length-$N'$ non-binary codeword ${\bm \Lambda}_{\rm NB}=(v_{\rm NB,1}, v_{\rm NB,2}, \ldots, v_{{\rm NB},N'})$, where $v_{{\rm NB},k} \in \{ 0,1,\ldots, M-1\}$ is the $k$-th non-binary coded symbol, with $w=\log_2 M$ and $k=1,2,\ldots,N'$. The $M$ possible values of non-binary symbol will be utilized to select their corresponding row indices of an $M$-order Walsh code, so as to guarantee the orthogonality of the $M$ different transmitted chaotic signals. In particular, a $2^w$-order orthogonal Walsh code is introduced, as \cite{7442517}\vspace{-1mm}
\begin{equation}
\overline{\bW}_{2^w} = \left [
\begin{array}{rr}
\overline{\bW}_{2^{w-1}} &  \overline{\bW}_{2^{w-1}} \cr
\overline{\bW}_{2^{w-1}} & -\overline{\bW}_{2^{w-1}}
\end{array}
\right] %\nonumber\\
= [{\bm \omega}_1,{\bm \omega}_2,\ldots, {\bm \omega}_M]^{\rm T},
\label{eq:Walsh-general}
\end{equation}
where $\overline{\bW}_{2^0}= \overline{\bW}_1 = 1, {\bm \omega}_\mu = [\omega_{\mu,1}, \omega_{\mu,2}, \ldots, \omega_{\mu,M}]$ is the $\mu$-th row vector of the Walsh code, $\mu=1,2,\ldots, M$.

Assume that $\bc_{k} = [c_{k,1},c_{k,2},\ldots,c_{k,\beta}]$ is the reference-chaotic fragment for the $M$-DCSK modulation with a spreading factor $\beta$. Then, the transmitted $M$DCSK-modulated signal $\bx_k$ corresponding to the non-binary symbol $v_{{\rm NB},k}$ can be formulated by multiplying the elements of the $(v_{{\rm NB},k}+1)$-th row vector ${\bm \omega}_{v_{{\rm NB},k}+1}$ in the Walsh code with the $M$ delayed replicas of reference-chaotic fragment $\bc_{k}$, which can be expressed as $\bx_k = [\omega_{v_{{\rm NB},k}+1,1} \bc_{k}, \omega_{v_{{\rm NB},k}+1,2} \bc_{k}, \ldots, \omega_{v_{{\rm NB},k}+1,M} \bc_{k}]$. As a consequence, a Walsh-code-based $M$DCSK signal consists of $M$ chaotic fragments, which will result in a global spreading factor of $M \beta$. At the receiver, a generalized-maximum-likelihood (GML) energy detector is employed to demodulate the ``corrupted" non-binary symbol output from the block-fading channel \cite{7442517}.
%Fig.28
\begin{figure}[tbp]
\centering
\includegraphics[width=2.8in,height=2.1in]{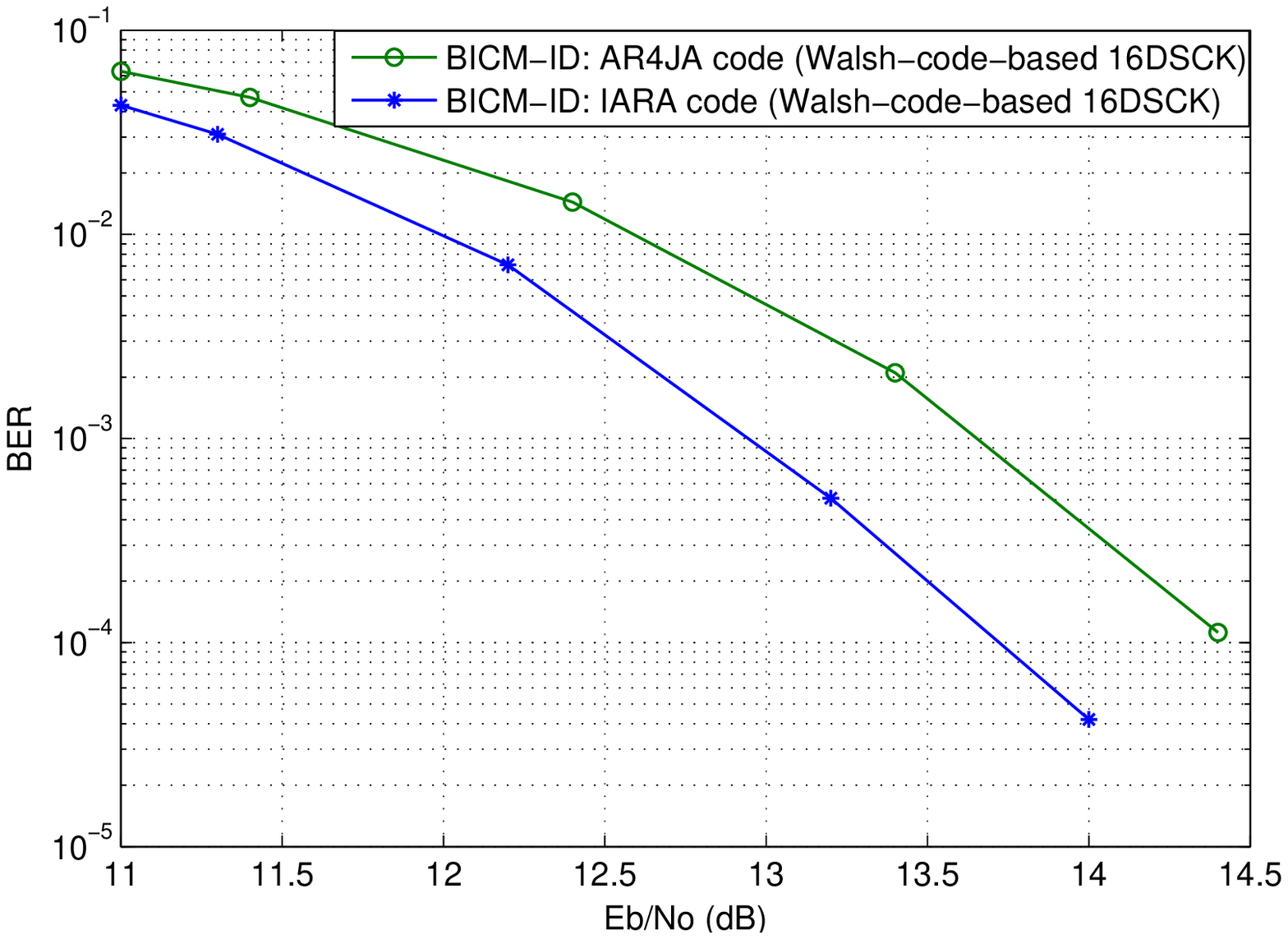}
\vspace{-0.2cm}
\caption{BER curves of the rate-$1/2$ IARA code and AR$4$JA code in the Walsh-code-based $16$DCSK-aided BICM-ID system over a Nakagami block-fading channel with $m=1, N_{\rm T}=4096, \beta=40, t_{\rm GL,max}=5,~{\rm and}~t_{\rm BP,max}=40$.}
\label{fig:Fig.28V}  %label for entire figure
\vspace{-3.5mm}
\end{figure}
%In the GML detector, $M$ energy metrics $z_1, z_2, \ldots, z_M$ are formulated to decide the retrieved symbol $\widetilde{g}_k$ for $v_{{\rm NB},k}$ \cite{7442517}.
%To elaborate a little further, we present the block diagrams of the modulator and demodulator for the Walsh-code-based $M$DCSK in Fig.~\ref{fig:Fig.26V}, where $\mu =v_{{\rm NB},k}+1$ for the modulator.
%%fig.26
%\begin{figure}[t]
%\center
%\includegraphics[width=3.6in,height=3.3in]{Fig.26.eps}
%\vspace{-0.5cm}
%\caption{Block diagrams of the (a) modulator and (b) demodulator in a Walsh-code-based $M$DCSK scheme.}
%\label{fig:Fig.26V}\vspace{-0.2cm}
%\end{figure}
% $\bs_{k, i} = [w_{2k-b'_i,1}  \bc_{k,i}, w_{2k-b'_i,2}  \bc_{k,i}, \ldots, w_{2k-b'_i,2N}  \bc_{k,i}]$
% where $b'_i = (b_i+1)/2$.

In $2015$, the AR$4$JA PLDPC codes were applied to the Walsh-code-based $M$DCSK system in \cite{7093149}, and the interaction between channel coding and modulation was investigated. Furthermore, a turbo-like iterative receiver was constructed, comprising a GML demodulator and a PLDPC decoder, to improve the performance of the BICM framework. Following the above work, a simple computer-search method was developed in \cite{8012533} to obtain a family of {\em improved ARA} (IARA) {\it codes} that possess desirable decoding thresholds and MHDGRs over block-fading channels with code rates ranging from 1$/2$ to $1-\epsilon$. The base matrix of a rate-$(e+1)/(e+2)$ IARA code is obtained as\vspace{-2mm}
\begin{equation}\label{eq:B-IARA}
\bB_{\rm{IARA}} =
\begin{bmatrix}
\begin{array}{cc}
\begin{array}{ccccc}
    1 & 2 & 0 & 0 & 0   \cr
    0 & 2 & 2 & 1 & 1   \cr
    0 & 1 & 1 & 2 & 1   \cr
\end{array}
\overbrace{\begin{array}{ccccc}
0 & 0 & \cdots & 0 & 0 \\
1 & 2 & \cdots & 1 & 2 \\
2 & 1 & \cdots & 2 & 1 \\
\end{array}}^{2e}
\end{array}
\end{bmatrix},
\end{equation}
where the variable node corresponding to the first column is punctured. Both analyses and simulations demonstrated that the IARA codes significantly outperform the AR$4$JA codes in the Walsh-code-based $M$DCSK-aided BICM-ID systems over Nakagami block-fading channels, as shown in Fig.~\ref{fig:Fig.28V}.
%%Fig.27
%\begin{figure}[tbp]
%\center
%\includegraphics[width=2.05in,height=2.0in]{Fig.27.eps}
%\vspace{-0.2cm}
%\caption{Protograph of the $r_{\rm P}=(e+1)/(e+2)$ IARA code.}\vspace{-0.2cm}
%\label{fig:Fig.27V}  %label for entire figure
%\end{figure}

%$\langle{\rm ii}\rangle$ Constellation-based $M$DCSK:
According to \cite{7442517}, the Walsh codes realize high-order $M$DCSK modulations by introducing $M-1$ radio-frequency delay lines and $M$ chaotic fragments (i.e., a length-$M\beta$ DCSK signal), which dramatically decrease the transmission throughput and increase the implementation complexity. To relax these limitations, a constellation-based $M$DCSK modulation scheme was proposed in \cite{7109922}, which can modulate $w$ coded bits into a length-$2 \beta$ DCSK signal for transmission. In other words, the constellation-based $M$-DCSK enables higher efficiency and lower complexity, as compared with the Walsh-coded-based counterpart, and turns out to be a more promising choice for bandwidth-limited wireless communication applications.

In the BICM-ID system with a constellation-based $M$DCSK modulation, a length-$N$ PLDPC codeword ${\bm \Lambda}=(v_1, v_2, \ldots, v_N)$ is first uniformly converted to a length-$N'$ complex-valued symbol $\bg=(g_1, g_2, \ldots, g_{N'})$ based on a given $M$-ary signal constellation, such as the $M$PSK/$M$QAM constellation shown in Fig.~\ref{fig:Fig.3V}, where $N' = N/\log_2 M =N/w$. The real part and imaginary part of the $k$-th $M$-ary symbol $g_k$ are denoted as $g_{{\rm RE},k}$ and $g_{{\rm IM},k}$, respectively, $k=1,2,\ldots, N'$. Based on a length-$\beta$ reference-chaotic fragment $\bc_k$ and its $\beta$-delayed signal $\bc_{{\rm R},k}$, one can construct its orthogonal signal $\bc_{\Gamma,k}$ by performing a Hilbert transform \cite{7442517,7109922}, where $\bc_{\Gamma,k} {\bm \cdot} (\bc_{{\rm R},k})^{\rm T} =0$. The information-bearing fragment can be generated through a linear combination of the quadrature chaotic signals $\bc_{{\rm R},k}$ and $\bc_{\Gamma,k}$ weighted by the fractions $g_{{\rm RE},k}$ and $g_{{\rm IM},k}$, respectively, i.e., $\bd_k = g_{{\rm RE},k} \bc_{{\rm R},k} + g_{{\rm IM},k} \bc_{\Gamma,k}$. As such, the length-$2\beta$ $M$-DCSK signal is obtained as $\bx_k=[\bc_k~\bd_k]$, and then transmitted over the block-fading channel. At the receiver, a modified energy detector is used to retrieve the original symbol $g_k$. Specifically, the two corrupted orthogonal signals $\widetilde{\bc}_{{\rm R},k}$ and $\widetilde{\bc}_{\Gamma,k}$ can be easily attained from the noisy reference-chaotic signal $\widetilde{\bc}_k$. These two signals correlate with the corrupted information-bearing signal $\widetilde{\bd}_k$, thereby generating two energy-based metrics $z_{{\rm R},k}$ and $z_{\Gamma,k}$, respectively, which will be finally utilized to make the decision and yield the retrieved symbol $\widetilde{g}_k$ for $g_k$ \cite{7109922}.
%As a graphical illustration, Fig.~\ref{fig:Fig.29V} gives the block diagrams of the modulator and demodulator for the constellation-based $M$DCSK.
%%fig.26
%\begin{figure}[t]
%\center
%\includegraphics[width=3.2in,height=2.6in]{Fig.29.eps}
%\vspace{-0.2cm}
%\caption{Block diagrams of the (a) modulator and (b) demodulator in a constellation-based $M$DCSK scheme.}
%\label{fig:Fig.29V}\vspace{-2mm}
%\end{figure}

In \cite{7956256}, the capacity-approaching AR$4$JA codes were applied to the constellation-based $M$-DCSK-aided BICM-ID system over block-fading channels. %For the sake of achieving high-reliability and high-efficiency transmissions, a modified PB-EXIT algorithm has been developed for the system, which substantially exploits the characteristics of the DCSK SS modulation and iterative turbo decoder.
According to the analyses and simulations, the proposed PLDPC-$M$DCSK-aided BICM-ID system significantly boosts the anti-noise and anti-multipath-fading capability compared to the BICM-NI counterpart. Furthermore, the impact of several critical parameters, such as the spreading factor and number of global iterations, on the performance of PLDPC-BICM-ID systems with constellation-based $M$DCSK was discussed, aiming to optimize the overall performance. However, this work did not focus on the code design, which deserves further exploration towards practical applications.

%{\em Remark:} The constellation-based $M$DCSK modulation is preferable to the Walsh-code-based $M$DCSK modulation in the PLDPC-BICM systems because the former benefits from higher efficiency and lower complexity.

\vspace{0.5mm}
\noindent{\bf (3) PLDPC-PNC BICM Design:} Inspired by the desirable advantages of PLDPC-BICM, its application was extended to other emerging systems, including TWR communication systems with PNC \cite{8052124}. Compared with the convolutional codes, PLDPC codes are more likely to approach the BICM capacity over PNC-aided TWR channels. For this reason, in \cite{7723910,8884235} the $M$PSK-aided PLDPC-BICM-ID was studied over PNC-aided TWR block-fading channels. This was perhaps the first attempt to apply the AR$3$A PLDPC-BICM-ID scheme to PNC-aided TWR communication systems, which can help boost the transmission throughput and error performance over block-fading channels. Besides, the effect of signal labelings on the performance of PLDPC-BICM-ID was investigated over block-fading channels. Interestingly, as illustrated through analyses and simulations, the Gray-labeled BICM-ID is able to achieve a considerable performance gain over its BICM-NI counterpart over PNC-aided TWR channels, which is in contrast to the observation in the point-to-point scenario. This phenomenon suggests that the Gray labeling has great potential for application in PLDPC-PNC BICM-ID systems over block-fading channels.

\subsubsection{Design of SC-PLDPC-BICM Systems} \label{sect:VI-B-2}

According to \cite{7112076,1564429}, the non-ergodic block-fading channel can be treated as a type of memory channel since its fading gain remains constant during the entire transmission period of each code block. As mentioned in \ref{sect:VI-B-1}, the code rate of a full-diversity root-PLDPC code is $r_{\rm P}=1/L$ over a BICM block-fading channel with $L$ fading blocks. However, how to improve the diversity order of their SC counterparts (i.e., SC-PLDPC codes) over such a channel is still a challenging issue for the coding community. In \cite{6875246}, the performance of SC-PLDPC codes over the BPSK-aided block-fading channel with $L=2$ was carefully evaluated. As a type of convolutional-like codes, SC-PLDPC codes may find their great potential for transmission over block-fading channels. In fact, the inherent memory feature of the SC-PLDPC codes makes them a powerful choice for slowly-varying fading wireless communications. It was demonstrated in \cite{6875246} that the diversity order of a TE-SC-PLDPC code can be improved by increasing the constraint length without a sophisticated design, which is determined by the coupling width and the codeword length of its original PLDPC code (i.e., $ \nu_{\rm SC}= (\varsigma+1)Zn_{\rm P} $). In other words, based on a length-$N$ PLDPC code, the diversity order of its corresponding TE-SC-PLDPC code can be significantly boosted by increasing the coupling width $\varsigma$ over a block-fading channel at the price of a small rate loss. More importantly, unlike the root-PLDPC codes \cite{8740906}, the TE-SC-PLDPC codes are capable of achieving full diversity without the need of any specific design. Although no high-order modulation has been considered, the above feature may be further exploited to boost the diversity order of SC-PLDPC-BICM systems over $M$PSK/$M$QAM-modulated block-fading channels.\vspace{-1mm}
% Due to the appealing advantages, the SC-PLDPC codes may stand out as a robust candidate for BICM systems to resist the distortion in slow-varying-fading wireless transmission scenarios.

\subsection{Summary}\label{sect:VI-C}\vspace{-0.5mm}

In this section, the recent development and contributions in PLDPC-BICM systems over wireless fading channels are reviewed. First, the code and bit-mapper design for such systems over ergodic fast-fading channels are described. Then, some typical research works concerning the design of PLDPC-BICM systems are introduced, under the $M$PSK/$M$QAM and $M$DCSK modulations over non-ergodic block-fading channels. Finally, the application of PLDPC codes to PNC BICM systems over TWR block-fading channels is discussed.

\section{Design of PLDPC-BICM over Poisson PPM Channels}\label{sect:VII}\vspace{-0.2mm}

Poisson channel is considered as a channel model that can suitably describe the statistic characteristics of optical deep-space and free-space links \cite{5439306,6581873}. PPM is a popular alternative for such applications, which not only allows simple bit-to-symbol mapping but also possesses near-optimal channel capacity. To realize power-efficiency transmissions in free-space optical communication systems, it makes sense to combine capacity-approaching codes (e.g., PLDPC codes) with PPM over Poisson channels. However, the PLDPC codes constructed for $M$PSK/$M$QAM-aided BICM systems over AWGN and fading channels do not work well for Poisson PPM channels.

During the past decade, a variety of code-design and performance-analysis methodologies have been developed for PLDPC-BICM systems over Poisson PPM channels, where the receiver includes an inner PPM demodulator and an outer PLDPC decoder \cite{6663748,8873472,6955112,7990043,5629496}. On the contrary, the application of SC-PLDPC codes to the Poisson PPM channels and their design approaches are relatively unexplored. For this reason, in the following only the PLDPC-BICM systems over Poisson PPM channels are discussed.

\subsection{Principles of PPM and Poisson Channels} \label{sect:VII-A}

%In this subsection, we briefly review the PPM-aided transmission mechanism over a Poisson channel.
Distinguished from conventional $M$PSK/$M$QAM modulations, PPM is a type of position-aware modulation scheme, in which each transmitted signal is represented by an $M$-slot PPM symbol. For each PPM symbol, only one single pulse is transmitted in one of the $M$ slots, while the remaining $M-1$ slots keep silent. In other words, an $M$-ary PPM symbol can be defined as a vector $\bx_k = [x_{{\rm P}, k,0}, x_{{\rm P}, k,1},\ldots, x_{{\rm P}, k,M-1}]$ of size $1 \times M$, where $x_{{\rm P}, k,\mu} \in \{0, 1\}, k=1,2,\ldots, N',\mu=0,1,2,\ldots,M-1,N'=N/w=N/\log_2 M,$ and each symbol has only one element equal to $1$ (i.e., one active pulse). In a PLDPC-BICM-ID system with $M$PPM, a length-$N$ PLDPC codeword ${\bm \Lambda}=(v_1, v_2, \ldots, v_N)$ is first converted to a length-$N'$ non-binary codeword ${\bm \Lambda}_{\rm NB}=(v_{\rm NB,1}, v_{\rm NB,2}, \ldots, v_{{\rm NB},N'})$, where $v_{{\rm NB},k} \in \{ 0,1,\ldots, M-1\}$. Subsequently, the non-binary codeword is modulated into an $M$-ary PPM symbol sequence $\bX = (\bx_1,\bx_2,\ldots,\bx_{N'})$, where the $k$-th PPM symbol $\bx_k$ is mapped from the $k$-th non-binary coded symbol $v_{{\rm NB},k}$ in accordance with its value. Specifically, the $\mu$-th element $x_{{\rm P}, k,\mu}$ of $\bx_k$ equals $1$ (i.e., $x_{{\rm P}, k,\mu}=1$) if and only if $v_{{\rm NB},k}=\mu$.

Consider a Poisson PPM channel. The received signal can be described by a $1 \times M$ vector $\by_k \hspace{-0.5mm}=\hspace{-0.5mm} [y_{{\rm P}, k,0}, y_{{\rm P}, k,1},\ldots, y_{{\rm P}, k,M-1}]$, where $ y_{{\rm P}, k,\mu} \hspace{-0.5mm}=\hspace{-0.5mm} \alpha_{{\rm P}, k,\mu} x_{{\rm P}, k,\mu} + n_{{\rm P}, k,\mu}$, $\alpha_{{\rm P}, k,\mu}$ and $n_{{\rm P}, k,\mu}$ are i.i.d. Poisson random variables with means $\lambda_{\alpha}$ and $\lambda_n$, respectively, where $\lambda_{\alpha}$ and $\lambda_n$ are the average power per pulsed slot and of background radiation, respectively. In consequence, the conditioned PDF of the $\mu$-th element in the $k$-th channel output is written as
%\begin{equation}
%f(y_{{\rm P}, k,\mu} | x_{{\rm P}, k,\mu}) \hspace{-0.6mm}=\hspace{-0.6mm}
%\frac{ (\lambda_{\alpha} x_{{\rm P}, k,\mu} \hspace{-0.7mm}+\hspace{-0.6mm}  \lambda_n) ^{y_{{\rm P}, k,\mu}}}  {y_{{\rm P}, k,\mu} !} \hspace{-0.3mm} \exp(-(\lambda_{\alpha} x_{{\rm P}, k,\mu} +  \lambda_n ) ),\nonumber
%\label{eq:PDF-PPM}
%\end{equation}
\begin{equation}
f(y_{{\rm P}, k,\mu} | x_{{\rm P}, k,\mu}) \hspace{-0.6mm}=\hspace{-0.6mm}
\frac{ (\lambda_{\alpha} x_{{\rm P}, k,\mu} \hspace{-0.7mm}+\hspace{-0.6mm}  \lambda_n) ^{y_{{\rm P}, k,\mu}}}  {y_{{\rm P}, k,\mu} !} \hspace{-0.3mm} \exp(-(\lambda_{\alpha} x_{{\rm P}, k,\mu} +  \lambda_n ) ),
\label{eq:PDF-PPM}
\end{equation}where the average SNR per symbol in the $M$PPM Poisson channel is defined as $\lambda_{\alpha}/(M \lambda_n )$.

\subsection{PLDPC-Code Design for PPM-aided BICM Systems} \label{sect:VII-B}

In $2010$, the first attempt to design and analyze the PLDPC codes in $M$PPM-aided BICM-ID systems over Poisson channels was carried out in \cite{5629496}. Specifically, the PEXIT algorithm was generalized to facilitate the optimization of the PLDPC codes in such transmission scenarios. Using the generalized PEXIT algorithm, one can get a rate-$1/2$ optimized PLDPC code, referred to as {\em PPM-PLDPC-A code}, for the $64$PPM-aided BICM-ID system based on certain constraints after a computer search. The protograph corresponding to the PPM-PLDPC-A code includes $88$ non-punctured variable nodes and $44$ check nodes. In detail, there are $66$ degree-$3$ variable nodes, $21$ degree-$2$ variable nodes, and $1$ degree-$24$ variable node among all the $88$ variable nodes. On the other hand, all the $44$ check nodes have a constant degree of $6$. Simulation results showed that at a BER of $10^{-5}$, the PPM-PLDPC-A code achieves a remarkable gain of about $0.8~{\rm dB}$ on the AR$4$A code in the BICM-ID system over a Poisson PPM channel with $M=64$ and $\lambda_n=0.2$. More importantly, the PPM-PLDPC-A code has a gap of only $1~{\rm dB}$ to the channel capacity. Although the PPM-PLDPC-A-coded BICM is slightly inferior to the convolutional-coded serially concatenated PPM, it is amenable to lower decoding latency (i.e., parallel decoding) with better rate compatibility.

As can be observed, the protograph size corresponding to the PPM-PLDPC-A code is a bit large, which leads to relatively complicated representation and implementation. Actually, it is preferred to design a smaller protograph tailored for BICM system over Poisson PPM channels. For this reason, another notable contribution related to PLDPC-code design over Poisson PPM channels was presented in \cite{6663748}, where the drawbacks of the $3 \times 5$ protograph corresponding to a rate-$1/2$ AR$4$A code (see Fig.~\ref{fig:Fig.11V}(a)) were identified. Then, a $2 \times 4$ protograph was optimized within the context of $64$PPM, in which the maximum number of edges connecting a variable node to a check node is limited to $3$. After a simple search, an optimized PLDPC code (referred to as {\em PPM-PLDPC-B code}) that exhibits the lowest decoding threshold can be obtained, with a base matrix %and structure are respectively presented in \eqref{eq:PPM-PLDPC-B} and Fig.~\ref{fig:Fig.30V}.
\begin{eqnarray}
{\bB}_{\rm PPM\textit{-}B} &=\left[\begin{array}{cccc}
1 & 1 & 0 & 1 \cr
1 & 2 & 2 & 2 \cr
\end{array}\right].
\label{eq:PPM-PLDPC-B}
\end{eqnarray}
%%Fig.30
%\begin{figure}[tbp]
%\centering
%\includegraphics[width=1.4in,height=0.56in]{{Fig.30.eps}}
%\vspace{-2mm}
%\caption{Protograph of the rate-$1/2$ PPM-PLDPC-B code for $64$PPM-aided BICM-ID system over a Poisson channel.}
%\label{fig:Fig.30V}  %label for entire figure
%\end{figure}

{\bf \em Example 21:} Considering a BICM-ID system over a Poisson PPM channel with $M=64$ and $\lambda_n=0.2$, one can compare the decoding thresholds and MHDGRs of the rate-$1/2$ PPM-PLDPC-B code and AR$4$A code by using the PEXIT and AWE analyses. It has been shown that the PPM-PLDPC-B code has a decoding threshold of $-14.98~{\rm dB}$, which is much smaller than that of the AR$4$A code (i.e., $-14.13~{\rm dB}$). Furthermore, the decoding threshold of the former is only about $0.6~{\rm dB}$ away to the channel capacity. However, both codes do not have effective MHDGRs because they have too many (i.e., more than $40\%$) degree-$2$ variable nodes, which implies that they may suffer from error-floor behavior in the high-SNR region \cite{7112076}.

\begin{figure}[tbp]
\centering
\includegraphics[width=2.82in,height=2.1in]{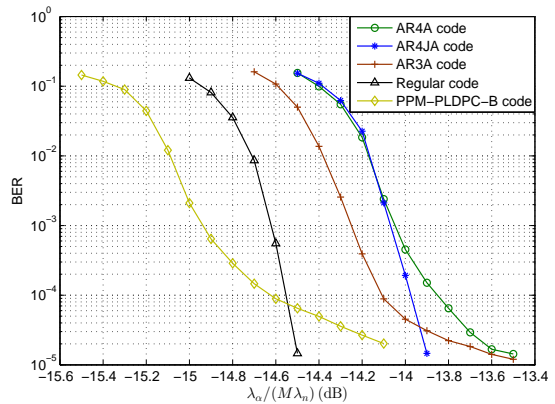}
\vspace{-0.25cm}
\caption{BER curves of the rate-$1/2$ PPM-PLDPC-B code, AR$4$A code, AR$4$JA code, AR$3$A code, and regular-$(3, 6)$ PLDPC code in the BICM-ID system over a Poisson PPM channel with $N_{\rm T}=8160, M=64, \lambda_n=0.2, t_{\rm GL,max}=8, ~{\rm and}~t_{\rm BP,max}=25$.}
\label{fig:Fig.31V}  %label for entire figure
\vspace{-3mm}
\end{figure}

To have more insight, the BER performances of the rate-$1/2$ PPM-PLDPC-B code and AR$4$A code in the BICM-ID system over a Poisson PPM channel are compared in Fig.~\ref{fig:Fig.31V}. %, where the transmitted codeword length and modulation order are set to be $N_{\rm T}=8160$ and $M=64$.
For reference, three classic rate-$1/2$ PLDPC codes, i.e., AR$4$JA code, AR$3$A code, and regular-$(3, 6)$ PLDPC code, are included in the same figure. It can be clearly seen that the PPM-PLDPC-B code significantly outperforms the other four types of PLDPC codes in the low-SNR region with respect to the capacity-approaching decoding threshold. Also, the PPM-PLDPC-B code, AR$4$A code, and AR$3$A code encounter severe error-floor issues in the high-SNR region. Particularly, the PPM-PLDPC-B code becomes inferior to the regular PLDPC code once the SNR exceeds $-14.55~{\rm dB}$. Therefore, it is desirable to design more robust PLDPC codes that have both excellent decoding thresholds and MHDGRs in the context of $M$PPM, to meet the ultra-high-reliability requirement of optical communication applications.

%Fig.31

More recently, a novel design method was proposed in \cite{7990043} to construct capacity-approaching punctured non-binary PLDPC codes for Poisson PPM channels with the usage of a surrogate erasure channel model. It was shown that the non-binary PLDPC-CM schemes outperform the binary PLDPC- and convolutional- BICM schemes. % with different modulation orders (i.e., different values of $M$).
Although the non-binary PLDPC codes can combine with $M$PPM in a more seamless fashion compared with the binary PLDPC codes, they have much higher implementation complexity. Thus, the binary PLDPC-BICM is usually more preferable for practical Poisson PPM applications with respect to the non-binary PLDPC-CM.\vspace{-1.5mm}
%thus are usually not preferred and suitable in actual implementation.

\subsection{Summary}\label{sect:VII-C}\vspace{-0.5mm}

In this section, the majority of contributions in the PLDPC-BICM design over Poisson PPM channels were reviewed. In particular, the structures of two types of optimized PLDPC codes, i.e., PPM-PLDPC-A and PPM-PLDPC-B codes, are described, which exhibit excellent decoding thresholds in the BICM-ID frameworks over Poisson PPM channels. However, both types of codes do not enjoy the benefit of the linear-minimum-distance-growth rates, and hence may suffer from severe error floors in the high-SNR region.

\section{Design of PLDPC-BICM over NAND Flash-Memory Channels}\label{sect:VIII}

As an excellent non-volatile memory technique, NAND flash memory indicates a revolutionized paradigm shift in the evolution of modern data-storage devices. With respect to conventional hard disk drives (HDDs), flash-memory-based solid state drives (SSDs) benefit from faster access speed, less power consumption, and higher storage capacity, and thus have become a mainstream data-storage scheme \cite{9000906,7920318}. Nowadays, flash-memory-based storage techniques have been ubiquitously used in most electronic and communication systems, including smart devices and enterprise data centers. To increase the data-storage density, the conventional single-level-cell (SLC) technique has been replaced by MLC, TLC, and QLC techniques, for which the storage density gradually increases from $1$\,bit/cell to $4$\,bit/cell \cite{8715674,8528505}. However, due to the continual increase in storage density, the flash-memory systems suffer from severer noise and interference, which give rise to more errors during the read/write procedure. %Thereby, the conventional algebraic codes are no longer able to offer satisfactory error performance \cite{9000906}.
As compared with the conventional algebraic codes \cite{9000906}, PLDPC codes can significantly enhance the reliability of such high-density flash-memory devices, which make capacity-approaching data storage attainable \cite{8278036,6804933,8277940,8047952}.

The voltage levels stored in the cell of an MLC/TLC/QLC flash memory can be regarded as a $4$PAM/$8$PAM/$16$PAM signal \cite{6804933}. In this sense, a multiple-level-cell flash memory deploying the PLDPC code can be regarded as a BICM system. % that has been widely used in a variety of wireless-communication scenarios.
Yet, due to the impact of various circuit-level noise, the conventional PLDPC codes designed for communication systems may not perform well in such environments. In addition, high code rate (i.e., $r \ge 4/5$) is required in order to meet the high-storage-efficiency demand of flash-memory devices.
%Based on the above features, the design and analysis methodologies of PLDPC-coded BICM must be re-investigated so as to improve the overall performance of flash-memory systems.
During the past few years, some research effort has been made to formulate robust design of high-rate PLDPC codes and bit-mapping schemes for MLC flash memory \cite{8509142}. However, relevant investigation in this field is still in its infancy, where many challenging problems have to be resolved. Here, several design paradigms are introduced for the PLDPC-BICM over MLC flash-memory channels in order to take advantage of the potential benefits of these frameworks. Within this context, the proposed designs may also be applicable to TLC and QLC scenarios with appropriate modifications.

In general, a flash-memory channel model can be modeled as  %\vspace{-1mm}%is quite different from the communication channels mentioned in the preceding sections, which
%\begin{equation}
$v_{\rm{th}} = v_{\rm w} + n_{\rm r} + n_{\rm c} + n_{\rm w} + n_{\rm u}  + n_{\rm p}$,
%\label{eq:noise}
%\end{equation}
where $v_{\rm{th}}$ is the threshold voltage representing multiple bits (e.g., $2$ bits) in a memory cell, $v_{\rm w}$ is the write-voltage level, $n_{\rm r}$ is the data retention noise; $n_{\rm c}$ is the cell-to-cell interference; $n_{\rm w}$ is the random telegraph noise; $n_{\rm u}$ is the incremental step pulse programming noise, and $n_{\rm p}$ is the programming noise \cite{9000906,7959126}.
To retrieve the stored data in flash memory, read voltages need to be applied to quantize the threshold voltages into serval voltage regions \cite{8805334,8648493}. In addition, the cell-to-cell interference can be mitigated by utilizing post-compensation techniques \cite{7959126,7416649}. As such, the flash-memory channel can be regarded as a {\em discrete memoryless channel} when establishing coding criteria \cite{6804933,8708250,8278036}.

\subsection{System Design for Regular Mapping} \label{sect:VIII-A}

The first works on specific design methodologies of unpunctured and punctured PLDPC-BICM systems over flash-memory channels were presented in \cite{8314735} and \cite{8708250}, respectively. Then, a novel combinatorial design approach, which consists of an optimal overlap partitioning and circulant power optimizer, was presented in \cite{8277940} to construct superb non-binary SC-PLDPC codes over flash-memory channels. However, this work did not consider the design of binary SC-PLDPC codes and their corresponding BICM schemes, which remains an unsolved issue today. In the following, the recent progress of PLDPC-BICM systems over flash-memory channels is briefly reviewed.%rather than SC-PLDPC-coded BICM systems over flash-memory channels.

\subsubsection{Design of Unpunctured PLDPC Codes} \label{sect:VIII-A-1}

In \cite{8314735}, a design and analysis of PLDPC-BICM-NI over MLC flash-memory channels was presented. As a representative contribution to this area, it first analyzed the raw BERs and MIs of the most significant bit (MSB) and least significant bit (LSB), which are located on different pages of the flash memory.\footnote{In an MLC flash memory, there are four possible symbols (i.e., $00, 01, 10, 11$) stored in each cell. Each symbol consists of two bits, where the left-most bit and the right-most bit are called {\em MSB and LSB}, respectively \cite{8013174}. The MSB and LSB within each stored symbol correspond to the high protection-degree labeling bit and the low protection-degree labeling bit within a $4$PAM symbol, respectively.} It was found that the MSB and LSB pages typically exhibit unbalanced raw BERs, which may be very useful for improving the decoding performance and convergence speed of the PLDPC-MLC flash-memory systems. More precisely, the MSB page has relatively better raw BER and MI than the LSB page. Accordingly, a new bit-mapping scheme, referred to as {\em variable-node-degree-based mapping} (VNDM) {\it scheme}, was proposed for the PLDPC-coded MLC flash-memory systems. In the VNDM scheme, the $n_{\rm P}$ variable nodes in a protograph are first re-ranked in a descending order by their degrees. Subsequently, the $n_{\rm P}/2$ higher-degree and $n_{\rm P}/2$ lower-degree variable nodes are respectively mapped to the MSB and LSB pages in order to accelerate the decoding convergence of the MLC flash-memory system. Simulation results showed that the VNDM enables a desirable performance gain over the conventional successive and random bit-mapping scheme. Note that the VNDM can be viewed as a generalization of the VDMM \cite{6133952} to the flash-memory channel, which takes into account the UEP feature between the MSB and LSB pages.

Assume a rate-$1/2$ unpunctured PLDPC code corresponding to a $3 \times 6$ base matrix, where the degrees of the first three variable nodes must be higher than those of the last three variable nodes, and the maximum value of each element is limited to $3$. Aiming to minimize the decoding threshold without losing the linear-minimum-distance-growth property under the VNDM scheme over MLC flash-memory channels, one can easily obtain an optimized PLDPC code, as shown in \cite[eq.\,(5)]{8314735}.%\vspace{-2mm}
%\begin{equation}\label{eq:B-VNDM-Ini}
%\bB_{\rm VNDM,\frac{1}{2}} =
%\begin{bmatrix}
%\begin{array}{ccc}
%\underbrace{\begin{array}{ccc}
%1 & 1 & 1 \\
%2 & 1 & 3 \\
%5 & 3 & 1  \\
%\end{array}}_{\rm MSB}
%\begin{array}{c}
%| \\
%| \\
%| \\
%\end{array}
%\underbrace{\begin{array}{ccc}
%0 & 0 & 0 \\
%2 & 2 & 1  \\
%1 & 1 & 1  \\
%\end{array}}_{\rm LSB}
%\end{array}
%\end{bmatrix},%\vspace{-0.5mm}
%\end{equation}

However, the VNDM scheme may not achieve the best BICM performance over MLC flash-memory channels. To find an optimal bit-mapping scheme for the resultant rate-$\frac{1}{2}$ PLDPC code in order to attain the lowest decoding threshold, one can estimate the decoding thresholds of all the possible permutations for the column patterns mapped to the MSB and LSB pages. Then, the optimized VNDM scheme for the PLDPC code in \cite[eq.\,(5)]{8314735} can be obtained, i.e.,\vspace{-2mm}
\begin{equation}\label{eq:B-VNDM-Opt}
\bB_{\rm OPT\text{-}VNDM,\frac{1}{2}} =
\begin{bmatrix}
\begin{array}{ccc}
\underbrace{\begin{array}{ccc}
1 & 0 & 1 \\
2 & 2 & 3 \\
5 & 1 & 1  \\
\end{array}}_{\rm MSB}
\begin{array}{c}
| \\
| \\
| \\
\end{array}
\underbrace{\begin{array}{ccc}
1 & 0 & 0 \\
1 & 2 & 1  \\
3 & 1 & 1  \\
\end{array}}_{\rm LSB}
\end{array}
\end{bmatrix},
\end{equation}
where the first three variable nodes and the last three variable nodes are mapped to the MSB and LSB pages, respectively.

According to \cite{8013174,8007178,8278036,9000906}, high-rate codes are always required in practical flash-memory applications. To meet this requirement, the rate-$\frac{1}{2}$ PLDPC code in \eqref{eq:B-VNDM-Opt} was extended to a family of optimized RC-PLDPC codes in \cite{8314735}, called {\em VNDM-PLDPC codes}, by repeatedly appending the same variable-node pattern to both the MSB and LSB pages. Consequently, the base matrix of a rate-$(6e+3)/(6e+6)$ VNDM-PLDPC code is obtained, i.e.,
\vspace{-2mm}
\begin{equation}\label{eq:B-VNDM-Opt-RC}
\hspace{-2mm}\bB_{\rm OPT\text{-}VNDM} \hspace{-0.7mm}=\hspace{-0.7mm}
\begin{bmatrix}
\begin{array}{ccc}
\underbrace{\begin{array}{ccccc}
1 & 0 & 1 & \overbrace{3 \ 2 \ 0 \ \cdots}^{3e} \\
2 & 2 & 3 & 1 \ 2 \ 1 \ \cdots \\
5 & 1 & 1 & 0 \ 1 \ 3 \ \cdots \\
\end{array}}_{\rm MSB}
\begin{array}{c}
~ \\
| \\
| \\
| \\
\end{array}
\underbrace{\begin{array}{cccc}
1 & 0 & 0 & \overbrace{3 \ 2 \ 0 \ \cdots}^{3e} \\
1 & 2 & 1 & 1 \ 2 \ 1 \ \cdots  \\
3 & 1 & 1 & 0 \ 1 \ 3 \ \cdots  \\
\end{array}}_{\rm LSB}
\end{array}
\end{bmatrix}\hspace{-1mm}.
\end{equation}
where the newly added variable-node pattern for the MSB/LSB page consists of three variable nodes with degrees $4,5,4,$ respectively. The VNDM-PLDPC codes cover various rates ranging from $1/2$ to $1-\epsilon$. For instance, one can construct a rate-$9/10$ VNDM-PLDPC code by setting $e=4$ in \eqref{eq:B-VNDM-Opt-RC}. The structure of such a variable-node pattern allows the resultant higher-rate VNDM-PLDPC codes to preserve low-decoding-threshold and linear-minimum-distance-growth benefits. The VNDM-PLDPC codes outperform the irregular LDPC codes in \cite{6804933}, optimized for the BICM-NI MLC flash-memory systems.

\subsubsection{Design of Punctured PLDPC Codes} \label{sect:VIII-A-2}

The code-design method developed in \cite{8314735} has only considered unpunctured PLDPC codes and the BICM-NI framework. In fact, incorporating some punctured variable nodes into the PLDPC codes and the ID framework into BICM can improve the transmission throughput and error performance in communication applications. This motivated the employment of the punctured PLDPC codes and BICM-ID in the flash-memory systems \cite{8278036}. However, due to the impact of the ID framework, the existing PLDPC codes designed for BICM-NI flash-memory systems may not preserve their superiority. For this reason, in \cite{8708250} a comprehensive study was proposed on the design of punctured PLDPC-codes for BICM-ID flash-memory systems with the anti-Gray labeling.

As is well known, the PEXIT algorithm can be used to predict the asymptotic convergence performance of PLDPC codes in terms of decoding thresholds. However, there exist some problems that make the standard PEXIT algorithm unsuitable for BICM-ID flash-memory channels, as follows.
\begin{itemize}
\item Different from communication channels, the flash memory needs to apply the read voltages that are selected for different PE cycles and retention time to dynamically adapt to the variation of the threshold voltages.
\item Due to the effect of quantization, the initial LLRs output from a flash-memory channel no longer follow a symmetric-Gaussian distribution.
\end{itemize}

To address the aforementioned issues, a voltage-sensing (VS) PEXIT algorithm was proposed for PLDPC-BICM-ID over flash-memory channels. Unlike the conventional PEXIT algorithm, the VS-PEXIT substantially takes into account the variation of the threshold voltages and the feature of the ID framework. Particularly, to deal with the variation of threshold voltages, a maximum MI (MMI) technique \cite{6804933} is utilized to obtain the read-voltage levels for different numbers of PE cycles in the VS-PEXIT algorithm. Furthermore, the VS-PEXIT algorithm applies the realistic LLRs to generate the initial MI of flash-memory channel based on Monte-Carlo simulations.
%Besides, considering the iterations between the inner detector and outer decoder, the VS-PB-EXIT algorithm traces the MI evolution not only between variable nodes and check nodes, but between the detector and decoder as well.
%With the above modification, the VS-PB-EXIT algorithm can be effectively used to evaluate the decoding thresholds of PLDPC-coded BICM-ID over flash-memory channels.

Based on the VS-PEXIT analysis, it was found \cite{8708250} that the conventional optimal PLDPC codes over AWGN channels cannot maintain excellent error performance over BICM-ID MLC flash-memory channels. Therefore, a new design method was proposed to construct high-rate PLDPC codes having low decoding thresholds and effective MHDGRs over such channels.

The design starts with a rate-$\frac{1}{2}$ PLDPC code. To obtain a good rate-$\frac{1}{2}$ protograph with the properties of a relatively low decoding threshold and an effective MHDGR, one can first impose some constraints on the protograph:
(1) incorporate a degree-$1$ variable node, a punctured highest-degree variable node, and a degree-$2$ variable node, into the rate-$\frac{1}{2}$ protograph;
(2) set the maximum number of degree-$2$ variable nodes to be less than the number of check nodes outside the precoding structure \cite{7112076}.
Through a VS-PEXIT-aided computer search, the base matrix corresponding to the optimized rate-$\frac{1}{2}$ PLDPC code with the lowest decoding threshold can be obtained, as\vspace{-1mm}
\begin{equation}\label{eq:OARA-0.5}
\bB_{\rm OPT\text{-}ID,\frac{1}{2}} =
\begin{bmatrix}
\begin{array}{ccccc}
    1 &\ 2 &\ 0 &\ 0 &\ 0   \\
    0 &\ 1 &\ 1 &\ 3 &\ 1   \\
    0 &\ 3 &\ 2 &\ 0 &\ 1   \\
\end{array}
\end{bmatrix},
\end{equation}
where the variable node corresponding to the second column is punctured. Furthermore, it was observed that repeatedly adding a degree-$3$ variable-node pattern (i.e., two degree-$3$ variable nodes) to the designed rate-$\frac{1}{2}$ protograph can yield higher-rate protographs with relatively low decoding thresholds.
\begin{figure}[t]
\center
\vspace{-1mm}
\includegraphics[width=2.8in,height=2.15in]{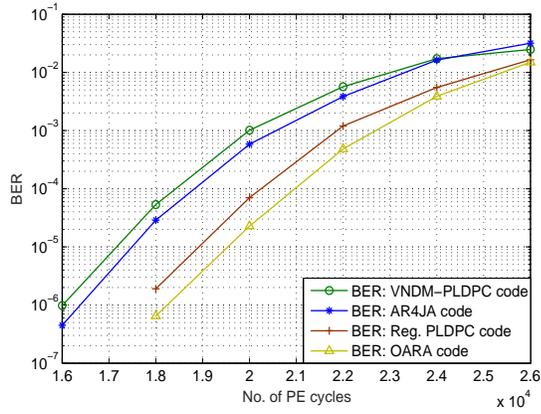}
\vspace{-0.2cm}
\caption{BER curves of four different PLDPC codes over a BICM-ID flash-memory channel with $r_{\rm P}=9/10, N_{\rm T}=4200, t_{\rm GL,max}=5,~{\rm and}~t_{\rm BP,max}=25$.}
\label{fig:Fig.34V}\vspace{-0.2cm}
\end{figure}

More importantly, appending VNs with degrees no less than $3$ to the rate-$\frac{1}{2}$ PLDPC code can ensure the linear-minimum-distance-growth property. Therefore, an optimized rate-$\frac{e+1}{e+2}$ PLDPC code, called {\em optimized ARA} (OARA) {\it code} \cite{8708250}, is constructed, with the base matrix formulated as
\vspace{-1.5mm}
\begin{equation}\label{eq:OARA-RC}
\bB_{\rm{OARA}} =
\begin{bmatrix}
\begin{array}{cc}
\begin{array}{ccccc}
    1 &\ 2 &\ 0 &\ 0 &\ 0   \\
    0 &\ 1 &\ 1 &\ 3 &\ 1   \\
    0 &\ 3 &\ 2 &\ 0 &\ 1   \\
\end{array}
\overbrace{\begin{array}{ccccc}
0 &\ 0 &\ \cdots &\ 0 &\ 0 \\
1 &\ 2 &\ \cdots &\ 1 &\ 2 \\
2 &\ 1 &\ \cdots &\ 2 &\ 1 \\
\end{array}}^{2e}
\end{array}
\end{bmatrix}.\vspace{-1mm}
\end{equation}
By means of AWE analysis, it can be verified that the OARA code has an effective MHDGR. Assuming that $e=9$, the decoding threshold of the rate-$\frac{9}{10}$ OARA code is obtained as $12.956$ dB, which is lower than that of the AR$4$JA code, regular column-weight-$3$ PLDPC code and the optimized VNDM-PLDPC code \cite{8314735}, and moreover has a gap of only about $0.6$ dB to the capacity limit of the BICM-ID flash-memory channel. Besides, Fig.~\ref{fig:Fig.34V} plots the BERs of four different rate-$\frac{9}{10}$ PLDPC codes over a BICM-ID flash-memory channel.
When the number of PE cycles is 18000, the OARA code achieves a BER of $7 \times 10^{-7}$, while the regular column-weight-3 PLDPC code, AR4JA code and optimized VNDM-PLDPC code only achieve BERs of $2 \times 10^{-6}$, $3 \times 10^{-5}$ and $5 \times 10^{-5}$, respectively.
This verifies that the OARA code outperforms the other three types of PLDPC codes.

%Additionally, Fig.~\ref{fig:Fig.35V} reveals that the OARA code obtains a more remarkable gain than the optimized VNDM-PLDPC code in the BICM-ID scenario with respect to that in the BICM-NI scenario.

%\begin{figure}[!tbp]
%\center
%%\includegraphics[width=2.5in,height=1.9in]{Fig.5.eps}
%\includegraphics[width=2.9in,height=2.3in]{Fig.35.eps}
%\vspace{-0.2cm}
%\caption{BER curves of the OARA code and VNDM-PLDPC code over BICM-NI and BICM-ID MLC flash-memory channels. The parameters used are $r_{\rm P}=9/10, N_{\rm T}=4200, t_{\rm GL,max}=5,~{\rm and}~t_{\rm BP,max}=25$.}
%\label{fig:Fig.35V}\vspace{-0.2cm}
%\end{figure}

\subsection{System Design for Irregular Mapping} \label{sect:VIII-B}

According to \cite{1112223,5960810,7589683}, IM is a useful technique to speed up the decoding convergence for BICM systems. However, the existing works have only considered the design of IM schemes over AWGN and fading channels, but not the flash-memory channel. To fill this gap, the IM technique was applied to the BICM-ID flash-memory systems in \cite{6777401}, with a specific code-design method developed for this framework.

\begin{figure}[!tbp]
\centering
%\vspace{-0.2cm}
\includegraphics[width=3.05in,height=1.68in]{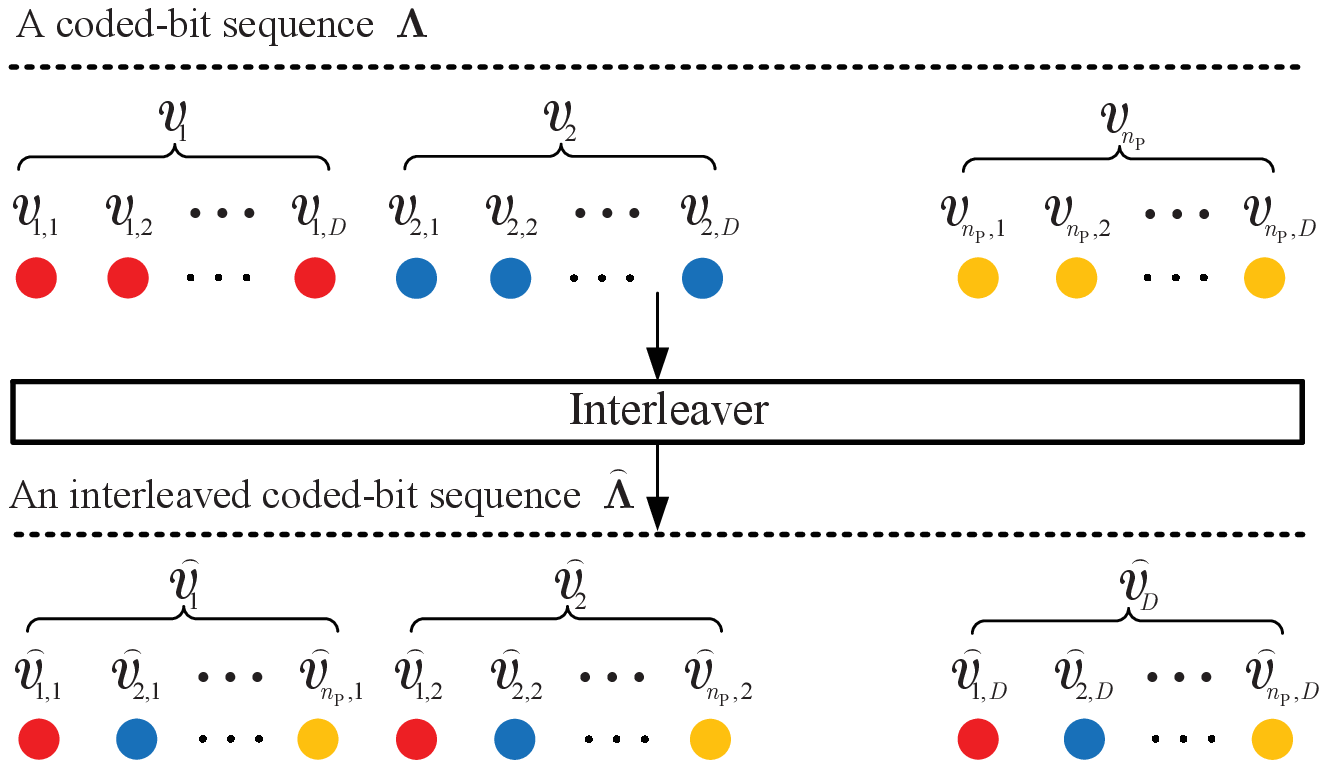}
\vspace{-0.2cm}
\caption{Bit-interleaving process for an IM-PLDPC-BICM-ID over a flash memory channel.}
\label{fig:Fig.36V}
\vspace{-0.3cm}
\end{figure}

Consider a PLDPC code corresponding to a protograph with $n_{\rm P}$ variable nodes and $m_{\rm P}$ check nodes. A length-$N_{\rm P}$ PLDPC codeword ${\bm\Lambda}=(v_1, v_2, \ldots, v_{N})$ is grouped into $n_{\rm P}$ blocks, i.e., ${\cal V}_1, {\cal V}_2, \ldots, {\cal V}_{n_{\rm P}}$, where ${\cal V}_j$ consists of $Z=N/n_{\rm P}$ replicas of $v_j$. Furthermore, each block ${\cal V}_j$ is divided into $D$ sub-blocks $\{ {\cal V}_{j,\kappa}\,|\,\kappa=1,2,\ldots,D \}$ if $D~(D \ge 2)$ different types of constellations are employed. The length of ${\cal V}_{j,\kappa}$ equals $\eta_\kappa Z$, where $\eta_\kappa \in(0,1)$ is the mixing ratio of the $\kappa$-th constellation and $\sum\nolimits_{\kappa=1}^D \eta_\kappa=1$. After processing by the interleaver, the interleaved sub-block $\widehat{\cal V}_\kappa$ is generated, as illustrated in Fig.~\ref{fig:Fig.36V}. Consequently, the written symbol sequence $\mathbf{x}_\kappa$ is produced by the $\kappa$-th constellation to modulate $\widehat{\cal V}_\kappa$.

{\bf {\em Example 22:}} Assume that two constellations (i.e., Gray-labeled constellation and anti-Gray-labeled constellation, $D=2$) with a mixing-ratio vector ${\bm \eta} = (\eta_1, \eta_2)=(1/2, 1/2)$ are applied in an IM-PLDPC-BICM-ID MLC flash-memory system. In this system, each block ${\cal V}$ is divided into $2$ sub-blocks. After processing by the interleaver, two interleaved coded-bit sub-sequences ($\widehat{\cal V}_1$ and $\widehat{\cal V}_2$) are produced. Then, the written symbol sequences ($\mathbf{x}_1$ and $\mathbf{x}_2$) are generated by the Gray- and anti-Gray-labeled constellations to modulate $\widehat{\cal V}_1$ and $\widehat{\cal V}_2$, respectively.

To satisfy the high-rate demand for flash-memory applications, high-rate PLDPC codes are constructed with two constellations (i.e., Gray-labeled constellation and anti-Gray-labeled constellation) in the IM-BICM-ID flash-memory systems in \cite{6777401}. Here, the mixing-ratio vector is assumed as ${\bm \eta} = (\eta_1, \eta_2)=(1/2, 1/2)$ without loss of generality. Note that the proposed construction approach can also be applied to other mixing ratios.

Next, assume a rate-$\frac{9}{10}$ PLDPC code corresponding to a $3 \times 21$ base matrix, which includes $1$ punctured variable node and $20$ unpunctured variable nodes. Then, some constraints are imposed on the base matrix so as to obtain a desirable decoding threshold and an effective MHDGR for IM-BICM-ID flash-memory systems. The procedure is as follows.
\begin{enumerate}[(1)]
\item Initialize a protograph with a precoding structure and a highest-degree punctured variable node.
\item Enable the linear-minimum-distance-growth property, where only one degree-$2$ variable node is allowed outside the precoding structure of the protograph.
\item Ensure the linear-minimum-distance-growth property by employing an extension variable-node pattern with two degree-$3$ variable nodes in the protograph.% are employed to constitute an extension variable-node pattern in the rate-$9/10$ protograph.
\end{enumerate}
%Here, to preserve linear-minimum-distance-growth property and low encoding complexity, two degree-$3$ variable nodes are employed to constitute an extension variable-node pattern in the rate-$9/10$ protograph.
%, and then the higher-rate protograph codes can be constructed via repeatedly adding a VN extension pattern to the resultant base matrix

Based on the above constraints, the optimized rate-$\frac{9}{10}$ PLDPC code, referred to as {\em irregular-mapped accumulate-repeat-accumulate} (IMARA) {\it code}, is constructed by a PEXIT-adided computer search.
%, whose corresponding base matrix is
According to \cite{5174517,7112076}, adding degree-$3$ VNs into a rate-$\frac{9}{10}$ protograph can retain the lowest complexity for its corresponding higher-rate counterparts without deteriorating the linear-minimum-distance-growth property. Hence, one can easily construct a higher-rate IMARA code by further appending the degree-$3$ variable-node pattern to the base matrix of rate-$\frac{9}{10}$ IMARA code. %can be easily extended to .%to the resultant base matrix.
The base matrix of a rate-$r_{\rm P}=\frac{e+1}{e+2}~(e \ge 8)$ IMARA code is given by
\vspace{-1.5mm}
\begin{equation}\label{eq:13}
\bB_{\rm{IMARA}} =
\begin{bmatrix}
\begin{array}{cc}
\begin{array}{ccccc}
1 &\ 2 &\ 0 &\ 1 &\ 0 \   \\
0 &\ 3 &\ 1 &\ 0 &\ 1 \   \\
0 &\ 1 &\ 2 &\ 3 &\ 1 \   \\
\end{array}
\overbrace{\begin{array}{ccccc}
0 &\ 0 &\ \cdots &\ 0 &\ 0 \\
1 &\ 2 &\ \cdots &\ 1 &\ 2 \\
2 &\ 1 &\ \cdots &\ 2 &\ 1 \\
\end{array}}^{2e}
\end{array}
\end{bmatrix},
\end{equation}
where the variable node corresponding to the second column is punctured.

Now, consider five different rate-$\frac{9}{10}$ PLDPC codes, i.e., IMARA code, AR$4$JA code, VNDM-PLDPC code, OARA code and regular column-weight-$3$ PLDPC code. The IMARA code not only possesses the lowest decoding threshold among the five codes (i.e., $11.483$~dB), but also has a $0.182$-dB gap to the capacity limit (i.e., $11.301$~dB) of an IM-BICM-ID flash-memory channel. Also, it is verified by AWE analysis that the IMARA code enables the linear-minimum-distance-growth property.
\begin{figure}[t]
\vspace{-2mm}
\centering
%\subfigure[\hspace{-0.7cm}]{ %% label for first subfigure
\includegraphics[width=2.78in,height=2.15in]{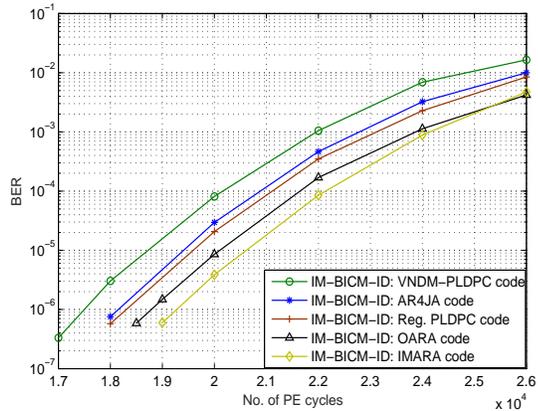}
%\subfigure[\hspace{-0.7cm}]{ %% label forsecond subfigure
%\includegraphics[width=2.8in,height=2.3in]{Fig.37b.eps}}
\vspace{-0.2cm}
\caption{BER curves of five different PLDPC codes in the IM-BICM-ID flash-memory system with $r_{\rm P}=9/10, N_{\rm T}=4000, t_{\rm GL,max}=6,~{\rm and}~t_{\rm BP,max}=40$.}
\label{fig:Fig.37V}  %label for entire figure
\vspace{-2mm}
\end{figure}

{\bf {\em Example 23:}} Fig.~\ref{fig:Fig.37V} compares the BER performance of the rate-$\frac{9}{10}$ IMARA code, AR$4$JA code, VNDM--PLDPC code, OARA code and regular PLDPC code in an IM-BICM-ID flash-memory system.
At a BER of $10^{-6}$, the IMARA code achieves performance gains of about $500$, $1000$, $1100$ and $1800$ PE cycles over the OARA code, regular PLDPC code, AR4JA code and VNDM-PLDPC code, respectively.
This demonstrates the excellent error performance of the IMARA code. Furthermore, it should be noted that the IM-IMARA-BICM-ID flash-memory system is obviously superior to its regular-mapped counterpart (Simulations are omitted here for simplicity).\vspace{-1mm}

%Also, Fig.~\ref{fig:Fig.38V} shows the average number of BP iterations (i.e., $t_{\rm total\text{-}BP,max}=240$) required to decode each codeword for the IM-BICM-ID schemes and regular-mapped BICM-ID schemes with the five different PLDPC codes versus the number of PE cycles. The results illustrate that the decoding latency (i.e., the average number of total BP iterations) of the IM-BICM-ID systems is much smaller than those of the regular-mapped BICM-ID systems for a fixed number of PE cycles. More importantly, the IMARA code requires the lowest number of BP iterations among all the PLDPC codes in the IM-BICM-ID scenario.

%\begin{figure}[!t]
%\centering
%%\includegraphics[width=1\linewidth]
%\includegraphics[width=2.8in,height=2.3in]{Fig.38.eps}
%\vspace{-0.2cm}
%\caption{Average number of BP iterations required to decode each codeword for the IM-BICM-ID schemes and regular-mapped BICM-ID schemes with five different PLDPC codes over an MLC flash-memory channel. The parameters used are $r_{\rm P}=9/10, N_{\rm T}=4000, t_{\rm GL,max}=6,~{\rm and}~t_{\rm BP,max}=40$.}
%\label{fig:Fig.38V}
%\vspace{-0.2cm}
%\end{figure}

\subsection{Summary}\label{sect:VIII-C}\vspace{-1mm}

In this section, the optimization methodologies for PLDPC-BICM flash-memory systems is reviewed.
%Since a multiple-level-cell flash-memory channel can be treated as an $M$PAM-aided DMC, the BICM technique appears to be a feasible solution to enhancing the storage reliability of such systems.
In particular, some design guidelines of PLDPC codes and bit mappers are presented for regular-mapped BICM flash-memory systems to meet the high-reliability and high-efficiency requirements. As a further advancement, a paradigmatic devise on the IM-PLDPC-BICM-ID flash-memory systems is introduced, which can achieve better error performance and lower decoding latency.
%It should be noted that the study of SC-PLDPC-coded BICM systems over Poisson PPM channels and flash-memory channels are relatively unexplored to date.

\section{Future Research Directions}\label{sect:IX}
%This paper has offered an overview of the recent research achievements of LDPC-code designs for MR systems.

In the above sections, a thorough survey is provided for the design of PLDPC-BICM over several typical channel models. The PLDPC-BICM has attracted considerable attention in recent years because it exhibits excellent performance with a simple design and easy implementation. Despite the fruitful results in the vast volume of literature on PLDPC-BICM, a number of challenging issues still require to be addressed. Here, some potential research directions are suggested for further investigation.\vspace{-1mm}

\subsection{Further Optimization of SC-PLDPC-BICM}

Most existing SC-PLDPC codes reported in the literature suffer from relatively low rates (i.e., $r_{\rm SC} \le \frac{1}{2}$), which to some extent limit their applications in digital communications and data-storage systems \cite{7460483,8883091,9057491}. Actually, high-rate codes are always preferred over low-rate codes, especially for data-storage systems. To satisfy the high-throughput requirement in practical applications, developing robust design approaches to construct higher-rate SC-PLDPC codes (e.g., couple a series of  higher-rate PLDPC codes) for BICM systems becomes an urgent task for the coding community.

Additionally, the investigation of SC-PLDPC-BICM systems has been restricted to the AWGN and fading channels. Similarly to conventional SC-LDPC codes \cite{9056068,8277940,9032092}, the SC-PLDPC codes will find more applications in optical communication systems, flash-memory systems, and underwater acoustic communication systems. Accordingly, it is expected to develop efficient SC-PLDPC-BICM frameworks enjoying the wave-like-convergence benefit in other channel models, which will support more diverse transmission/storage services.%\vspace{-1mm}

Different form the conventional point-to-point communications, multiple access is a critical technology to the future massive IoT network. The existing orthogonal multiple access technologies divide the resource elements (e.g., time and frequency) orthogonally to avoid the inter-user interference, but sacrifice the spectrum efficiency. To circumvent this limitation, NOMA [130]-[132] becomes a promising solution, which can assign limited resource elements to support more users at the cost of inducing inter-user interference. Therefore, SC-PLDPC-BICM will be an appealing technique for boosting the error performance of NOMA-aided wireless communication systems.

\subsection{Receiver Design of PLDPC-BICM}

To date, most relevant work has been devoted to tackling design issues from the transmitter perspective for PLDPC-BICM systems. However, the optimization and implementation of receivers, including detection and decoding algorithms, have been largely ignored during the past decade. In fact, a detection/decoding algorithm is an inevitable technique to ensure high performance and hardware-friendly implementation of the BICM systems. Although there are some attempts to develop novel receiver designs for BICM systems \cite{schreckenbach2007iterative,7105928,7589683,7887724,7339431}, these algorithms do not consider the salient features of protograph and spatial coupling. Aiming to improve the error performance and to reduce the computational complexity, high-efficiency detection/decoding algorithms for the PLDPC-BICM are worth exploring. Besides, as receiver does not know the CSI in actual implementation, channel estimation becomes a critical issue in detector design.
Then, based on the characteristics of PLDPC codes and interleaver, using the soft information (i.e., bit-level LLRs) output from the decoder to assist the channel estimator is expected to further improve the system performance.
An emerging solution is to incorporate learning techniques \cite{9427170} into receiver designs.

The hybrid-automatic-repeater-request (HARQ) \cite{DCai3} PLDPC-BICM will be a promising technique to realize high-reliability and high-throughput transmissions over non-ergodic channels. The transmitter first sends the symbols associated with the coded bits that are relatively easier to be recovered (i.e., the variable nodes with relatively high degrees in the parity check matrix). The receiver exploits the cyclic-redundancy check and HARQ to implement decoding, then the undecodable codeword in the previous transmission slot can combine with new incremental parity check bits to decode again.

The decoding performance of BICM flash-memory systems is highly dependent on the reliability of read voltages that produce initial LLRs for the PLDPC decoder. However, the existing detection schemes, such as MMI and voltage-entropy-based detection schemes \cite{6804933,7416649}, require the knowledge of threshold-voltage distributions in order to optimize the read voltages, and thus dramatically increase the implementation complexity. As a remedy, novel learning-aided detection schemes (e.g., read-voltage optimization schemes), which can achieve excellent performance without requiring threshold-voltage distributions, are needed to be developed for such systems.\vspace{-1mm}

\subsection{New Application Scenarios of PLDPC-BICM}

\subsubsection{Ultra-High-Density Flash-Memory Systems}

Although the design methodologies for high-rate PLDPC codes and bit mappers over BICM MLC flash-memory channels with regular mapping and IM have been extensively investigated, as discussed in Section~\ref{sect:VIII}, the optimization of PLDPC-BICM TLC/QLC flash-memory systems have not been completely studied. Since these higher-density flash-memory systems suffer from severer circuit-level noise, more errors will occur. Consequently, it is indispensable to develop excellent code-construction methods and bit-mapping schemes to guarantee the storage reliability for PLDPC-BICM TLC/QLC flash-memory systems.

Recently, three-dimensional ($3$D) NAND flash memory has emerged as a new non-volatile storage medium. This type of flash-memory devices have already been adopted in some electronic and communication devices. Compared with the planar flash memory, $3$D flash memory can significantly increase the storage capacity and decrease the storage cost, but it is prone to severer interference and noise. LDPC codes have been demonstrated to have outstanding error-correction capability in $3$D TLC flash-memory systems \cite{8936476,8634908}, which may inspire more research efforts on the design of low-complexity PLDPC codes as well as robust BICM schemes. %Some related issues are further discussed below.

\subsubsection{Underwater Acoustic Communication Systems}

Underwater acoustic channel represents another typical channel model for practical wireless communications. The underwater acoustic channel, which is commonly described as a doubly-selective fading channel, is viewed as one of the most complicated wireless-communication environments. It is extremely difficult to establish design criteria for capacity-approaching FEC codes together with their BICM schemes in such scenarios. Recently, PLDPC
codes were applied to significantly enhance the robustness against fading and noise incurred by the underwater acoustic channel \cite{8242532,9056068}. To achieve higher throughput, the PLDPC-BICM-ID is expected to outperform the algebraic-coded counterparts \cite{5887355} and to be a more favorable transmission solution for underwater acoustic communications.\vspace{-1.5mm}

\subsection{New Types of BICM Frameworks}

\subsubsection{Spatial-Modulation-aided PLDPC-BICM}

Spatial modulation can realize high spectral and energy efficiency with relatively simple design, which has been explosively investigated since $2008$. In contrast to the conventional modulations, like $M$PSK/$M$QAM, the space-domain and other signal-domain (e.g., frequency/time/code/angle-domain) resources have been integrated into the spatial modulation, thereby achieving more desirable performance \cite{8765384,8315127}. Although the LDPC coding techniques have been applied to high-order spatial-modulation systems \cite{6920539,8290977,7383250}, there is a lack of systematic design guideline for the BICM systems involving PLDPC codes and spatial modulations.
Actually, the joint design and optimization of PLDPC code, bit-to-symbol mapping and index scheme has a significant effect on the performance of spatial-modulation systems.
Therefore, it is important to conceive some sophisticated methods for intelligently integrating the PLDPC codes into spatial-modulation systems, which may yield new types of capacity-approaching BICM schemes.

\subsubsection{Joint PLDPC-and-PNC BICM}

Relay-aided cooperation is a crucial enabling technique for future wireless-communication applications, especially for PNC systems and IoT. Specifically, this technique can be exploited to enhance the spatial diversity, thus it has been widely applied in low-power and low-complexity wireless-communication systems. Inspired by these advantages, several robust constructions were developed for the PLDPC codes in BPSK-aided relay systems \cite{6497022,7080854,8544026,8019817}. While PLDPC codes have recently been deployed to PNC BICM systems over TWR channels \cite{7723910}, their optimization is still unexplored. Especially, how to develop capacity-approaching joint PLDPC-and-PNC algorithms tailored for BICM is still a challenging issue to be resolved.

\subsubsection{Polar-Coded BICM}

As the first type of capacity-achieving FEC codes, polar codes are superior to LDPC codes in certain cases, hence providing a promising error-correction solution for communication and storage \cite{8962344}. In $2018$, polar codes were selected as the coding scheme for the control channel in $5$G radios. Some research efforts have been made to establish a series of polar-coded BICM systems so as to achieve bandwidth-efficient transmissions in communication and storage systems \cite{7121013,6952075}. In this direction, a number of research issues (e.g., encoder/decoder design, constellation and bit-mapper optimization) are waiting for further investigation to fully exploit the polarization feature towards optimal BICM performance.\vspace{-0.5mm}

%\iffalse
\section{Concluding Remarks}\label{sect:IXI}%\vspace{-0.5mm}

%\subsection{Concluding Remarks}\label{sect:IX-A}

This article provides an overview of an promising bandwidth-efficiency BICM technique, namely the PLDPC-BICM. During the past fifteen years, the PLDPC-BICM has drawn a great deal of attention from both academic and industrial communities as it is capable of achieving capacity-approaching performance with low implementation complexity. To offer a comprehensive review, the evolution of the PLDPC-BICM was traced from the deployment of PLDPC codes to the application of their SC-PLDPC relatives. These two types of codes can be conveniently represented by small-size protographs, which lead to easy design, analysis and implementation. The inherited benefits allow the PLDPC codes to be seamlessly combined with various high-order modulations, thus making the resultant BICM attractive for modern communication and storage systems including deep-space communication systems, satellite broadcasting systems, optical communication systems, wireless communication systems, and flash-memory-aided data-storage systems.

In order to give a full picture of this research field, most of the representative design and analysis methodologies were presented in relation to the PLDPC-BICM with many illustrative examples. The code construction, constellation shaping and bit-mapper optimization were discussed in several typical transmission environments, including AWGN, fading, Poisson PPM, flash-memory channels, assuming a serially concatenated decoding framework which comprises a soft-decision demodulator and a BP decoder at the receiver. More than two hundred and thirty technical papers were carefully selected, so as to substantially elucidate the recent research progress and emerging development trends in the regard of PLDPC-BICM systems.
%
%It should be noted that the study of SC-PLDPC-coded BICM systems over Poisson PPM channels and flash-memory channels are relatively unexplored to date.

As one type of structured LDPC codes, the PLDPC codes exhibit capacity-approaching decoding thresholds and desirable linear-minimum-distance-growth property in BICM systems under different channel conditions. Although the presentation is limited to several typical channel models, the corresponding design and analysis techniques offer certain universal principles effective for more generalized channel models. Owing to the superb performance and simple implementation, PLDPC-BICM has been widely recognized as a competitive transmission scheme for a myriad of digital-communication and data-storage applications. Beyond any doubt, the PLDPC-BICM will be adopted by more emerging practical use and industry standards.% related to communication and storage applications.
\vspace{-0.3mm}
%\newpage

\section{Acknowledgements}\label{sect:IXI}%\vspace{-0.5mm}

This work was supported in part by the NSF of China under Grants 62071131, the Guangdong Basic and Applied Basic Research Foundation under Grant 2022B1515020086, the International Collaborative Research Program of Guangdong Science and Technology Department under Grant 2022A0505050070, in part by the Open Research Fund of the State Key Laboratory of Integrated Services Networks
under Grant ISN22-23, and the Industrial R\&D Project of Haoyang Electronic Co., Ltd. under Grant 2022440002001494, in part by Singapore University of Technology Design under ``Advanced Error Control Coding for 6G URLLC and mMTC'' Grant FCP-NTU-RG-2022-020, and A*STAR under ``Campus Wide V2X Integrated System Platform'' Grant SERC A19D6a053.

%\bibliographystyle{IEEEtran}
%\bibliography{IEEEabrv,bib/Reference}

%\begin{thebibliography}{100}
%\providecommand{\url}[1]{#1}
%\csname url@samestyle\endcsname
%\providecommand{\newblock}{\relax}
%\providecommand{\bibinfo}[2]{#2}
%\providecommand{\BIBentrySTDinterwordspacing}{\spaceskip=0pt\relax}
%\providecommand{\BIBentryALTinterwordstretchfactor}{4}
%\providecommand{\BIBentryALTinterwordspacing}{\spaceskip=\fontdimen2\font plus
%\BIBentryALTinterwordstretchfactor\fontdimen3\font minus
%  \fontdimen4\font\relax}
%\providecommand{\BIBforeignlanguage}[2]{{%
%\expandafter\ifx\csname l@#1\endcsname\relax
%\typeout{** WARNING: IEEEtran.bst: No hyphenation pattern has been}%
%\typeout{** loaded for the language `#1'. Using the pattern for}%
%\typeout{** the default language instead.}%
%\else
%\language=\csname l@#1\endcsname
%\fi
%#2}}
%\providecommand{\BIBdecl}{\relax}
%\BIBdecl

%%%%%%%%%%%%%%%%%%%%%%%%%%%%%%%%%%%%%%%%%%%%%%%%%%%%%%%%%%%%%%%%%%%%%%%%%%%%%

\end{document}